\documentclass[preprint,12pt]{elsarticle}

\usepackage{amssymb}
\usepackage{threeparttable}
\usepackage{float}
\usepackage{graphicx}
\usepackage{subfig}
\biboptions{numbers,square,sort&compress}
\journal{Physics Reports}

\begin{document}

\begin{frontmatter}



\title{Neutron-Antineutron Oscillations: Theoretical Status and Experimental Prospects}


\author[ncu,tri]{D. G. Phillips II}
\author[iub,ceem]{W. M. Snow\corref{cor1}}
\ead{wsnow@indiana.edu}
\author[osu]{K. Babu}
\author[snp]{S. Banerjee}
\author[iub,ceem]{D. V. Baxter}
\author[inf,lau]{Z. Berezhiani}
\author[ucd]{M. Bergevin}
\author[snp]{S. Bhattacharya}
\author[cun]{G. Brooijmans}
\author[utn]{L. Castellanos}
\author[uci]{M-C. Chen}
\author[utk]{C. E. Coppola}
\author[wau]{R. Cowsik}
\author[orl]{J. A. Crabtree}
\author[vec]{P. Das}
\author[ncu,tri]{E. B. Dees}
\author[itp,nov,ufi]{A. Dolgov}
\author[orl]{P. D. Ferguson}
\author[utk]{M. Frost}
\author[utk]{T. Gabriel}
\author[huj]{A. Gal}
\author[orl]{F. Gallmeier}
\author[csu]{K. Ganezer}
\author[inr]{E. Golubeva}
\author[utk]{G. Greene}
\author[csu]{B. Hartfiel}
\author[nnu]{A. Hawari}
\author[utn]{L. Heilbronn}
\author[iub]{C. Johnson}
\author[utk]{Y. Kamyshkov}
\author[itp,mip]{B. Kerbikov}
\author[nuj]{M. Kitaguchi}
\author[utf]{B. Z. Kopeliovich}
\author[inr,mip]{V. B. Kopeliovich}
\author[inr]{V. A. Kuzmin}
\author[iub,ceem]{C-Y. Liu}
\author[lnl]{P. McGaughey}
\author[lnl]{M. Mocko}
\author[umd]{R. Mohapatra}
\author[fnl]{N. Mokhov}
\author[lnl]{G. Muhrer}
\author[nis]{H. P. Mumm}
\author[itp]{L. Okun}
\author[ncu,tri]{R. W. Pattie Jr.}
\author[fnl]{C. Quigg}
\author[fnl]{E. Ramberg}
\author[vec]{A. Ray}
\author[iua]{A. Roy}
\author[utn]{A. Ruggles}
\author[prl]{U. Sarkar}
\author[lnl]{A. Saunders}
\author[spn]{A. P. Serebrov}
\author[nuj]{H. M. Shimizu}
\author[sny]{R. Shrock}
\author[vec]{A. K. Sikdar}
\author[lnl]{S. Sjue}
\author[fnl]{S. Striganov}
\author[utn]{L. W. Townsend}
\author[fnl]{R. Tschirhart}
\author[umm]{A. Vainshtein}
\author[iub]{R. Van Kooten}
\author[lnl]{Z. Wang}
\author[ncu,tri]{A. R. Young}
\cortext[cor1]{Corresponding author}
\address[csu]{California State University, Dominguez Hills, Department of Physics, 1000 E. Victoria St., NSMB-202, Carson, CA 90747, USA}
\address[ceem]{Center for the Exploration of Energy and Matter, 2401 Milo B. Sampson Lane, Bloomington, IN 47408, USA}
\address[cun]{Columbia University, Department of Physics, 538 West 120th St., 704 Pupin Hall, MC 5255, New York, NY 10027, USA}
\address[fnl]{Fermi National Accelerator Laboratory, P. O. Box 500, Batavia, IL 60510, USA}
\address[iub]{Indiana University, Department of Physics, 727 E. Third St., Swain Hall West, Room 117, Bloomington, IN 47405, USA}
\address[inr]{Institute for Nuclear Research, Russian Academy of Sciences, Prospekt 60-letiya Oktyabrya 7a, Moscow 117312, Russia}
\address[itp]{Institute of Theoretical and Experimental Physics, Bolshaya Cheremushkinskaya ul. 25, 113259 Moscow, Russia}
\address[iua]{Inter University Accelerator Centre, 110067 Aruna Asaf Ali Marg, Vasantkunj Sector B, New Dehli, DL, India}
\address[inf]{Istituto Nazionale de Fisica Nucleare, Laboratori Nazionali Gran Sasso, 18, Strada Statale 17 Bis, 67100 Assergi, L'Aquila, Italy}
\address[lnl]{Los Alamos National Laboratory, P. O. Box 1663, Los Alamos, NM 87545, USA}
\address[mip]{Moscow Institute of Physics and Technology, 9 Institutskiy per., Dolgoprudnyi, Moscow Region, 141700, Russia.}
\address[nuj]{Nagoya University, Department of Physics, School of Science, Furo-cho, Chikusa-ku, Nagoya, Aichi 464-8602, Japan}
\address[nis]{National Institute of Standards and Technology, 100 Bureau Dr., Stop 1070, Gaithersburg, MD 20899, USA}
\address[nnu]{North Carolina State University, Department of Nuclear Engineering, 2500 Stinson Drive, 3140 Burlington Engineering Labs, Raleigh, NC 27695, USA}
\address[ncu]{North Carolina State University, Department of Physics, 2401 Stinson Drive, Box 8202, Raleigh, NC 27695, USA}
\address[nov]{Novosibirsk State University, 2 Pirogova Str., Novosibirsk, 630090, Russia}
\address[orl]{Oak Ridge National Laboratory, Spallation Neutron Source, Bear Creek Rd., Oak Ridge, TN 37830, USA}
\address[osu]{Oklahoma State University, Department of Physics, 145 Physical Sciences Bldg., Stillwater, OK 74078, USA}
\address[prl]{Physical Research Laboratory, Navrangpura, Ahmedabad, Gujarat, 380009, India}
\address[huj]{Racah Institute of Physics, The Hebrew University, Jerusalem 91904, Israel}
\address[snp]{Saha Institute of Nuclear Physics, Block - AF, Sector - 1, Bidhannagar, Kolkata, 700064, West Bengal, India}
\address[spn]{St. Petersburg Nuclear Physics Institute, Orlova Roscha, Gatchina, Leningrad District, 188300, Russia}
\address[sny]{Stony Brook University, Physics and Astronomy Department, Stony Brook, NY 11794, USA}
\address[tri]{Triangle Universities Nuclear Laboratory, P. O. Box 90308, Durham, NC 27708, USA}
\address[lau]{Dipartimento di Fisica, Universit\`{a} dell'Aquila, Via Vetoio, 67100 Coppito, L'Aquila, Italy}
\address[ucd]{University of California, Department of Physics, One Shields Ave., Davis, CA 95616, USA}
\address[uci]{University of California, Department of Physics, 4129 Frederick Reines Hall, Irvine, CA 92697, USA}
\address[ufi]{Dipartimento de Fisica e Scienze della Terra, Universit\`{a} degli Studi di Ferrara, Polo Scientico e Tecnologico - Edificio C, Via Saragat 1, 44122 Ferrara, Italy}
\address[umd]{University of Maryland, Department of Physics, 1117 John S. Toll Building $\#$082, College Park, MD 20742, USA}
\address[umm]{William I. Fine Theoretical Physics Institute, School of Physics and Astronomy, 116 Church St. S.E., University of Minnesota, Minneapolis, MN 55455, USA}
\address[utf]{Departamento de F\'{i}sica, Universidad T\'{e}cnica Federico Santa Mar\'{i}a, Centro de Estudios Subat\'{o}micos, and Centro Cient\'{i}fico-Tecnol\'{o}gico de Valpara\'{i}so, Av. Espa\~{n}a 1680, Casilla 110-Valpara\'{i}so, Chile}
\address[utn]{University of Tennessee, Department of Nuclear Engineering, 315 Pasqua Nuclear Engineering, Knoxville, TN 37996, USA}
\address[utk]{University of Tennessee, Department of Physics, 401 Nielsen Physics Building, 1408 Circle Drive, Knoxville, TN 37996, USA}
\address[vec]{Variable Energy Cyclotron Centre, Block - AF, Sector - 1, Bidhannagar, Kolkata, 700 064, West Bengal, India}
\address[wau]{Washington University, Department of Physics, One Brookings Drive, Campus Box 1105, St. Louis, MO 63130, USA}

\begin{abstract}
The observation of neutrons turning into antineutrons would constitute a
discovery of fundamental importance for particle physics and cosmology. Observing the $n-\bar n$ transition would show that baryon number ($\mathcal{B}$) is violated by two units  and that matter
containing neutrons is unstable. It would {provide a clue to how} the matter in
our universe might have evolved from the $\mathcal{B} = 0$ {early universe}. {If seen at rates observable in foreseeable next-generation
experiments,} it might well help us understand the observed baryon asymmetry of the universe. A
demonstration of the violation of $\mathcal{B-L}$ by 2 units would have a
profound impact on our understanding of phenomena beyond the Standard Model of
particle physics.  

%A null result at the anticipated sensitivity would
%correspond to {a lower} limit on {the} matter{-}instability {lifetime} of about
%$10^{35}$ years{.}  {C}ombined with data from the Large Hadron Collider and
%other experimental searches for rare processes it {would severely constrain} the possibility
%of baryogenesis below the electroweak phase transition involving
%first{-}generation quarks, one of the few experimentally-testable ideas for
%baryogenesis.

{Slow neutrons have kinetic energies of a few meV.  By exploiting new slow neutron sources and
optics technology developed for materials research, an optimized search for oscillations
using free neutrons from a slow neutron moderator could improve existing limits on the free oscillation probability by at least three orders of magnitude. Such an experiment would deliver
a slow neutron beam through a magnetically-shielded vacuum chamber to a thin
annihilation target surrounded by a low-background antineutron annihilation
detector.} {Antineutron annihilation in a target downstream of a free neutron beam is such a spectacular experimental signature that an essentially background-free search is possible.}  {An authentic} positive
 {signal} can be  {extinguished} by a very small change in the ambient
magnetic field in such an experiment. It is also possible to improve the sensitivity of neutron oscillation searches in nuclei using large underground detectors built mainly to search for proton decay and detect neutrinos. 

This paper summarizes the relevant theoretical developments, outlines
some ideas  {to improve} experimental searches  {for} free neutron oscillations, and suggests avenues both for theoretical investigation and for future improvement in the experimental sensitivity.
\end{abstract}


\end{frontmatter}

\tableofcontents
% \linenumbers

\section{Introduction}
\label{nnbar:sec:intro}
%%%%%%%%%%%%%%%%%%%%%%%%%%%%%%%%%%%%%%%%%%%%%%%%%%%%%%%%%%%%

Oscillations of electrically neutral particles into other  {species} are no longer a surprising phenomenon in particle physics. Neutrino and {flavored neutral meson oscillations ($K^0, D^0, B^0, B_s$)} behave like natural interferometers which continue to teach us about aspects of physics (lepton{-}number violation, $CP$-violation, neutrino mass)  {that} are not otherwise {readily} accessible experimentally. The high sensitivity of oscillations to rare processes arises from the interference between the relative phases for the relevant amplitudes. Since neutrons and antineutrons are electrically neutral, as is indicated by all
available data~\cite{Bressi11,Baumann88}, {only}  the conservation of baryon number forbids a neutron ($\mathcal{B}=1$) from
transforming into an antineutron ($\mathcal{B}=-1$). The long $\beta$-decay lifetime of the free neutron makes a sensitive search for neutron-antineutron ($n - \bar n$) oscillations possible. In this paper we discuss the methods by which searches for $n - \bar n$ oscillations can be
conducted and the possibility for significant improvement in the experimental sensitivity to this process.

 {The decay of} the lightest known charged particle, {the} electron{,} into
 other particles {is forbidden by} conservation of electric charge{.} {Many
   channels are open for proton decay, so electric charge conservation does not
   account for proton stability.} {T}o understand the stability of the hydrogen
 atom, the existence of a conserved {additive} quantum number, called baryon
 number, for both protons and neutrons (with their antiparticles {assigned}
 opposite baryon number) was proposed long ago~\cite{Stueckelberg38}.  {All
   laboratory experiments to date support exact baryon number conservation
   under the experimental conditions probed (which are at temperatures $T <<
   v_{EW}$, so SU(2) instanton/sphaleron violation of $\mathcal{B}$ is completely
   negligible). This conservation law does not follow from any known physical principle
   but is inferred from experiment. At the level of the Standard Model (SM)
   Lagrangian, baryon number conservation is ``accidental,''} {in the sense
   that it depends on the specific gauge groups and matter content of the
   theory.} {Lepton number, $\mathcal{L}$, is also accidentally conserved in
   the same sense in the SM Lagrangian. Note that baryon number and
   lepton number are exactly conserved only in perturbation theory;
 ``sphaleron'' effects involving non-perturbative
   electroweak gauge field configurations can lead to the creation or
   destruction of baryon and lepton number while preserving the difference
   $\mathcal{B} - \mathcal{L}$.} {However, this non-perturbative process is so
   strongly suppressed in our present vacuum that there is absolutely no hope
   that it will induce a $n - \bar n$ transition in the laboratory.}

{Does anything stand in the way of explicit, Lagrangian-level $\mathcal{B}$ or
  $\mathcal{L}$-violating interactions? In quantum field theory, Noether's
  theorem links conservation laws to continuous global symmetries. In the
  example we know best, conservation of electric charge is linked to a global
  $\mathrm{U(1)_{em}}$ phase symmetry of the Lagrangian. Promoting
  $\mathrm{U(1)_{em}}$ to a local (gauge) symmetry leads to quantum
  electrodynamics, in which the massless photon couples minimally to the
  conserved current mandated by the global symmetry. A global
  $\mathrm{U(1)_{\mathcal{B}}}$ symmetry could enforce $\mathcal{B}$
  conservation, but we have no experimental evidence for the long-range interaction coupled
  to baryon number that would arise from a local $\mathrm{U(1)_{\mathcal{B}}}$
  symmetry~\cite{Schlamminger08}.}  Besides, there are many examples where
extensions of the SM, such as grand unified theories, lead to
violation of baryon number.  {This expectation meshes} with our present
{conception} of {early universe} cosmology. Even if the universe began with no
net baryon number, the 60 - 70 {rapid} $e$-foldings of the expansion of {the
  cosmological scale parameter, by which the inflationary paradigm accounts for
  the remarkable homogeneity of the cosmic microwave background,} {would}
tremendously dilute the number density of any particles which carry conserved
quantum numbers{.}  {Yet observations indicate that in our region of the
  universe, the density of antibaryons is negligible, whereas the average
  density of matter, characterized by the baryon-to-photon ratio $\eta$ in the
  present universe, also known as the baryon asymmetry,}
\begin{equation}
5.7 \times 10^{-10} \le \eta \le  6.7 \times 10^{-10} \quad\hbox{(95\% C.L.),}
\label{eq:etab2014}
\end{equation}
{is small but nonzero~\cite{Fields12}.}  {T}he possibility of calculating this
number directly from the laws of physics and cosmology and thereby explain{ing}
the ultimate origin of most of the (baryonic) matter around us is very
exciting. It would be the cosmological equivalent of our understanding of the
stellar fusion origin of most nuclei starting mainly from hydrogen and helium.
This circumstance provides a new impetus both to the search for laboratory
evidence of baryon-number violation and to theories which can explain the
dynamics which lead to the present size of the cosmological baryon asymmetry.

This review summarizes the state of theoretical knowledge and experimental
constraints on $n - \bar n$ oscillations{,} and discusses recent developments
in slow neutron technology {that} make it possible to improve the present limit
on the free neutron oscillation probability by a few orders of magnitude.  {A}
null result {at this level} would place the most stringent limit on this
possible mode of matter instability. As {we shall discuss} in \S2.3.2, the
practical experimental figure of merit for a free neutron $n - \bar n$ search
is $N_n t^2$, where $N_n$ is the total number of free neutrons observed in the
experiment and $t$ is the observation time for free neutron propagation.  {The experiment would deliver} a high flux of free neutrons from
a slow neutron source through a {long} vacuum vessel with good magnetic
shielding to a thin target to absorb the antineutron{,} surrounded by an
antineutron annihilation detector. We argue that an improved experiment with a
scientifically interesting reach requires a dedicated beamline optimized for
the production of slow neutrons and their delivery to an antineutron
annihilation target using modern neutron moderator and {neutron} optics
technology.

The delivery rate of slow neutrons to the annihilation target {represents the single most important factor in the experimental sensitivity.} {It} can be increased by maximizing the phase space acceptance for
neutron extraction around the cryogenic converter with advanced neutron
optical supermirrors, whose performance (especially the neutron phase space
acceptance) far exceeds that available to previous free neutron experiments.
 {Optimizing} the flux and brightness of the cold neutron source
fed by the spallation target, the supermirror optics, the vacuum and magnetic shielding, and the annihilation detector for an oscillation experiment
 {will need} detailed R\&D studies, but no new technology is required. The
speed of the neutrons falls within the regime where gravity has a significant
influence on their trajectories, and therefore the orientation of
the apparatus relative to {E}arth's gravitational field is important.

It is also possible for neutrons in nuclei to oscillate into antineutrons, which would immediately annihilate inside the nucleus. Searches for $\Delta \mathcal{B}=2$ processes using free neutron oscillations and neutron conversion in nuclei in underground detectors are complementary on many levels. Of course both processes require the presence of some effective operator which violates $\mathcal{B}$ by 2 units. For free neutron oscillations, however, there is the additional requirement that the neutron and antineutron energies are degenerate within the limit set by the Heisenberg uncertainty principle over the observation time of their flight from source to detector (this is the so-called \lq\lq quasi free\rq\rq condition). In the case of neutron conversion inside the nucleus, this degeneracy is removed by the large difference in neutron and antineutron optical potentials which partly arises from the strong-interaction antineutron annihilation 
occurring in matter. This means that any practical measurement under these conditions is 
insensitive to any sources of much smaller effects which can break this degeneracy. This makes the free neutron oscillation experiment sensitive to extremely small effects from exotic physics which can lift the neutron-antineutron energy degeneracy such as $CPT$/Lorentz violation~\cite{Babu2015}. A free neutron oscillation experiment is sensitive only to specific $\Delta \mathcal{B}=2$ operators at the nucleon level which lead to a free antineutron in the final state, whereas a search for neutron conversion in nuclei can access other $\Delta \mathcal{B}=2$ operators. For these reasons, a scientific program which aims to discover or constrain $\Delta \mathcal{B}=2$ processes should endeavor to search in both of these channels. We therefore also include a summary of the present limits on neutron-antineutron oscillations in nuclei from analysis of data from underground detectors and some discussion of future opportunities in this area in both theory and experiment. 

 {Three main observations guide the selection of material in this review.}
 First, a free $n - \bar n$ oscillation experiment offers a greatly improved
 sensitivity to $\mathcal{B}$-violation combined with the possibility of robust
 background suppression. We argue that this combination of features is unique
 among all other known experimental approaches to $\mathcal{B}$-violation using
 existing technology for the foreseeable future and therefore deserves to be
 brought to the attention of the broader scientific community. Second, there is a
 realistic possibility {that a stringent lower bound on the $n - \bar n$
   oscillation rate, combined with Large Hadron Collider (LHC) results and constraints on rare
   processes from other experiments, could place significant constraints on that subset of
   baryogenesis scenarios which lie within the reach of experiment, such as
   electroweak baryogenesis and post-sphaleron baryogenesis. {Third,} the set
   of slow neutron technologies {that enable} this more sensitive search for
   free neutron oscillations is not widely known {among particle physicists,}
   as most of the {development has occurred at} neutron scattering user facilities {that focus on}
   materials research, so there is a need for a presentation which introduces a
   broader set of scientists to the physics behind these developments.
 
The rest of this article is organized as follows:  In \S2 we summarize the scientific motivation for the search for $n - \bar n$ oscillations.  A review of previous searches and prospects for improved searches in free neutron and intranuclear experiments are given in \S3 and \S4 respectively.  The cold neutron source and optics needs for an improved free neutron search are detailed in \S5.  We address the important design issues for a new detector in \S6.  \S7 outlines research opportunities that can advance $n - \bar n$ physics. \S8 is a compact summary.  {Primers} on
  slow neutron sources, {neutron} moderation, and {neutron} optics  {are given in three appendices.}

%%%%%%%%%%%%%%%%%%%%%%%%%%%%%%%%%%%%%%%%%%%%%%%%%%%%%%%%%%%%
\section{Physics Motivation for $n - \bar n$ Searches}
\label{nnbar:sec:physics}
%%%%%%%%%%%%%%%%%%%%%%%%%%%%%%%%%%%%%%%%%%%%%%%%%%%%%%%%%%%%

There are many compelling reasons to think that fundamental particle
interactions violate baryon number. Arguably the most powerful reason is that
generating the observed matter-antimatter asymmetry in the universe
requires that baryon number must be violated. The observed baryon asymmetry of
the universe is often taken as indirect evidence for $\mathcal{B}$-violating
processes in nature, since it is strongly suspected on theoretical grounds that
the baryon number in the very early universe was effectively zero. One might
naively expect that one can simply insert any needed $\mathcal{B}$ asymmetry
into the initial conditions of the universe by hand as an arbitrary initial
condition. This possible explanation is in
conflict with the apparent need for an inflationary period in the early history
of the universe~\cite{Dolgov92, Dolgov98}. In this case inflation would drive the
number density of any particles carrying a globally-conserved quantum number
like $\mathcal{B}$ or $\mathcal{L}$ to zero at very early times. Subsequent
$\mathcal{B}$-violating processes are then needed to generate the baryon
asymmetry of the universe as it expands and cools. For this reason
the present predominance of matter over antimatter in the universe is widely
interpreted as indirect evidence for $\mathcal{B}$-violation.

As Sakharov pointed out long ago~\cite{Sak67}, this circumstance gives one hope
that the baryon asymmetry is calculable from first principles.  He noted
that a simultaneous combination of $\mathcal{B}$-violation, $C$ and $CP$ violation (or equivalently time reversal violation, assuming $CPT$ invariance), and
a deviation of the number densities of particle species in the universe from
thermal equilibrium could do it. Subsequent theoretical work showed that there are many other ways to generate the baryon asymmetry~\cite{Dolgov92}. For example, baryogenesis with broken $CPT$ has been considered~\cite{Zeldovich81, Dolgov10} and constrained in a subsequent analysis~\cite{Dolgov14}.  

This argument does not specify a mechanism for
the generation of the asymmetry. Within this framework, any proposed
baryogenesis mechanism can be characterized in part by the point in time since
the Big Bang (or the temperature of the universe) at which this asymmetry
arises. In cosmological models this matter-antimatter asymmetry is
parameterized by the dimensionless baryon to photon ratio $\eta = n_{B}/
n_{\gamma}=(6.19 \pm 0.14) \times 10^{-10}$ determined independently from Big
Bang Nucleosynthesis (BBN) and measurements of the power spectrum of
fluctuations in the cosmic microwave background~\cite{Bennett03}.

The fact that the SM allows for nonperturbative processes involving
SU(2) instantons (sphalerons)  which
violate $\mathcal{B}$~\cite{Hooft76,Hooft78,Kuzmin85} is of
fundamental importance in trying to sort out the possible mechanisms to
generate the baryon asymmetry~\cite{Ramsey_Musolf12}.  These nonperturbative
tunneling transitions in the electroweak sector can change
$\mathcal{B}$ into $\mathcal{L}$ and vice-versa, but conserve the difference
$\mathcal{B}-\mathcal{L}$. Since the violation of ${\cal B}$ and ${\cal L}$ by
sphalerons in the SM is exponentially suppressed at temperatures $T << v_{EW}$, these global symmetries
are effectively preserved in the SM at zero temperature. Even though the rates
for nonperturbative processes which lead to $\mathcal{B}$-violation are
exponentially suppressed in the present ground state of the universe, they
should have been thermally excited in the early universe and can become
important for baryogenesis at temperatures near the electroweak phase
transition, which is expected to occur at $T\approx 170$ GeV~\cite{Kuzmin85,
  Shaposhnikov87}.  Unfortunately it seems that this known $\mathcal{B}$-violation mechanism combined with the $CP$-violation known to exist in the
SM, along with the degree of deviation from thermal equilibrium at
the phase transition, yields a prediction for $\eta$ that is many orders of
magnitude smaller than the observed value, when calculated within the standard
Big Bang scenario. We therefore suspect that there must be some additional
physics beyond the SM that greatly increases $\eta$, adding a new
mechanism to one or more of the Sakharov conditions. Beyond the requirement
that all three ingredients must be present and active at the same time since
the Big Bang, however, this argument does not specify the mechanism
responsible.

The existence of these sphaleron processes naturally divides
models of baryogenesis into three qualitatively different regimes, according to
whether or not the $\mathcal{B}$ generation comes from (a) a post-inflation
$\mathcal{L}$ asymmetry from a higher scale that is converted by sphalerons
into a $\mathcal{B}$ asymmetry (this general idea is referred to as
leptogenesis~\cite{Fukugita86}), (b) the phase transition dynamics at the electroweak scale
itself in the absence of a pre-existing $\mathcal{L}$ asymmetry, coupled with
new sources of $CP$-violation at or near this scale (this general idea is called
electroweak baryogenesis), or (c) some new $\mathcal{B}$-violating process at a
scale below the electroweak phase transition (referred to as post-sphaleron
baryogenesis). The leptogenesis scenario that violates $\mathcal{L}$ at high energy scales is favored at present by many theories, but it seems difficult to test using laboratory measurements.  More attention has been devoted recently to scenarios wherein leptogenesis could occur at the TeV scale where it can be testable.  The last two scenarios occur
at energy scales low enough to be probed experimentally through searches for
nonstandard $\mathcal{B}$-violation and $CP$-violation.  Electroweak
baryogenesis, which provides an example of models of class (b), remains quite viable: see this recent review of electroweak baryogenesis models~\cite{Ramsey_Musolf12}. Some versions of electroweak baryogenesis are now under pressure from the recent discovery of the Higgs boson at
the relatively low mass of 125 GeV. (Electroweak baryogenesis in the
minimal supersymmetric SM (MSSM) requires that the scalar
partner of the top quark must have a mass below 120 GeV for which there seems
to be no evidence yet at the LHC).  An example of class (c) models is the
scenario of post-sphaleron baryogenesis, which is most likely to occur below
the electroweak phase transition temperature.  Such a low temperature
essentially requires that baryogenesis must proceed via higher dimensional
$\mathcal{B}$-violating operators, the most likely example of which comes from
the six-quark operator that leads to a $\Delta\mathcal{B}=2$ process; constraints on $\Delta\mathcal{B}=1$ processes from proton-decay
searches are too stringent to generate nonzero $\mathcal{B}$ at such a low
scale, and $\Delta\mathcal{B}=3$ processes would be in conflict with
nucleosynthesis. Neutron-antineutron oscillations, which come from
$\Delta\mathcal{B}=2$ operators, are therefore especially interesting as a
probe for low-scale $\mathcal{B}$-violation relevant to the baryon asymmetry.

Other arguments support the idea that $\mathcal{B}$ may be violated through
additional mechanisms specific to the structure of the SM
of particle physics. In the SM both $\mathcal{B}$ and $\mathcal{L}$
are globally conserved quantities in perturbation theory. Since the proton is
the lightest baryon, these conservation laws explain why the proton does not
decay. However it is obvious upon inspection that the SM cannot
possibly represent the final word on particles and forces. In particular there
is no reason why one would expect either $\mathcal{B}$ or $\mathcal{L}$ quantum
numbers to be conserved on general grounds, even perturbatively.  There is no direct experimental evidence for the corresponding gauge field that one introduces
for all other charges in the SM in order to enforce the local
conservation of charge demanded by relativity. This circumstance is sometimes
expressed by saying that $\mathcal{B}$ and $\mathcal{L}$ are ``accidental''
symmetries of the SM. One therefore expects generically that
neither $\mathcal{B}$ nor $\mathcal{L}$ is conserved perturbatively in the
real underlying theory. A true understanding of the physics of baryon number
violation requires comprehensive knowledge of the underlying symmetry
principles, with distinct selection rules corresponding to different
complementary scenarios for grand unification and for the generation of baryon
asymmetry of the Universe.

Grand unified theories of matter and forces, which are prime candidates for
physics beyond the SM, predict violation of baryon number. A grand unified theory is one
in which the Standard-Model gauge group, $G_{\rm{SM}} = {\rm SU}(3)_{\rm c}
\otimes {\rm SU}(2)_{\rm L} \otimes {\rm U}(1)_Y$, is embedded in a simple
group, $G$, which thus has only one gauge coupling. The minimal example of this
idea is the SU(5) theory~\cite{Georgi74}; a recent review with references to the
literature has appeared~\cite{Raby08}. Proton decay with the selection rule
$\Delta\mathcal{B} = 1$ would imply the existence of new physics at an energy
scale of $\sim 10^{16}$ GeV, while $n-\bar n$ oscillations (which entail
$|\Delta\mathcal{B}| = 2$) would correspond to new physics near or above the
TeV scale. These two processes would therefore lead to qualitatively different
baryogenesis mechanisms.

Proton decay has not yet been seen despite three
decades of concerted effort to find it: typical lower limits for the proton lifetimes for various decay modes are longer than
$10^{32}-10^{34}$ years~\cite{Beringer12}. There exist many models (including
those with extra space dimensions at the TeV scale, for example) with local or
global $\mathcal{B}$ or $\mathcal{B}-\mathcal{L}$ symmetry that do not allow
proton decay, but would admit $n - \bar n$ oscillation as a baryon
number violating process. The presence of sphalerons also means that the
discovery of a new source of $\mathcal{B}$-violation might or might not be
relevant for baryogenesis since sphaleron processes might erase any
$\mathcal{B}$ asymmetry generated at a high scale. For example, it was
originally thought that proton decay predicted by grand unified theories could
generate the matter-antimatter asymmetry. However, since sphaleron processes in
the SM violate $\mathcal{B}+\mathcal{L}$ number, any baryon
asymmetry produced at the GUT scale would be washed out before the electroweak
phase transition. Processes that violate $\mathcal{B}-\mathcal{L}$ are
therefore especially interesting to investigate in order to gain insight into
baryogenesis.

The discovery of neutrino masses and mixing provided direct evidence for
physics beyond the SM. This discovery may also be relevant for $n - \bar
n$ oscillation physics. A simple way to understand the small neutrino masses is
by the seesaw mechanism~\cite{seesaw}, which predicts that the neutrino is a
Majorana fermion, which breaks lepton number by two units. Even if the
Majorana nature of the neutrino is established through observation of
neutrinoless double beta decay, we would still need to understand at what scale
this dynamics occurs.  In general, a theory may violate ${\mathcal L}$ by two units
without violating ${\mathcal B}$; for example, the (zero-temperature) Standard
Model, augmented by the addition of electroweak-singlet Majorana neutrino mass
terms, violates ${\mathcal L}$ by two units, but conserves ${\mathcal B}$.  However in
many extensions of the SM these two symmetries are connected.  For example,
the left-right-symmetric extension of the SM has the gauge group ${\rm SU}(3)_c
\otimes {\rm SU}(2)_L \otimes {\rm SU}(2)_R \otimes {\rm U}(1)_{B-L}$, i.e., the
combination ${\mathcal B}-{\mathcal L}$ is gauged. Insofar as this U(1)$_{B-L}$ symmetry remains
exact, $\Delta \mathcal{B} = \Delta \mathcal{L}$.  When the left-right symmetric model is embedded in an SO(10) GUT, it follows that at the level where the ${\mathcal B}-{\mathcal L}$ is exact, the existence of a $\Delta \mathcal{L}=2$ Majorana neutrino mass term is connected with the existence of a  $\Delta \mathcal{B}=2$ operator, although these have different Maxwellian dimensions (3 versus 9 in mass units)~\cite{marshak80}.  Similarly, in an SU(5) GUT with only 5- and 24-dimensional Higgs fields, ${\mathcal B}-{\mathcal L}$ is a symmetry, albeit a global rather than local one (which can be removed by adding a 15-dimensional Higgs field). A search for $n\to\bar n$ oscillations might
therefore supplement the search for neutrinoless double beta decay by
establishing a common mechanism for these processes. In particular, an
observation of $n\to\bar n$ could provide a hint that the small observed
neutrino masses might not be a signal of physics at the GUT scale, as is widely
assumed, but instead possible new physics at much lower scales.  Several
experiments are planned and in progress to search for neutrinoless double beta
decay~\cite{Elliott12}: any discovery of this process would strongly suggest
that $n - \bar n$ oscillations must also exist. From all of this it is clear
that processes that can lead to baryon number and lepton number violations
possess interesting interrelationships that are worth considering together as one analyzes various scenarios beyond the SM.

To summarize: once we accept the possibility that baryon number is not a good symmetry of nature, there are many
questions that must be explored to decide the nature of physics associated with $\mathcal{B}$-violation:

\begin{itemize}

\item

Is (a non-anomalous extension of) baryon number, $\mathcal{B}$, a global or local symmetry?

\item

Does baryon number occur as a symmetry by itself~\cite{Duerr13} or does it
appear in combination with lepton number, $\mathcal{L}$,
i.e. $\mathcal{B}-\mathcal{L}$, as the SM might hint?

\item

What is the scale of baryon number violation and the nature of the associated
physics that is responsible for it? For example, is this physics characterized
by a mass scale not too far above the TeV scale, so that it can be probed in
experiments already searching for new physics in colliders as well as 
low-energy rare processes?

\item

Can the physics responsible for baryon-number violation also 
explain the observed matter-antimatter asymmetry?

\end{itemize}
In the rest of this section we will emphasize extensions to the SM
that can produce $n - \bar n$ oscillations 
\footnote{Before doing so, we briefly
discuss circumstances that can suppress such oscillations. For example, if the
neutron and antineutron masses are different as can (but need not necessarily)
happen if $CPT$ symmetry is somehow violated, then $|n\rangle$ and
$|\bar{n}\rangle$ are no longer degenerate states, even in the absence of
matter and external fields, and the rate for $n - \bar n$ oscillations
even in the presence of $\Delta\mathcal{B} = 2$ interactions can be greatly
suppressed~\cite{Abov84, Lamoreaux91}. Experimentally the neutron and
antineutron are known to possess the same mass to a precision of only
$10^{-4}$~\cite{Yao06}. Any such idea to suppress $n - \bar n$
oscillations would also have to explain why other types of observed particle
oscillations are not suppressed, but no one, to our knowledge, has shown that
such an idea is impossible. Recent theoretical work on $CPT$/Lorentz violation in neutron-antineutron oscillations~\cite{Babu2015} can be used to
quantitatively analyze this question. Furthermore, if $n - \bar n$
oscillations are observed experimentally, it should be possible to set
stringent limits on the possible size of $CPT$-violating effects.}. 

% ========================================================================

\subsection{Some Historical Background Concerning Theories of Baryon-Number 
Violation}
\label{nnbar:subsec:theorybkgd}

Shortly after the development of the SM it was observed that in a
model with a left-right symmetric electroweak group, $G_{\rm LR} = 
{\rm SU}(2)_{\rm L} \otimes {\rm SU}(2)_{\rm R} \otimes 
{\rm U}(1)_{\mathcal{B-L}}$, baryon and lepton 
numbers in the combination $\mathcal{B}-\mathcal{L}$ can be gauged in an
anomaly-free manner. The resultant U(1)$_{\mathcal{B-L}}$ can be combined with
color SU(3)$_{\rm c}$ in an SU(4) gauge group \cite{Pati74}, giving rise to the group
$G_{422} = {\rm SU}(4) \otimes {\rm SU}(2)_{\rm L} \otimes {\rm SU}(2)_R$
\cite{Pati74,Mohapatra75rm,Mohapatra75jp}. 
A higher degree of unification involved models that embed either the Standard
Model gauge group $G_{SM}$ or $G_{422}$ in a simple group such as SU(5) or
SO(10) ~\cite{Georgi74,Raby08}.  The motivations for these grand unification
theories (GUTs) are well known and include the unification of gauge
interactions and their couplings, the related explanation of the quantization
of weak hypercharge and electric charge, and the unification of quarks and
leptons. While the SM gauge couplings do not precisely unify in a
nonsupersymmetric SU(5) framework, they do approach a common value at the GUT
scale of $\sim 10^{16}$ GeV in the minimal supersymmetric extension of the
SM as well as in models with low scale local symmetries such as $\mathcal{B}-\mathcal{L}$, the latter being relevant to the discussion of neutron oscillations.

The unification of quarks and leptons in grand unified theories generically
leads to the decay of the proton and the decay of neutrons bound in nuclei. In the simplest GUTs, these decays obey the
selection rule $\Delta
\mathcal{B} = -1$ and $\Delta \mathcal{L} = -1$, whence $\Delta({\cal B}-{\cal
L})=0$. Limits from experimental searches for decays of protons and otherwise
stably bound neutrons helped to exclude nonsupersymmetric GUTs and are also in
tension with some supersymmetric GUTs. The general possibility of a different
kind of baryon-number violating process, namely the $|\Delta \mathcal{B}|=2$
process of $n - \bar n$ oscillations, was suggested~\cite{Kuzmin70} even before
the advent of GUTs. (One personal view of some of this history is given in 
\cite{Okun2013}). This possibility was further discussed and studied after the
development of GUTs in~\cite{marshak80,Glashow79} and in 
subsequent
models~\cite{Kuo80,Chang80,Mohapatra80rn,Cowsik81,Rao82,Misra83,Rao84,Huber01,Babu01,Nussinov02,Mohapatra05,Babu06,Dutta06,Berezhiani06,Babu09,Mohapatra09,Gu11,Babu13,Arnold13}.
Recently, other models have been constructed that predict $n - \bar n$
oscillations at levels within reach of an improved search,
e.g.~\cite{Babu01,Nussinov02,Dutta06,Babu13}.  We proceed to discuss some of
these models.

% =====================================================================

\subsection{Some Models that Produce $n - \bar n$ Oscillations}
\label{nnbar:subsec:models}

On dimensional grounds, the existing experimental limit on $n-\bar n$ oscillations probes a mass scale near $\sim 10^2$ TeV. A new experiment to improve the $n-\bar n$ oscillation time sensitivity by two orders of magnitude beyond the existing limits would probe an effective mass scale well above the scales probed
directly by the LHC.  This shows that such an experiment would explore possible new
physics beyond the SM in an interesting regime.  The specific models described below illustrate this.

It was pointed out in 1980 that a class of unified theories for Majorana
neutrino mass where the seesaw mechanism operates in the TeV mass range
predicts $n - \bar n$ oscillation transition times which are accessible to experiments~\cite{marshak80}. This model
was based on the idea that $\mathcal{B}-\mathcal{L}$ is a local rather than
a global symmetry. This idea is incorporated in the electroweak gauge group
$G_{\rm LR}$ and accommodates right-handed neutrinos and an associated seesaw
mechanism.  Once the model is embedded into a $G_{422}$ model that unifies
quarks with leptons, it predicts the existence of $n - \bar n$ oscillations
with a transition time within the reach of experiments.  The essential reason for this is the
existence of TeV-scale color sextet bosons in this model, which can also be sought at
the LHC.

By requiring that the model
also explain the observed matter-antimatter asymmetry, the range of neutron oscillation times allowable in
this class of models is quite restricted. The basic idea is that
since $n - \bar n$ oscillations are a TeV-scale ${\cal B}$-violating
phenomenon, they will remain in equilibrium in the thermal plasma down to very
low temperatures in the early universe.  Hence, in combination with
Standard-Model baryon-number-violating processes they will erase any
pre-existing baryon asymmetry in the universe. Therefore in models with
observable neutron oscillation, one must search for new ways to generate
matter-antimatter asymmetry near or below the weak scale. Such a mechanism was
proposed in a few recent papers~\cite{Babu06,Babu09,Gu11}, where it was shown
that the high-dimensional operators that lead to processes such as neutron
oscillation can indeed generate a baryon asymmetry via a mechanism called
post-sphaleron baryogenesis. This mechanism specifically applies to the class
of $G_{422}$ models for neutron oscillations discussed in
Ref.~\cite{marshak80}, as well as to other models for neutron oscillations.
(For some other recent discussions of baryogenesis
at and near the electroweak scale; see, e.g.,~\cite{Ramsey_Musolf12,Canetti2013}
and references therein.) 

In order to give an overview of how post-sphaleron baryogenesis works, we note
that because of quark-lepton unification, the field responsible for the seesaw
mechanism now has colored partners which are color sextets. The neutral scalar
field, which breaks $\mathcal{B}-\mathcal{L}$ gauge symmetry to generate
neutrino masses, has couplings to these colored scalars and decays slowly to 
six-quark states via the exchange of virtual color sextet fields. This decay in
combination with $CP$-violation is ultimately responsible for baryogenesis. Due
to its slowness the decay cannot compete with the Hubble expansion
until the universe cools below the weak scale. The cosmological requirements
for baryogenesis then impose strong constraints on the parameters of the model
and predict that there must be an upper limit on the free neutron oscillation
time of $5\times 10^{10}$ s~\cite{Babu13}, while for most of the parameter
range it is below $10^{10}$ s.  Essentially what happens is that if the neutron
oscillation time exceeds this bound, then the magnitude of the baryon asymmetry
becomes smaller than its observed value or the color symmetry of the model
breaks down, neither of which is acceptable for a realistic theory.  It may
therefore be concluded that if the search for $n - \bar n$ oscillation up to a
transition time of $10^{10}$ s yields a null result, this scenario for baryogenesis will be ruled out.

A different type of model that predicts $n - \bar n$ oscillations at a rate
close to current limits involves a theoretical
framework including extra dimensions of spacetime~\cite{Nussinov02}. Although current experimental data are fully
consistent with a four-dimensional Minkowski spacetime, it is useful to explore
the possibility of extra dimensions, both from a purely phenomenological point
of view and because the main candidate theory for quantum gravity, namely
string theory, suggests the existence of higher
dimensions. Ref.~\cite{Nussinov02} focuses on theories where SM fields can
propagate in extra dimensions and the wavefunctions of SM fermions have strong
localization at various points in this extra-dimensional space. The effective
size of the extra dimension(s) is denoted $L$; the associated mass parameter
$\Lambda_L=L^{-1}$ can be $\sim 50-100$ TeV.  Such models are of interest
partly because they can provide a mechanism for obtaining a hierarchy in
fermion masses and quark mixing.  In generic models of this type, excessively
rapid proton decay can be avoided by arranging that the wavefunction centers of
the $u$ and $d$ quarks are separated far from those of the $e$ and $\mu$.
However, as was pointed out in Ref.~\cite{Nussinov02}, this does not guarantee
adequate suppression of $n -
\bar n$ oscillations, since this process does not involve any lepton fields.
Indeed, for typical values of the parameters of the model, it was shown that
$n - \bar n$ oscillations occur at levels that are in accord with the current
experiment limit but not too far below this limit.  One of the interesting
features of this model is that it is an example of a theory in which proton
decay is negligible, while $n - \bar n$ oscillations could be observable at
levels close to current limits~\cite{Nussinov02}.  An effective-field-theory
study of operators that can suppress proton decay but allow $n-\bar n$ 
oscillations at observable levels was given in~\cite{Arnold13}.  
These operators make use of hypothetical scalar fields in addition to the 
Standard-Model Higgs boson.

If $n-\bar n$ oscillation is experimentally observed in the foreseeable future, the associated scale $M_X$ is much less than $M_{\rm GUT} \sim 10^{16}$ GeV.  Such low scales can be consistent with grand unification only within a restricted class of SO(10) theories consistent with gauge coupling unification where the $n - \bar{n}$ 6 quark operator arises following breakdown of the SO(10) symmetry~\cite{Babu12}. In the context of SU(5) GUT theories this would point to different mechanisms for neutrino mass generation than the higher dimensional operators with same dimension as the $n - \bar{n}$ operator~\cite{Rao84, Degouvea14}, which will have implications for neutrino model building. Neutron-antineutron oscillations due to non-conservation of baryons by virtual black hole evaporation (i.e. by the Zeldovich process~\cite{zeld-B}) have been studied in ref.~\cite{Bambi07}. It was argued in this work that the oscillation time would be $1-2$ orders of magnitude above the existing bounds if the fundamental  gravity scale is a few TeV. 

We include some further remarks on models relevant to $n-\bar n$ oscillations
here.  In addition to the non-accelerator experiments discussed in this paper,
there are also attempts to find direct or indirect evidence for $\mathcal{B}$
and $\mathcal{L}$-violation in the accelerator based experiments such as those
based at the LHC.  It is interesting to note that the simplest hint for
possible $\mathcal{B}-\mathcal{L}$ violating interactions may be the
$\mathcal{B}-\mathcal{L}$ conserving signal of the right-handed charged gauge
boson ($W_{{\rm R}}$).  Searches for such a signal are being carried out at the
LHC \cite{cms,deppisch}.  If evidence for a $W_{{\rm R}}$ were to be
established by the LHC, this would be consistent with the inference that at
energies above the TeV scale, the applicable gauge group of Nature could be ${\rm
  SU}(3)_{\rm c} \otimes {\rm SU}(2)_{\rm L} \otimes {\rm SU}(2)_{\rm R}
\otimes {\rm U}(1)_{\mathcal{B} - \mathcal{L}}$. The ${\rm SU}(2)_{\rm R}
\otimes {\rm U}(1)_{\mathcal{B} - \mathcal{L}}$ would be broken to U(1)$_Y$
around the TeV scale. This would have important implications for an experiment
searching for $n - \bar n$ oscillations.  Since possible light diquark scalar
fields with the appropriate Standard-Model quantum numbers and couplings could
mediate $n-\bar n$ oscillations, limits on their masses are relevant here.
Searches for diquarks are currently being conducted by the ATLAS and CMS
experiments at the LHC.  Another relevant line of theoretical research concerns
spontaneous breaking of global baryon number~\cite{baryo-majoron}.

% =======================================================================

From this summary it is clear that a number of candidate theories beyond the SM lead to neutron-antineutron oscillations. We next present a general phenomenological discussion of $n - \bar n$
oscillations as background to the later sections on current experimental
limits and future possibilities.

% =======================================================================

\subsection{General Formalism for Analysis of $n- \bar n$ Oscillations}
\label{nnbar:subsec:formalism}

% ======================================================================

Before developing the formalism for $n - \bar n$ oscillations
step-by-step for various physical circumstances, we first make a few general
points. Here we assume $CPT$ symmetry (we note that a new study of $\Delta \mathcal{B}=2$ Lorentz invariance violating operators has recently appeared~\cite{Babu2015}).  We denote the effective Hamiltonian that
is responsible for $n- \bar n$ oscillations as $H_\mathrm{eff}$.  The
transition matrix elements are taken to be real and are denoted
\begin{equation}
\langle \bar n | H_\mathrm{eff} | n \rangle =
\langle n | H_\mathrm{eff} | \bar n \rangle \equiv \delta m \ .
\label{nnbtransition}
\end{equation}
We label the diagonal matrix elements of $H_\mathrm{eff}$ as 
\begin{equation}
\langle n | H_\mathrm{eff} | n \rangle \ = \ M_{11}, \quad 
\langle \bar n | H_\mathrm{eff} | \bar n \rangle \ = \ M_{22}
\label{mdiag}
\end{equation}
with ${\rm Im}(M_{jj})=-i\lambda/2$ for $j=1,2$, where $\lambda^{-1} = \tau_n =880$ s is the mean lifetime of a free neutron.  We define
\begin{equation}
\Delta M \equiv M_{11}-M_{22}
\label{deltam}
\end{equation}
The matrix of $H_\mathrm{eff}$ in the basis $(n,\bar n)$ 
thus has the general form
\begin{equation}
\cal{M}=\left(\begin{array}{cc}
M_{11}   & \delta m \\
\delta m & M_{22}  \end{array}\right)
\label{mgeneral}
\end{equation}
The diagonalization of this matrix yields the mass eigenstates
\begin{equation}
\left(\begin{array}{cc}
|n_1 \rangle \\
|n_2 \rangle \end{array}\right) =
\left(\begin{array}{cc}
 \cos\theta & \sin\theta \\
-\sin\theta & \cos\theta \end{array} \right)
\left(\begin{array}{cc}
|n \rangle \\
|\bar n \rangle \end{array}\right)
\label{nnbtransformation}
\end{equation}
where
\begin{equation}
\tan(2\theta) = \frac{2\delta m}{\Delta M}
\label{tan2theta}
\end{equation}
The real energy eigenvalues are
\begin{equation}
E_{1,2} = \frac{1}{2}\bigg [ M_{11}+M_{22} \pm \sqrt{(\Delta M)^2  +
4(\delta m\ )^2} \ \bigg ]
\label{generaleigenvalues}
\end{equation}
We define 
\begin{equation}
\Delta E = E_1-E_2 = \sqrt{ (\Delta M)^2 +4(\delta m)^2}
\label{deltae}
\end{equation}

If one starts with a pure $|n\rangle$ state at $t=0$, then there is a
finite probability $P(n(t)=\bar n)$ for it to evolve to an 
$|\bar n\rangle$ at $t \ne 0$ given by
\begin{eqnarray}
P(n(t)=\bar n) & = & |\langle \bar n|n(t) \rangle|^2 = 
\sin^2(2\theta) \, \sin^2 [(\Delta E)t/2] \, e^{-\lambda t} \cr\cr
& = & 
\Bigg [ \frac{(\delta m)^2}{ (\Delta M /2)^2 + (\delta m)^2} \Bigg ] \,
\sin^2  \Big [ \sqrt{ (\Delta M/2)^2 + (\delta m)^2 } \ t \Big ] \, 
e^{-\lambda t}
\cr\cr
& & 
\label{pgen}
\end{eqnarray}
where $\lambda^{-1}=\tau_{n}=880$ s is the mean life of a free neutron. The difference $\Delta M$ incorporates any interaction effects that
are different for the neutron and the antineutron. For example, the
neutron-nucleus and antineutron-nucleus strong interactions are very different,
so that the proximity of matter suppresses any oscillation that might otherwise
be induced by the off-diagonal matrix elements. For low-energy neutrons one can
parametrize the neutron-nucleus interaction in terms of a (complex) $s$-wave
scattering amplitude, and this scattering amplitude enters the Schr\"{o}dinger
equation for the neutron and antineutron propagation in a medium through the
neutron(antineutron) optical potential. In addition, any ambient external
magnetic field splits the energy of neutron and antineutron states since they
possess magnetic moments of opposite sign.

It is important to understand from the beginning that the existing upper bound on the
magnitude of the off-diagonal term, $|\delta m|$, in the $n - \bar n$
effective Hamiltonian is known to be less than approximately $10^{-29}$ MeV,
which is a very small energy, approximately 32 orders of magnitude smaller than
the mass of the neutron. This constraint means in practice that $|\Delta M|$
will be orders of magnitude larger than $|\delta m|$ for any
conceivable experimental environment. For example, in free space with an ambient magnetic
field ${\vec B}$, $\Delta M = -2 {\vec \mu}_n \cdot {\vec B}$, and
\begin{equation}
|{\vec \mu}_n| B = (6.03 \times 10^{-23} \ {\rm MeV}) \, 
\bigg ( \frac{B}{10^{-9} \ {\rm Tesla}} \bigg ) 
\label{mubsize}
\end{equation}
where $B \equiv |{\vec B}|$.  Hence, the value of $|\Delta M|$ resulting from
even a very small $1$ nT external magnetic field is $0.6 \times 10^{-22}$
MeV. This is much larger than $|\delta m|$; in this case $|\Delta M/\delta m|
\simeq 10^{7}$.\footnote{
The very small value of $|\delta m|$, combined with the limited
time of observation of a slow-neutron ensemble in the gravitational field of
the Earth, means that attempts to amplify the transition rate by either
resonant excitation or some sort of adiabatic level-crossing mechanism, as in
the well-analyzed case of neutrino oscillations in matter, seem to be
impractical to implement. As is evident from Eq.~(\ref{tan2theta}), the
magnitude of the mixing angle is very small, i.e., $|\theta| \ll 1$.  We can
also consider the possibility of an adiabatic level crossing in a slowly
varying magnetic field tending to zero and reaching the regime $|\Delta M| \ll
|\delta m|$ over some distance as a means to resonantly enhance the
oscillations. This is already a regime that corresponds to the need to establish magnetic fields which are several orders of magnitude smaller than ever created in a laboratory, and
therefore it is difficult to imagine how it could be done in practice. There
are other methods for forcing the neutron and antineutron levels in a magnetic
field to cross: for example one can apply a time-dependent RF magnetic field to
produce the well-known ``dressed'' neutron states known from quantum optics,
which can cross for specially chosen amplitudes and frequencies of the RF
field, or one could apply an electric field with a slow spatial gradient that
changes sign and is effectively like a small $\vec{v} \times \vec{E}$ magnetic
field in the rest frame of the neutron. Even if any of these methods could
reliably produce a level crossing, however, the well-known condition for
reaching the adiabatic limit in this case, $|d\theta/dt| \ll 2\sqrt{(\Delta
M)^{2}+(\delta m)^2 }$, would, unfortunately, require the neutrons to spend an
amount of time near this level crossing region of almost zero field which is
many orders of magnitude larger than the neutron beta decay lifetime.~\cite{Wietfeldt:2009}} 

The search for oscillations is therefore
best conducted in the so-called ``quasi-free'' limit corresponding to
$|\Delta E| \, t \ll 1$; in this limit, one Taylor-expands the sine function
in Eq.~(\ref{pgen}), and the resultant factor of $(\Delta M)^2+ (\delta m)^2$ 
cancels with the prefactor.  Consequently, in this regime one has 
\begin{equation}
P(n(t) = \bar n) = [(\delta m)\, t]^2 \, e^{-\lambda t} = 
(t/\tau_{n- \bar n})^2 \, e^{-\lambda t}
\label{pgenquasifree}
\end{equation}
where 
\begin{equation}
\tau_{n- \bar n} = \frac{1}{|\delta m|}
\label{taunnb}
\end{equation}
(Here and elsewhere, we use natural units, $\hbar = c = 1$.)  The physical
meaning of this so-called ``quasi-free'' limit should be clear; it corresponds
to the application of the energy-time uncertainty principle to the
transition. The point is that, in the presence of an amplitude that causes
oscillations between energy states, the free neutron is not an energy eigenstate of $H$. Also, for states like those of the neutron and antineutron which are degenerate in energy in the absence of an interaction that leads to oscillations,  the
admixture of the antineutron component of the state grows linearly for times short compared to
the time needed for an observation to distinguish between the different neutron
and antineutron energies in the ambient field, in accordance with the usual nonrelativistic dynamics of two state systems in quantum mechanics.  It is perhaps a bit
imprecise to use the term ``oscillations'' in this case, since in practice only
a small fraction of an oscillation period can be accessed by any conceivable
experiment, and one can argue that the search for $n - \bar n$
oscillations might be better characterized conceptually for many purposes as
essentially a rare decay search. However, we will not use such terminology,
since the oscillation language has been established in the extensive literature
on the subject.  In any case, it is clear that in this situation the
experimental challenge is to observe a large enough number of neutrons for a
long enough time under conditions in which the oscillation rate is not (yet)
suppressed by these inevitably larger and difficult to control diagonal terms.

\subsubsection{$n - \bar n$ Oscillations in Field-Free Vacuum}
\label{nnbar:subsubsec:vacuum}

In vacuum that is free of electromagnetic fields (assuming $CPT$ symmetry), 
\begin{equation}
\langle n | H_\mathrm{eff} | n \rangle \ = \
\langle \bar n | H_\mathrm{eff} | \bar n \rangle \ = m_n 
- \frac{i \lambda}{2} \ ,
\label{diagonal}
\end{equation}
Here, the matrix of $H$ in the $(n,\bar n)$ basis takes the form 
\begin{equation}
\cal{M}_{\rm f}=\left(\begin{array}{cc}
m_n - i\lambda/2 & \delta m \\
\delta m         & m_n - i\lambda/2 \end{array}\right)
\label{nnbmatrix}
\end{equation}
(where the subscript f stands for ``free'').  
Diagonalizing this matrix ${\cal M}_{\rm f}$ yields the mass eigenstates 
\begin{equation}
|n_\pm \rangle =\frac{1}{\sqrt{2}}( |n \rangle \pm |\bar n \rangle )
\end{equation}
with mass eigenvalues
\begin{equation}
m_{\pm} = (m_n \pm \delta m) - \frac{i\lambda}{2} \ .
\label{meigfree}
\end{equation}
Evidently, in this case there is maximal mixing, i.e., $\theta=\pi/4$, where
the mixing angle $\theta$ was defined in Eq. (\ref{tan2theta}). 
The oscillation probability is simply 
\begin{equation}
P(n(t)=\bar n) = [\sin^2(t/\tau_{n- \bar n})] \, e^{-\lambda t} \ ,
\label{pfree}
\end{equation}
where $\tau_{n- \bar n}$ is defined in Eq. (\ref{taunnb}). 
As will be discussed in greater detail below, current limits give
$\tau_{n- \bar n} \gtrsim 10^8$ s, so
\begin{equation}
\tau_{n- \bar n} \gg \tau_n.
\label{tnnbtrel}
\end{equation}

% ===================================================================

\subsubsection{$n - \bar n$ Oscillations in a Static Ambient Magnetic Field ${\vec B}$}
\label{nnbar:subsubsec:bfield}

We next review the formalism for the analysis of $n - \bar n$ oscillations in
an ambient magnetic field~\cite{Mohapatra80rn,Cowsik81}. This
formalism is relevant for an experiment searching for $n - \bar n$ oscillations
using neutrons that propagate some distance in a vacuum pipe since in practice
it is impractical to make the energy splitting $2\mu B$ of the neutron and
antineutron states in an external field small compared to $\delta m$. This
formalism is relevant for a discussion of the Institut Laue-Langevin (ILL)
experiment at Grenoble, which yielded the best bound on $n - \bar n$
oscillations in a free propagation experiment, using neutrons from a
reactor~\cite{BaldoCeolin94}, and for possible future experiments.

The $n$ and $\bar n$ interact with the external $\vec B$ field via their
magnetic dipole moments, ${\vec \mu}_{n,\bar n}$, where $\mu_n = -\mu_{\bar n}
= -1.91 \mu_N$ and $\mu_N = e/(2m_N) = 3.15 \times 10^{-14}$ MeV/Tesla.  Hence,
the matrix ${\cal M}_{\rm B}$ now takes the form
\begin{equation}
\cal{M}_{\rm B}=\left(\begin{array}{cc}
m_n - {\vec \mu}_n \cdot {\vec B} - i\lambda/2 & \delta m \\
\delta m         & m_n + {\vec \mu}_n \cdot {\vec B} -  i\lambda/2
\end{array}\right)
\end{equation}
The diagonalization of this mass matrix yields the mass eigenstates 
$|n_1 \rangle$ and $|n_2 \rangle$ given by Eq. (\ref{nnbtransformation}) with 
\begin{equation}
\tan(2\theta) = -\frac{\delta m}{{\vec \mu}_n \cdot {\vec B}} \ .
\end{equation}
From the general result Eq. (\ref{generaleigenvalues}), the eigenvalues are
\begin{equation}
E_{1,2} = m_n \pm \sqrt{({\vec \mu}_n \cdot {\vec B})^2 + (\delta m)^2} \ -
 \frac{i\lambda}{2} \ .
\label{eigenvalues_magneticfield}
\end{equation}
The ILL experiment reduced the magnitude of the magnetic field to $|\vec B|
\sim 10^{-4} \ {\rm G} = 10^{-8}$ T, so $|{\vec \mu}_n \cdot {\vec B}| \simeq
10^{-21}$ MeV.  It is expected that a future experiment could achieve a
reduction to $|\vec B| \sim 10^{-9}$ T. 
Since one knows from the experimental bounds that $|\delta m|
\lesssim 10^{-29} \ {\rm MeV}$, which is much smaller than $|{\vec \mu}_n \cdot
{\vec B}|$, it follows that $|\theta| \ll 1$.  Thus,
\begin{equation}
\Delta E = 2 \sqrt{({\vec \mu}_n \cdot
{\vec B})^2 + (\delta m)^2} \simeq 2 |{\vec \mu}_n \cdot {\vec B}| \ .
\end{equation}
As discussed above, in order to avoid the suppression of the oscillation rate
from this breaking of the degeneracy of the diagonal parts of the
$n - \bar n$ effective Hamiltonian in a quasi-free-propagation
experiment, one can arrange an observation time $t$ such that $|{\vec
\mu}_n \cdot {\vec B}|t \ll 1$ and also $t \ll \tau_n$.  Then the oscillation
probability reduces to 
\begin{equation}
P(n(t) = \bar n) \simeq
(2\theta)^2 \Big ( \frac{\Delta E t}{2} \Big )^2 \simeq
\bigg ( \frac{\delta m}{{\vec \mu}_n \cdot {\vec B}} \bigg )^2
\bigg ({\vec \mu}_n \cdot {\vec B} \, t \bigg )^2 = [(\delta m) \, t]^2 =
(t/\tau_{n- \bar n})^2 \ .
\label{pmagnetic}
\end{equation}
The number of $\bar n$'s produced by the $n - \bar n$ oscillations is given
essentially by $N_{\bar n}=P(n(t)=\bar n)N_n$, where $N_n = \phi T_{\rm run}$,
with $\phi$ the neutron flux and $T_{\rm run}$ the running time.  The
sensitivity of the experimental signal depends on the product $N_n t^2$, so,
with adequate magnetic shielding, one wants to maximize $t$, subject to the
condition that $|{\vec \mu}_n \cdot {\vec B}|t \ll 1$.  A different approach to $n - \bar n$ propagation via spin-flip transitions in an ambient magnetic field has recently been proposed in~\cite{Gardner15}.

% =======================================================================

\subsubsection{$n - \bar n$ Oscillations in Matter}
\label{nnbar:subsubsec:matter}

In matter, the matrix ${\cal M_{\rm A}}$ takes the form~\cite{Kuo80}
\begin{equation}
\cal{M}_{\rm A}=\left(\begin{array}{cc}
m_{n, {\rm eff}}  & \delta m \\
\delta m         & m_{\bar n, {\rm eff}} \end{array}\right)
\label{mat}
\end{equation}
with
\begin{equation}
m_{n,{\rm eff}} = m_n + V_n \ , \quad 
m_{\bar n, {\rm eff}} = m_n + V_{\bar n} \ ,
\end{equation}
where the nuclear potential $V_n$ is practically real, $V_n = V_{n {\rm R}}$,
but $V_{\bar n}$ has a large imaginary part representing the antineutron 
annihilation with another nucleon,
\begin{equation}
V_{\bar n} = V_{\bar n {\rm R}} - i V_{\bar n {\rm I}} \ ,
\end{equation}
with~\cite{Dove83,Friedman08}
\begin{equation}
V_{n {\rm R}}, \ V_{\bar n {\rm R}}, \ 
V_{\bar n {\rm I}} \sim O(100) \ {\rm MeV} \ .
\end{equation}
The mixing is thus strongly suppressed; $\tan(2\theta)$ is determined by
\begin{equation}
\frac{2\delta m}{|m_{n,{\rm eff}} - m_{\bar n, {\rm eff}}|} =
\frac{2\delta m}{\sqrt{(V_{n {\rm R}}-V_{\bar n {\rm R}})^2 + 
V_{\bar n {\rm I}}^2}} \ll 1 \ .
\end{equation}

The eigenvalues from the diagonalization of ${\cal M}_{\rm A}$ are
\begin{equation}
E_{1,2} = \frac{1}{2} \bigg [ m_{n,{\rm eff}} +  m_{\bar n, {\rm eff}} \pm
\sqrt{ (m_{n,{\rm eff}} -  m_{\bar n,{\rm eff}})^2 
+ 4(\delta m)^2 } \ \bigg ] \ .
\end{equation}
Expanding $m_1$ for the mostly $n$ mass eigenstate $|n_1\rangle \simeq
|n\rangle$, one obtains
\begin{equation}
E_1 \simeq m_n + V_n - i \frac{(\delta m)^2 \, V_{\bar n {\rm I}}}
{(V_{n {\rm R}}-V_{\bar n {\rm R}})^2 + V_{\bar n {\rm I}}^2} \ .
\end{equation}
The imaginary part leads to matter instability
via annihilation of the $\bar n$, producing mainly pions (with mean
multiplicity $\langle n_\pi \rangle \simeq 4-5$).  The rate for this is
\begin{equation}
\Gamma_{\rm m} = \frac{1}{\tau_{\rm m}} = 
\frac{2(\delta m)^2 |V_{\bar n {\rm I}}|}
{(V_{n {\rm R}} - V_{\bar n {\rm R}})^2 + V_{\bar n {\rm I}}^2} \ .
\end{equation}
where the subscript ${\rm m}$ stands for ``matter''~\cite{Albe82}. 
Thus, $\tau_{\rm m} = 1/\Gamma_{\rm m} \propto (\delta m)^{-2}$. Writing
\begin{equation}
\tau_{\rm m} = R \, \tau_{n- \bar n}^2 \ ,
\end{equation}
one has $R \simeq 10^2$ MeV, or equivalently, 
\begin{equation}
R \simeq 10^{23} \ {\rm s}^{-1} \ .
\end{equation}
The value of $R$ depends on the nucleus; for example, detailed calculations 
yield $R \simeq 1 \times 10^{23} \ {\rm s}^{-1}$ for $^{56}$Fe and 
$R \simeq 0.5 \times 10^{23} \ {\rm s}^{-1}$ for $^{16}$O \cite{Friedman08}.

Before delving into the more detailed discussions below, we pause to make an interesting
observation.  If one takes the lower bound on $\tau_{n- \bar n}$ from
$n - \bar n$ searches in free neutron experiments, one can estimate a lower
bound on $\tau_{\rm m}$ and vice versa. Numerically, 
\begin{equation}
\tau_{\rm m} > (1.6 \times 10^{31} \ {\rm yr}) \Big ( \frac{\tau_{n- \bar n}}
{10^8 \ {\rm s}} \Big )^2
\Bigg ( \frac{R}{0.5 \times 10^{23} \ {\rm s}^{-1}} \Bigg )
\label{ttrel}
\end{equation}
Hence, with $R \simeq 0.5 \times 10^{23}$ s$^{-1}$, the lower limit $\tau_{n-
\bar n} > 0.86 \times 10^8$ s from the ILL experiment yields $\tau_{\rm m}
\gtrsim 10^{31}$ yr.  An interesting feature of this limit is that it is
the same order of magnitude as the present sensitivity of experiments conducted
in large underground detectors.  There is no fundamental physics reason why this
should turn out to be the case, but it is true.  We present an overview of the status and prospects of both free neutron experiments in \S\ref{nnbar:sec:freeneutron} and experiments using neutrons bound in nuclei in \S\ref{nnbar:sec:intranuclear}, before concentrating on details concerning future free neutron searches in the following sections.

Finally it is worth mentioning that there is another baryon system in which a search for oscillations have been considered: the $\Lambda- \bar \Lambda$ system. The formalism presented above  is also valid for this system. The presence of a valence strange quark in the $\Lambda$ means that a search in this system is in principle sensitive to different sources for $\Delta \mathcal{B}=2$ interactions.~\cite{Kang2010}
The shorter lifetime of the $\Lambda$ and the lower intensities of $\Lambda$ beams imply that a dedicated fixed target experiment devoted to such a search is much less sensitive in terms of $\delta m$ than a search using neutrons. A dedicated experiment to search for $\Lambda- \bar \Lambda$ oscillations  could be expected to reach a sensitivity to $\delta m_{\Lambda}$ of about $10^{-11}$ eV based on the capabilities of existing accelerator facilities.~\cite{Sokoloff2014}

 % =====================================================================

\subsection{Operator Analysis and Estimate of Matrix Elements}
\label{nnbar:subsec:analysis}

In addition to the phenomenological analysis of $n - \bar n$ oscillations
under different circumstances at the effective Hamiltonian level presented
above, one can also discuss some general characteristics of the relation
between the off-diagonal term and the microscopic quark-level operator
responsible for the transition. At the quark level, the $n \to \bar n$
transition is $(u d d) \to (u^c d^c d^c)$.  This is mediated by six-quark
operators ${\cal O}_i$, so the effective Hamiltonian is
\begin{equation}
H_\mathrm{eff} = \int d^3x {\cal H}_\mathrm{eff}  ,
\end{equation}
where the effective Hamiltonian density is
\begin{equation}
{\cal H}_\mathrm{eff} = \sum_i c_i {\cal O}_i \ .
\end{equation}
In 4D spacetime this six-quark operator has dimension $9$ in mass units, so the coefficients have dimension $-5$. We write
them generically as
\begin{equation}
c_i \sim \frac{\kappa_i}{M_X^5}
\end{equation}
If the fundamental physics yielding the $n - \bar n$ oscillation is
characterized by an effective mass scale $M_X$, then with $\kappa_i \sim O(1)$
and after absorbing dimensionless numerical factors into the effective scale
$M_X$, the transition amplitude is
$\delta m = \langle \bar n | H_\mathrm{eff} | n \rangle$.  This is determined
by 
\begin{equation}
\langle \bar n | {\cal H}_\mathrm{eff} | n \rangle  = \frac{1}{M_X^5}
\sum_i \kappa_i \langle \bar n |{\cal O}_i  | n \rangle
\end{equation}
Hence,
\begin{equation}
\delta m \sim \frac{\kappa \Lambda_{QCD}^6}{M_X^5} \ ,
\end{equation}
where $\kappa$ is a generic $\kappa_i$ and $\Lambda_{\rm QCD} \simeq 180$ MeV
arises from the matrix element $\langle \bar n | {\cal O}_i | n \rangle$.
Numerically, 
\begin{equation}
\tau_{n- \bar n} = (2 \times 10^8 \ {\rm s}) \,
\bigg ( \frac{M_X}{4 \times 10^5 \ {\rm GeV}}\bigg )^5 \,
\bigg ( \frac{3 \times 10^{-5} \ {\rm GeV}^6}
{|\langle \bar n | \sum_i \kappa_i {\cal O}_i | n \rangle | } \bigg )
\label{tomx}
\end{equation}
Here we have used the illustrative value $3 \times 10^{-5}$ GeV$^6$ for the 
matrix elements of the operators ${\cal O}_i$ because this is a
typical value obtained in a calculation \cite{Rao82,Rao84}. 

With this input and with $M_X \sim {\rm few} \ \times 10^5$ GeV, one has
$\tau_{n- \bar n} \simeq 10^8 \ {\rm s}$. However it is important to understand 
that the effective scale $M_X$ derived in this way can be misleading as $M_{X}$ can be a
function of several particle mass scales and dimensional and dimensionless
couplings.  For example, one type of Feynman diagram contributing to a $n-\bar n$
transition amplitude involves ingoing $uud$ quarks and outgoing $u^c d^c d^c$
quarks on three $q$ ($q^c$) propagator lines, with three Higgs propagators
coupling to these, each transforming $q \to q^c$. Let us denote these Higgs
couplings to the quarks generically as $y_i$ and the Higgs masses as $m_{H_i}$.
These three Higgs propagators meet at a triple-Higgs vertex whose coupling $g_H$ has mass dimension $1$.  Say these contribute to a given ${\cal O}_i$. Then
\begin{equation}
c_i = \frac{\kappa_i}{M_X^5} =
\frac{y_1 \, y_2 \, y_3 \, g_H}{m_{H_1}^2 m_{H_2}^2 m_{H_3}^2}
\label{ciexample}
\end{equation}
For illustrative Feynman diagrams that give rise to such a coefficient $c_i$,
see, e.g., Fig. 3 of~\cite{Rao84}.  As Eq. (\ref{ciexample}) shows, some mass
scales contributing to $n-\bar n$ oscillations may be substantially lower than
$10^5$ GeV, as is the case in some recent models. These lower mass
scales may be in the same regime that can be probed experimentally both in
future LHC measurements and in the large variety of other searches for rare
processes whose sensitivity to new physics also reaches into this regime. These
latter measurements include searches for possible new sources of $CP$-violation
in electric dipole moment searches, searches for lepton flavor number violation
in muon decay and muon-to-electron conversion, precision parity violation
measurements, etc.  As the physics community has absorbed the results from 
the LHC run at 7 TeV and 8 TeV which (so far) show no clear evidence for
physics beyond the SM, and as we await the results from the LHC run at 13 TeV and 14 TeV, 
the time is now ripe for a more general
effective-field-theory analysis of the overall landscape of possibilities for
new physics in this regime, including their various interrelationships. Any
such global analysis should include the possibility of $n - \bar n$
oscillations into the mix.

The operators ${\cal O}_i$ must be singlets under color ${\rm SU}(3)_{\rm c}$ and, for
$M_X$ larger than the electroweak symmetry-breaking scale, they must also be
singlets under ${\rm SU}(2)_{\rm L} \times {\rm U}(1)_Y$.  An analysis of these
operators was carried out in~\cite{Rao82}, and the $\langle \bar n | {\cal O}_i
| n \rangle$ matrix elements were calculated in the MIT bag model.  Further
results were obtained varying MIT bag model parameters in~\cite{Rao84}.  These
calculations involve integrals over sixth-power polynomials of spherical Bessel
functions from the quark wavefunctions in the bag model.  As expected from the
general arguments above, it was found that
\begin{equation}
|\langle \bar n | {\cal O}_i | n \rangle | \sim O(10^{-4})
\ {\rm GeV}^6 \simeq (180 \ {\rm MeV})^6 \simeq \Lambda_{\rm QCD}^6
\label{mitmatrixelements}
\end{equation}
An exploratory effort has recently begun to calculate these matrix elements
using lattice gauge theory methods~\cite{Buchoff12}.  As the earlier MIT
bag-model calculations showed \cite{Rao82,Rao84}, different operators ${\cal
O}_i$ have different matrix elements.  When reliable lattice calculations of
the matrix elements of the various ${\cal O}_i$ become available, they will be
useful in connecting a possible observed $n - \bar{n}$ oscillation rate to the
corresponding effective mass scale $M_X$, as in Eq. (\ref{tomx}) (assuming that
one has determined the coefficients $\kappa_i$ in a given model).

Finally we briefly discuss the symmetry properties of $\delta m$~\cite{Berezhiani2015}. These are clarified if one writes the term in the Lagrangian which generates $\delta m$ as

$$\Delta L_{B}=-{\epsilon \over 2}[n^{T}Cn+\bar nC\bar n^{T}]$$

where $C$ is the charge conjugation matrix, $C=i\gamma^2\gamma^0$, and $\epsilon$ can be chosen to be real. 
This term is even under charge conjugation, $n \to n^c=C\bar n^{T}$, and odd under $P$ reflection,
$n \to \gamma^0 n$. This is in correspondence with the opposite parities of $n$ and $\bar n$. However,
one can define $P_z$ as $n \to i\gamma^0 n$. Then the $P_z$ parities of $n$ and $\bar n$ are the same and 
equal to $i$ with $P_z^2=-1$ while $P^2=-1$. Now the baryon charge breaking term preserves both $C$ and $P_z$ symmetries. Its local form guarantees $CPT$ conservation, so time inversion $T$ is preserved as well. 
Breaking of $CP_z$, or equivalently $T$, can appear only due to interactions.

% =====================================================================

\subsection{Summary of Motivations and Phenomenology}
\label{nnbar:subsec:theorysummary}

We argue that this discussion shows there is strong scientific motivation to pursue a
higher-sensitivity $n - \bar n$ oscillation search experiment that can achieve
a lower bound of $\tau_{n-\bar n}$ of $ \sim 10^9 - 10^{10}$ s. Whether or not
$n - \bar n$ oscillations are observable in the next 
generation of experiments depends critically on the scale at which the ${\cal
B}-{\cal L}$ symmetry of the SM is broken. If there is no new
physics between the scale of $\sim 10^3$ GeV now being probed
directly at the LHC, and the commonly-accepted Grand Unification scale of
$\approx 10^{16}$ GeV, then it would seem that $n - \bar n$ oscillations
are unlikely to be observed.  However many recent theories
populate this range in various ways and introduce new scales which can
generate $n - \bar n$ oscillations. 

The existence of $n - \bar n$ oscillations at relatively low scales
could also affect baryogenesis. They could erase any $\mathcal{B}$ asymmetry
produced at higher scales at early times in the universe (such as from the
currently-popular leptogenesis mechanisms) and therefore refocus the search for
the origin of the $\mathcal{B}$ asymmetry of the universe on later times after
the electroweak phase transition. Theoretical mechanisms for 
post-sphaleron baryogenesis~\cite{Babu07} and other low-scale
baryogenesis mechanisms~\cite{Dolgov06, Bambi07} have recently appeared.  If
the effective scale of the relevant new physics is around 10$^{4}$ GeV - 10$^{6}$
GeV, as predicted by various theoretical models~\cite{Babu06,Babu09}, the
possible range of $n - \bar n$ oscillation time is $\tau_{n\bar{n}}\sim$
10$^{9}$ s - 10$^{10}$ s and a $n - \bar n$ oscillation search
experiment is likely to detect new physics~\cite{Mohapatra09}. Observation of
$n-\bar{n}$ oscillations at currently-achievable sensitivity would illuminate
physics that affords a mechanism to regenerate the matter-antimatter asymmetry
at scales below this transition. Some existing theories describing such processes
also predict colored scalars within the reach of the LHC.

% =======================================================================

\subsection{Neutron - Mirror Neutron Oscillation}
\label{nnbar:subsec:mirrorneutrons}

Although this report focuses on $n - \bar{n}$ oscillations, we mention here another type of transition involving neutrons that
might occur and has recently been sought experimentally, namely neutron oscillations into ``mirror neutrons.''  This
possibility occurs in a theoretical framework in which one envisions the
possibility of what has been termed a ``mirror'' world.  According to the usual
definition, this world is assumed to be almost identical to our world, but to
differ in some respects, including a reversal in the sign of parity violation.
Thus, instead of left-handed chiral components of fermions being doublets under
an SU(2) weak isospin gauge group, the right-handed chiral components of the
mirror fermions would be subject to a mirror SU(2) gauge group. The physics of
the hypothetical mirror world could also be relevant at some level for $n - \bar
{n}$ oscillations.  In particular, one could envision a different type of
experiment, namely a regeneration experiment that could test for transitions of
neutrons into mirror neutrons.  In this experiment, a neutron beam would be
incident on a thick absorber.  If some fraction of the neutrons were to undergo
transitions to mirror neutrons, then they would pass through the absorber,
since particles in the hypothetical mirror world do not have direct
interactions with particles in our world.  The mirror neutrons would then
emerge from the absorber and could undergo further oscillations back to regular
neutrons, which could be detected. For some recent theoretical and 
experimental work on possible neutron-mirror neutron oscillations, see 
\cite{Berezhiani06,EPJ-More,Ban,Serebrov,Altarev,Bodek,Serebrov2,Nesti,Foot,magnetic,Mohapatra_Nussinov_Teplitz2002}
and references therein. 

%%%%%%%%%%%%%%%%%%%%%%%%%%%%%%%%%%%%%%%%%%%%%%%%%%%%%%%%%%%%%%%%%%%%%%%%%%%%%%%%%%%%%%%%
\section{Free Neutron Searches for $n - \bar n$ Oscillations}
\label{nnbar:sec:freeneutron}
%%%%%%%%%%%%%%%%%%%%%%%%%%%%%%%%%%%%%%%%%%%%%%%%%%%%%%%%%%%%%%%%%%%%%%%%%%%%%%%%%%%%%%%%

Experiments which utilize free neutrons to search for $n - \bar n$
oscillations have a number of remarkable features.  The basic idea for
these experiments (greater detail is given in \S\ref{nnbar:sec:coldneutrons} and \S\ref{nnbar:sec:sensitivity})
is to prepare a beam of slow neutrons (kinetic energies of a few meV and below:~\ref{nnbar:apdx:nmod} discusses the physics of slow neutron moderation) which propagate
freely from the exit of a neutron optical guide to a distant antineutron annihilation target. (Neutrons in this low energy range possess an index of refraction in matter which is less than one. This corresponds to total external reflection from material surfaces and makes neutron optical guides possible: see~\ref{nnbar:apdx:noptics}). During
the time in which the neutron propagates freely, a $\mathcal{B}$-violating interaction can
produce oscillations from an initial neutron state to one with an admixture of neutron and antineutron amplitudes. Antineutron appearance can be sought experimentally through annihilation in a thin material target, which generates a star pattern of final state momenta with $4-5$ pions on average. This signature can be seen with a tracking detector and calorimeter enclosing the target region.  This rather unique signature
strongly suppresses potential backgrounds. As noted above to observe this
signal, the \lq\lq quasi-free\rq\rq\ condition $\Delta E\cdot t \ll 1$ must hold. This creates an experimental requirement for low gas pressures in the vacuum chamber (below roughly $10^{-5}$ Pa) and very small ambient magnetic fields
(between 1 and 10 nT) for any proposed new experiment using free neutrons
in order to prevent the energy splitting between the neutron and antineutron in matter and external fields from damping the oscillations. 

Some attractive features of a new $n - \bar n$ oscillation search experiment on a free ensemble of slow neutrons operated in this mode include:
\begin{itemize}
\item detection of antineutron appearance using annihilation events with a sharply-localized vertex that can approach a \lq\lq zero\rq\rq background condition to maximize the discovery potential,
\item the capability to turn off any non-zero $n - \bar n$ signal with a modest increase in the magnetic field to lift the $n - \bar n$ energy degeneracy and thereby strongly  damp out the oscillations,
\item the opportunity to achieve orders of magnitude improvement in sensitivity over the current free neutron limit through the use of existing neutron optics technology to greatly increase the integrated neutron fluence and average free observation time to the annihilation target.
\end{itemize}
We believe that these advantages provide a strong experimental motivation to search for $n - \bar n$ oscillations in a dedicated experiment.

%%%%%%%%%%%%%%%%%%%%%%%%%%%%%%%%%%%%%%%%%%%%%%%%%%%%%%%%%%%%%%%%%%%%%%%%%%%%%%%%%%%%%%%%
\subsection{Previous Free Neutron Searches for $n - \bar n$ Oscillations}
\label{nnbar:subsec:previousfree}
%%%%%%%%%%%%%%%%%%%%%%%%%%%%%%%%%%%%%%%%%%%%%%%%%%%%%%%%%%%%%%%%%%%%%%%%%%%%%%%%%%%%%%%%

The current best limit for an experimental search for free $n - \bar n$
oscillations was performed at the Institut Laue-Langevin (ILL) in
Grenoble in the early 1990's~\cite{BaldoCeolin94} and two previous measurements at ILL and Pavia University's Triga Mark II reactor in the mid-1980's and early 1990's~\cite{Fidecaro, Bressi90} (see Fig.~\ref{ill:fig:logo}).
The ILL experiment used a cold neutron beam from their 58 MW research reactor
with a neutron current of 1.25$\times$10$^{11} {\it n}/{\rm s}$
incident on the annihilation target and achieved a limit of
$\tau_{n-\bar{n}} > 0.86\times10^{8}$ $\rm{s}$~\cite{BaldoCeolin94}.
The average velocity of the cold neutrons was $\sim$ 600 ${\rm m/s}$
and the average neutron observation time was $\sim$ 0.1 ${\rm s}$.
A vacuum of $P\simeq 2\times10^{-4}$ ${\rm Pa}$ maintained in the
neutron flight volume and a magnetic field of $|{\vec B}| < 10$ ${\rm nT}$
satisfied the quasi-free conditions for oscillations to occur~\cite{Bitter91,Kinkel92,Schmidt92}.
Antineutron appearance was sought through annihilation with a $\sim$
130 ${\rm \mu m}$ thick carbon film target which generated at least
two tracks (one due to a charged particle) in the tracking detector with
a total energy above 850 MeV in the surrounding calorimeter. In one year of
operation the ILL experiment saw zero candidate events with zero
background~\cite{BaldoCeolin94} using a tracking detector with crude (several cm) spatial resolution for the annihilation vertex compared with present technology. We feel that this impressively clean 
experiment offers strong encouragement to believe that the sensitivity of an upgraded 
experiment along the same lines could be improved given the great progress made in slow neutron optics since this experiment was performed.

\begin{figure}
  \begin{center}
    \scalebox{0.64}{\includegraphics{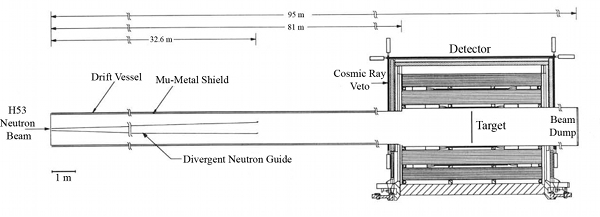}}
    \caption{The ILL $n - \bar n$ oscillation search experiment. Details of the detector are included in Fig.~\ref{nnbarx:fig:illdetector}~\cite{BaldoCeolin90}. }
  \end{center}
\end{figure}

\begin{figure}
    \centering \includegraphics[width=0.99\textwidth]{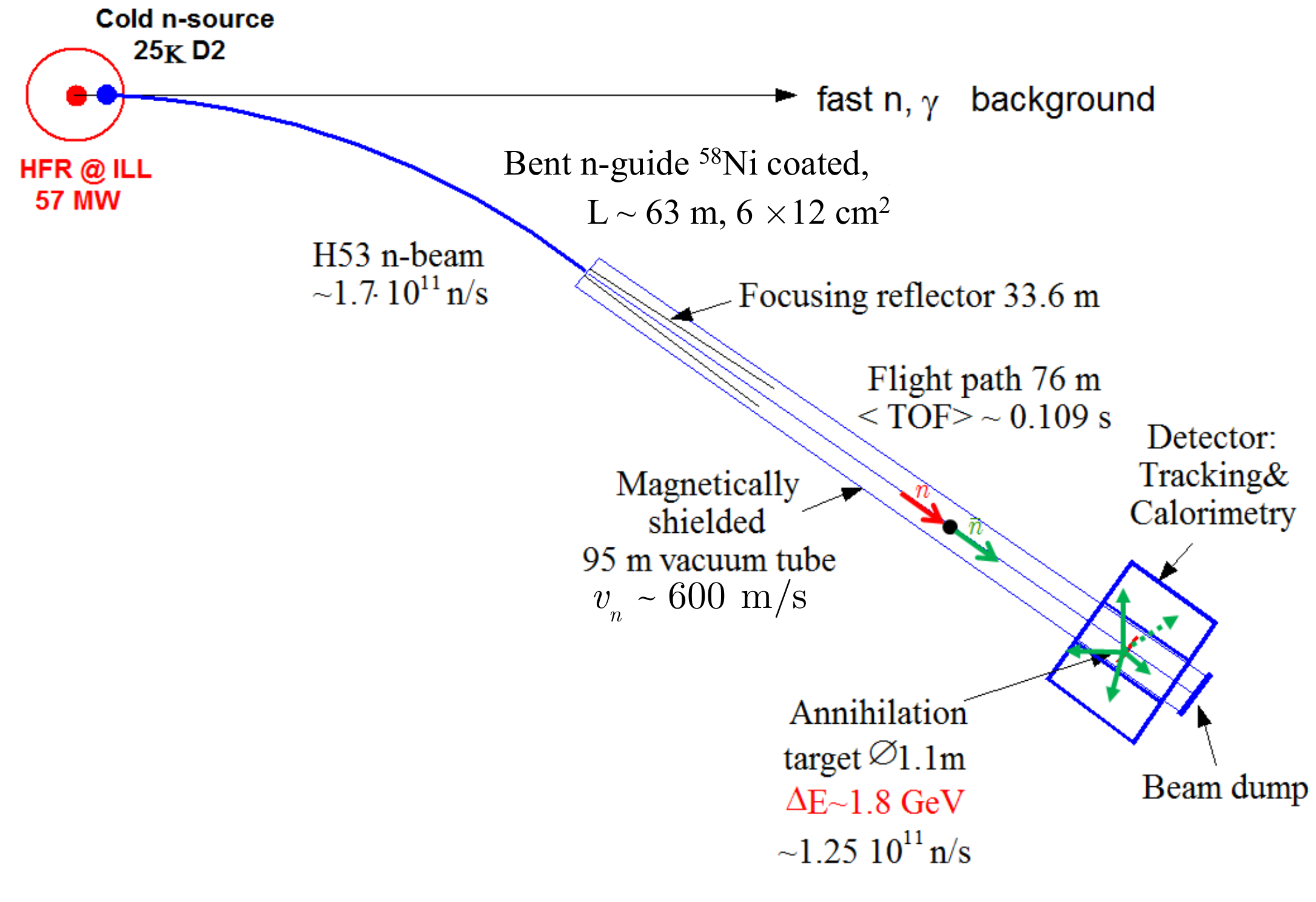}
    \caption{Configuration of the horizontal $n - \bar n$ search
    experiment at ILL/Grenoble~\cite{BaldoCeolin94}.}
    \label{ill:fig:logo}
\end{figure}

%%%%%%%%%%%%%%%%%%%%%%%%%%%%%%%%%%%%%%%%%%%%%%%%%%%%%%%%%%%%%%%%%%%%%%%%%%%%%%%%%%%%%%%%
\subsection{Improved Free Neutron Searches for $n - \bar n$ Oscillations}
\label{nnbar:subsec:improvedfree}
%%%%%%%%%%%%%%%%%%%%%%%%%%%%%%%%%%%%%%%%%%%%%%%%%%%%%%%%%%%%%%%%%%%%%%%%%%%%%%%%%%%%%%%%
The last two decades have seen considerable advances in the technologies of neutron transport/optics and neutron moderation, and these developments now make it possible to improve considerably upon the limits established by the ILL experiment described in the previous section.  Below we describe how these technological advances could be incorporated into a new free $n - \bar n$ oscillation search experiment provided that this new experiment is sited at a facility where its special requirements can be incorporated into the source design at the outset.

The first thing to realize about any attempt to conduct an improved $n - \bar n$ oscillation search with free neutrons is that there will be essentially no help from improved neutron source brightness at the initial point of liberation of the neutrons from nuclei.~\ref{nnbar:apdx:nsources} briefly discusses some aspects of intense neutron sources based on fission and spallation. Unlike the 1 GW power reactors sometimes used for neutrino oscillation studies, the relevant parameter for intense slow neutron sources is not total flux (and therefore total power) but rather neutron brightness (and therefore power density). The power density near the core of research reactors like the Institut Laue-Langevin (ILL) is already close to the present engineering limits beyond which the reactor core is in danger of melting.  The most ambitious design project known to us for a next-generation reactor-based slow neutron facility, the Advanced Neutron Source considered at ORNL in the early 1990's, was able with difficulty to achieve an increase of about a factor of $5$ in the neutron brightness near the core at the cost of a rather complex design. In the end this project was not pursued, and no similar sources are under consideration at present. 

The other technique which can supply bright neutron sources uses proton spallation in heavy nuclei. The 1 MW Spallation Neutron Source (SNS) and the 1 MW Japanese Spallation Neutron Source (JSNS) are about one order of magnitude more efficient in the number of neutrons produced per unit dissipation of power in the target but still possesses a time-averaged neutron brightness which is about a order of magnitude below that from the ILL.  Only the 5 MW European Spallation Source (ESS) planned for early 2020 foresees a time-averaged neutron brightness equal to the ILL.  Therefore, the opportunities for improving the measurement precision rely mainly on advances in the efficiency and effectiveness of neutron moderation near the primary source, on the increased phase space acceptance of the devices which deliver the neutrons from the moderator to the annihilation target, and on the arrangement of the experimental geometry to maximize the observation of the free flight time of the neutron while maintaining the quasi-free condition. Furthermore, present neutron technology does not yet possess the capability to increase the phase space density of the slow neutrons during their delivery from the moderator to the annihilation target, which means that we must assume that the motion of the neutrons after leaving the moderator conserves phase space density according to Liouville's theorem. This constraint fundamentally limits how the phase space of the beam can be manipulated. We therefore keep these constraints in mind in presenting the opportunities for improvement which we describe below. 

\begin{figure}[hbtp]
  \centering
    \scalebox{0.9}{\includegraphics{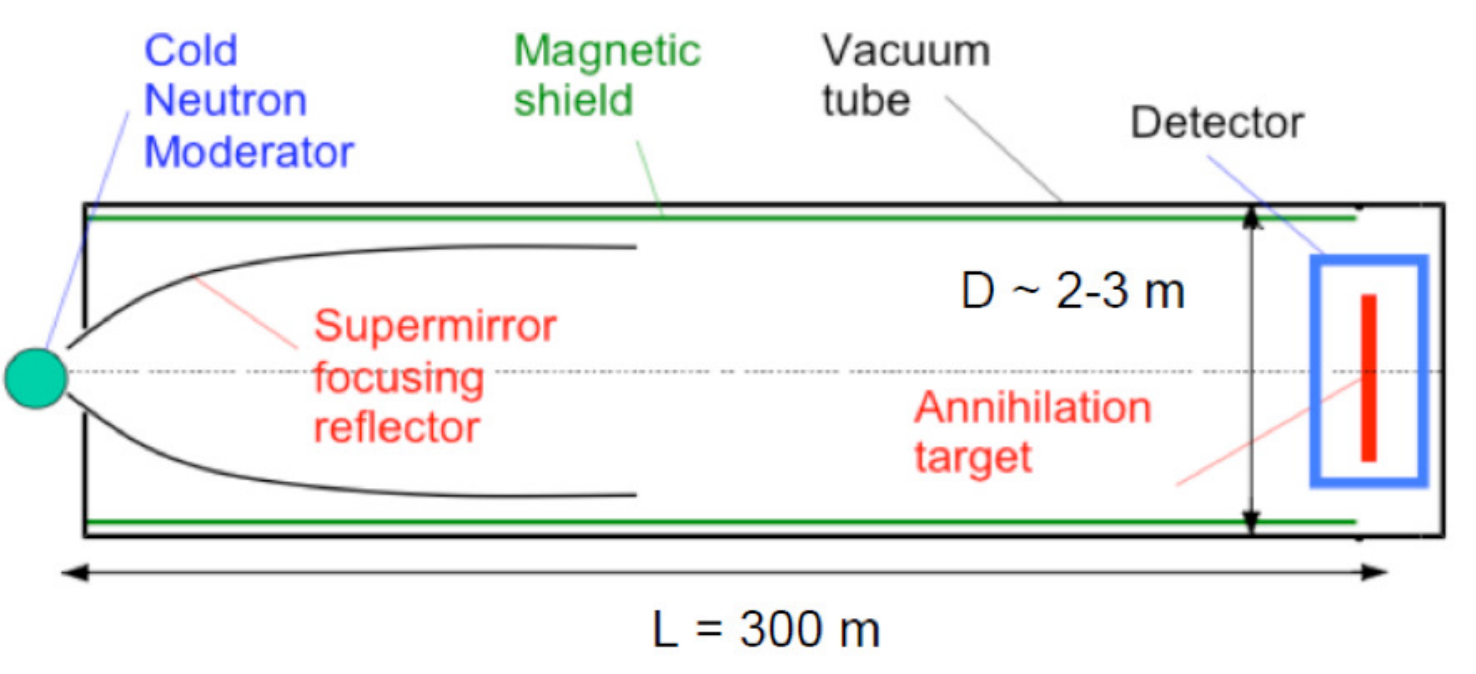}}
    \caption{Schematic view of a $n - \bar n$ oscillation search experiment with horizontal layout.}
  \label{nnbarx:fig:horiz}
\end{figure}

The single most important improvement in the sensitivity of a $n - \bar n$ search can be achieved by focusing neutrons 
with neutron reflecting surfaces, e.g. of elliptical shape. If a point-like neutron source is 
placed in the focus of an ellipsoid, neutrons specularly reflected from the inner surface of the ellipsoid
will arrive at the second focus of the ellipse.  Supermirror surface reflecting technology is discussed
in \S\ref{nnbar:subsec:noptics} of this paper. Using a truncated large elliptical focusing reflector~\cite{Kamyshkov95} around 
the beam axis makes it possible to intercept neutrons emitted from a larger solid 
angle of the source and direct them by single reflection to the desired position along the beam.  Despite the finite dimensions of the source and the limited phase space acceptance of the supermirrors, which reflect only neutrons with incident transverse velocities $\lesssim$ 40 m/s, very effective focusing can still be achieved 
in the focal plane where the image intensity can be 2-3 orders of magnitude higher 
than without a focusing device based on the simulations presented in section 6. Only the slow part of the neutron spectrum can be efficiently transported in this 
way, but this part is most valuable for optimizing the $Nt^{2}$ figure of merit for $n - \bar n$.  

If no focusing optics are present, the solid angle subtended by the annihilation target (and number of observed neutrons $N$) is proportional to $L^{-2}$ and the square 
of neutron time-of-flight $t^{2}$ is proportional to $L^{+2}$, so the sensitivity $Nt^{2}$ does not 
depend on the distance $L$ between source and annihilation target in the absence of a focusing reflector.
The use of a focusing reflector allows enhancement of the sensitivity roughly proportional to 
$L^{2}$.  Earth's gravity can significantly reduce the sensitivity of focused cold neutron beams because 
the most valuable slowest neutrons will fall and miss the annihilation target.
Optimization of the sensitivity for a given energy spectrum of source neutrons can be achieved by 
adjustment of the length $L$ as well as parameters of reflector, source, and annihilation target.

The neutron spallation target/moderator/reflector system and the experimental
apparatus need to be designed together in order to optimize the sensitivity of
the experiment. For this work, we consider a spallation target system based on
a 1 GeV proton beam linac operating at 1 mA. The easiest version of the
experiment to realize from a civil engineering point of view assumes an
experimental apparatus in a horizontal geometry in a configuration similar to
the ILL experiment~\cite{BaldoCeolin94}, but employing modernized technologies
including an optimized slow neutron target/moderator/reflector system and an
elliptical supermirror neutron focusing reflector. Based on our studies,
presented in part below, we estimate that such an arrangement appears to be
able to improve the sensitivity to the probability for $n - \bar n$
oscillations by 2-3 orders of magnitude beyond the limits obtained in the ILL
experiment~\cite{BaldoCeolin94}. This level of sensitivity would surpass the $n
- \bar n$ oscillation limits obtained in the Super-Kamiokande, Soudan-II, and
SNO intranuclear searches~\cite{Chung02,Abe15,Bergevin10}.  The correspondence
between free neutron and intranuclear $n - \bar n$ transformations is discussed
in \S\ref{nnbar:subsubsec:matter} and \S\ref{nnbar:sec:intranuclear} of this
paper.

As described in \S\ref{nnbar:subsubsec:bfield}, the figure of merit for the sensitivity of a free
$n - \bar n$ search experiment is $N_{n}\cdot t^{2}$, where $N_{n}$
is the number of free neutrons observed and ${\it t}$ is the neutron
observation time. The initial intensity of the
neutron source was determined in the ILL experiment by the brightness
of the liquid deuterium cold neutron source and the transmission of
the curved neutron guide.  Although in principle one expects the
sensitivity to improve as the average velocity of
neutrons is reduced, it is
not practical in a horizontal geometry to use very cold ($<$ 200 ${\rm m/s}$) and ultracold
neutrons (UCN, defined roughly as neutrons with speeds of $<$ 7 ${\rm m/s}$).  Earth's gravity will not allow free transport of very
slow neutrons over significant distances in the horizontal direction without having then eventually bounce from the vacuum chamber wall and therefore bring to an end the free oscillation time by subjecting them to the large $n - \bar n$ optical potential difference in matter.  These very slow neutrons will not reach the annihilation target and will be lost in the search for $n - \bar n$.  As a result,
the ultimate sensitivity of a free neutron oscillation experiment on Earth can only be reached
with a vertical orientation for the flight path from source to detector. We estimate that this approach could achieve an additional factor of $\sim$ 100 in experimental sensitivity, corresponding to limits for the oscillation time parameter of $\tau_{n-\bar{n}} > 10^{10}$ $\rm{s}$. However we do not present a specific design for that option in this paper, in part because of the increased expense required for the magnetic shielding and vacuum, in part due to the expenses associated with excavating a vertical hole deep enough to take advantage of this operational mode (we have not yet been able to identify an existing vertical hole which would fulfill the needed requirements), and in part due to the possible complications involved in putting an intense neutron source directly over a large hole. 

It is clear that any new search for oscillations in a free neutron beam
will require a very intense source of cold neutrons. Such neutron beams are available at
facilities optimized for condensed matter studies focused on neutron
scattering. These sources may be based on high flux reactors such as
the ILL, the High Flux Isotope Reactor (HFIR), and the Dubna IBR-2 pulsed reactor, or on accelerator
based spallation sources such as SNS, JSNS, or SINQ.  However, all such existing neutron
sources are already designed and optimized to serve a large number of neutron
scattering instruments that each require relatively small neutron beams compared to what would be interesting for a neutron oscillation experiment. There are no beams to our knowledge suitable for this experiment constructed at existing sources as the beam size is limited to provide the energy or momentum resolution
necessary for virtually all neutron scattering spectroscopy instruments suitable for materials
research. A fully optimized neutron source for an $n - \bar n$ oscillation
experiment would require a beam having a larger cross section and
larger solid angle in comparison with those available at neutron scattering facilities. The creation of such a beam at an existing facility would
typically require very major modifications to the source/moderator/shielding
configuration that would seriously impact its efficacy for neutron
scattering. This access issue is one important reason there has been no improvement
in the limit on free neutron $n - \bar n$ oscillations since the ILL
experiment of 1994. Below we argue that a beam at a MW-scale spallation source designed 
specifically for delivering intense cold neutrons for nuclear and
particle physics experiments is an essential element of any slow neutron experiment capable of extending the current limit on
$\tau_{n-\bar n}$ by many orders of magnitude.

In passing we also note that relatively modest improvements in the magnetic
field and vacuum levels reached for the ILL experiment would still
assure satisfaction of the quasi-free condition for a horizontal
experiment, but that more stringent limits on the magnetic field
($|{\vec B}|\leq 1$ ${\rm nT}$ in the whole free
flight volume) and vacuum (better than $P\sim 10^{-5}$ ${\rm Pa}$) by about an order of magnitude
would be needed for a vertical experiment. The practical difficulties of realizing these somewhat more stringent goals does not seem to be an issue of principle which makes the experiment impossible. Still this will be one of many considerations that would have to be taken into account
in designing a future experiment.

\begin{figure}
    \centering 
    \subfloat[]{\includegraphics[width=2.4in]{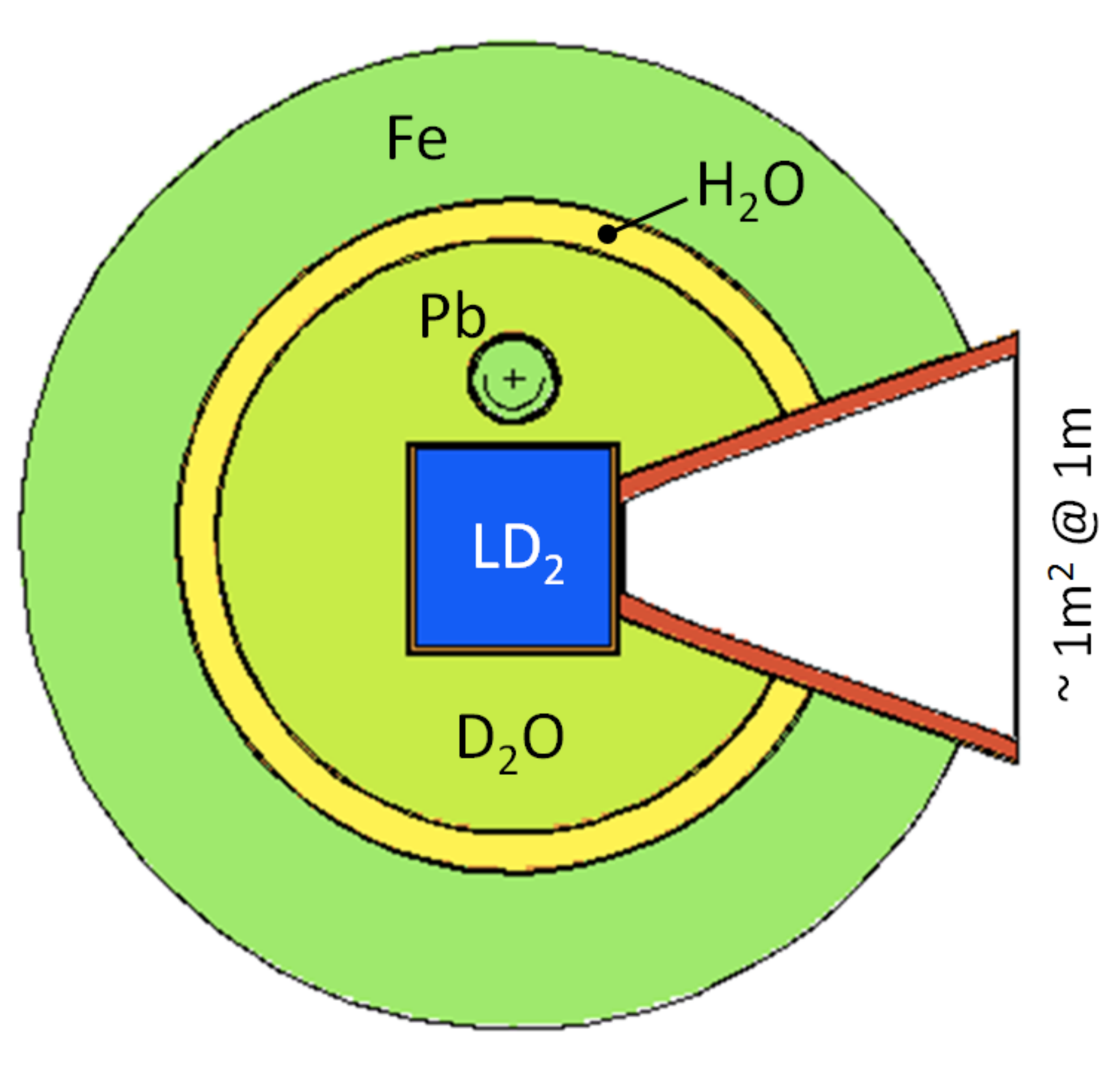}}
    \subfloat[]{\includegraphics[width=3.2in]{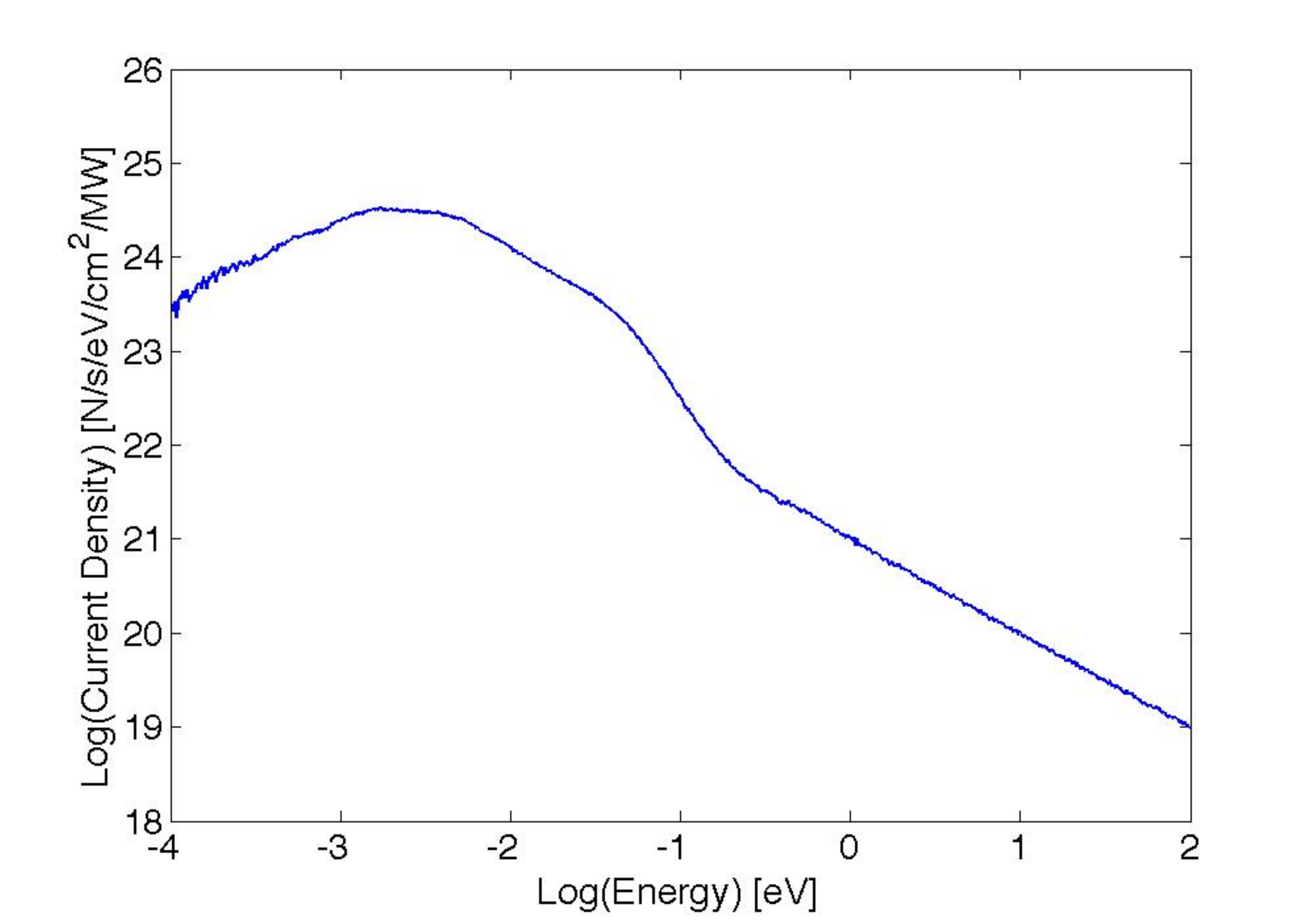}}
    \caption{a) Depiction of a layout of a baseline
cold neutron source geometry.  b) Depiction of a MCNP simulation of the cold neutron
spectrum entering the neutron optical system.}
    \label{nnbarx:fig:source}
\end{figure}

Target assemblies for producing neutrons at MW power levels present significant
engineering challenges that range from very large heat flows from relatively
small volumes to extreme radiation damage in components. 
Special challenges relevant to a neutron oscillation experiment at a new source include the mitigation of radiation damage in delicate neutron optical
components in close proximity to the moderator and the provision of adequate high energy neutron shielding for a beam that is much larger in size than any available at these existing sources.  In \S\ref{nnbar:sec:coldneutrons}, \S\ref{nnbar:sec:sensitivity} and \S\ref{nnbar:sec:randd}, we review some of the specifications for operating 1 MW spallation neutron sources, our
strategy to increase the number of neutrons directed to the annihilation target, and the sensitivity improvements relative to the ILL experiment which can be expected from an experiment in a horizontal geometry.

%%%%%%%%%%%%%%%%%%%%%%%%%%%%%%%%%%%%%%%%%%%%%%%%%%%%%%%%%%%%%%%%%%%%%%%%%%%%%%%%%%%%%%%%
\section{Intranuclear Searches for $n - \bar n$ Oscillations}
\label{nnbar:sec:intranuclear}
%%%%%%%%%%%%%%%%%%%%%%%%%%%%%%%%%%%%%%%%%%%%%%%%%%%%%%%%%%%%%%%%%%%%%%%%%%%%%%%%%%%%%%%%

In this section we discuss neutron to antineutron oscillation searches for neutrons bound 
inside nuclei. Theoretical treatment of the suppression mechanism of $n - \bar n$ oscillations
for this case is described in \S\ref{nnbar:sec:physics}.

%%%%%%%%%%%%%%%%%%%%%%%%%%%%%%%%%%%%%%%%%%%
\subsection{Previous Intranuclear Searches for $n - \bar n$ Oscillations}
\label{nnbar:subsec:previousintra}
%%%%%%%%%%%%%%%%%%%%%%%%%%%%%%%%%%%%%%%%%%%

Indirect limits on $\tau_{n - \bar{n}}$ and therefore $\delta m$ have been set by the absence of evidence for spontaneous $n - \bar n$ transitions in nuclei in large underground detectors built mainly for proton decay and neutrino oscillation studies. In large underground detectors the very large suppression in mixing from the difference of the neutron and antineutron optical potential in the nuclear medium is compensated by the large number of nucleons available for observation in a nucleon decay detector such as Soudan-2~\cite{Chung02} or Super-Kamiokande~\cite{Abe15}, the latter of which contained $\sim 10^{33}$ neutrons in the fiducial volume of the detector. Limits on matter instability due to $n - \bar n$ oscillations have therefore been reported by several nucleon decay experiments~\cite{Beringer12}.  The striking signature from an antineutron annihilation event, consisting of $\sim 2$ GeV of energy shared by $\sim 5$ pions with zero net momentum, can be clearly visible in even a large volume detector. However these reaction products are emitted from a point within the nucleus
($^{16}$O in a water Cherenkov detector and mainly $^{56}$Fe in the Soudan-2 detector) and interact as they propagate through the nucleus.  This is a complicated process to model and it introduces some extra uncertainty in the interpretation of the results. 

Neutron-antineutron transformations inside nuclei are greatly suppressed by the different interactions neutrons and antineutrons experience in nuclei. This suppression is parametrized by a dimensional factor $R$ that relates $\tau_{m}$ to the free $n-\bar{n}$ oscillation time $\tau_{n- \bar n}$ as 
$\tau_{m}=R\tau_{n- \bar n}^2$~\cite{Dove83,Friedman08,Albe82,Huef98}.  One of the recent experimental result for such a search was obtained by the Soudan-2 Collaboration~\cite{Chung02}.  They extracted a limit of $1.3\times 10^8$ s for the free $n - \bar n$
oscillation time from the measured lifetime limit 
for the absence of $^{56}$Fe nuclei decaying into multi-pion final states of $\tau_{m} > 0.72\times 10^{32}$ years. This limit and the limits from other previous and more recent 
bound neutron experiments are given in Table~\ref{nnbar:table:limits} along with the best limit for free 
neutron oscillation time obtained in the ILL reactor experiment~\cite{BaldoCeolin94}.

\begin{table}
\begin{minipage}{\textwidth}
\begin{center}
\begin{tabular}{|c|c|c|c|c|}
  \hline
Experiment & 10$^{32}$ {\it n}-yr &$\tau_{m}$(10$^{32}$ yr) & $R$(10$^{23}$/s) & $\tau_{n- \bar n}$(10$^{8}$ s) \\ \hline
ILL (free-$n$)~\cite{BaldoCeolin94} & n/a & n/a & n/a & 0.86 \\ \hline
IMB ($^{16}$O)~\cite{Jones84} & 3.0 & 0.24 & 1.0 & 0.88 \\ \hline
Kamiokande ($^{16}$O)~\cite{Takita86} & 3.0 & 0.43 & 1.0 & 1.2 \\ \hline
Frejus ($^{56}$Fe)~\cite{Berger90} & 5.0 & 0.65 & 1.4 & 1.2 \\ \hline
Soudan-2 ($^{56}$Fe)~\cite{Chung02} & 21.9 & 0.72 & 1.4 & 1.3 \\ \hline
SNO ($^{2}$H)~\cite{Bergevin10} & 0.54 & 0.30 & 0.25 & 1.96 \\ \hline
Super-K ($^{16}$O)~\cite{Abe15} & 245 & 1.9 & 0.517 & 2.7 \\ \hline
\end{tabular}
\caption{Neutron-antineutron lifetime lower limits (90$\%$ CL).}
\label{nnbar:table:limits}
\end{center}
\end{minipage}
\end{table}

All experiments in this table possess detection efficiencies from 10-50$\%$ and suffer from irreducible backgrounds generated by the interactions of 
atmospheric neutrinos in the underground detectors. In the presence of irreducible backgrounds it 
is possible to set higher limits for the nuclear lifetime, but it is impossible to discover a new effect 
unambiguously with a modest improvement of detector mass or exposure time.
    
Table~\ref{nnbar:table:limits} also gives limits obtained recently by the SNO~\cite{Bergevin10} (preliminary result based 
on a fraction of total statistics) and Super-K~\cite{Abe15} collaborations.  The limit from the SNO detector is close to the limit obtained by the much larger 
Super-K experiment partially due to smaller nuclear suppression factor for deuterium.  A more recent evaluation of this factor from a field-theoretical approach
is given in recent publications~\cite{Kopel11}, where the role of the possible spin
structure of the $p-\bar{n}$ annihilation amplitude has been studied as well.  The SNO 
result can be improved using the complete SNO data set, which is a factor of 4 times larger than the sample used in the existing analysis.  A careful analysis of 
the SNO data is especially important in this context as the deuteron is a simple enough system 
to imagine the possibility of a much more reliable theoretical calculation of $R$.

The limit on the free $n-\bar{n}$ oscillation time from the recent Super-Kamiokande experiment is $2.7 \times 10^8$ s~\cite{Abe15}.  
The Super-Kamiokande limit was derived from $24$ observed candidate events with a selection 
efficiency of 12.1\% and with an estimated background of $24.1$ events from atmospheric neutrino 
interactions in the detector. This inherent atmospheric neutrino background makes further improvement 
of the $n-\bar{n}$ search difficult in water-Cherenkov detectors larger than Super-Kamiokande, such as the proposed Hyper-Kamiokande detector~\cite{Abe11}. The expansion of the size and granularity of these large water Cherenkov-based 
underground detectors, combined with the presence of significant atmospheric neutrino backgrounds 
which already affects the extraction of the present limits, seems to indicate that it will be difficult 
in the future to use these detectors to significantly improve the present limits without the development 
of a new detector technology which can be deployed in large volumes. The large new proposed 
liquid Argon detectors (LBNF/DUNE, GLACIER) ~\cite{LBNE14, GLACIER} may have better suppression of atmospheric neutrino backgrounds than water-Cherenkov detectors due to their detailed tracking information, particle ranging, and particle identification capabilities.  Whether background-free
operation in these detectors corresponding to decay lifetimes of $\sim 10^{33} - 10^{35}$ yr will be possible has not yet been demonstrated.

Theoretical models used to predict the nuclear suppression factor have improved with time.  The most recent 
calculation for the $^{16}$O nucleus of $R = 0.5\times 10^{23}$ ${\rm s}^{-1}$~\cite{Friedman08}, 
yields an oscillation time lower limit of $2.7\times 10^{8}$ s. Existing theoretical 
calculations of the relationship between the rate of $n - \bar n$ oscillation in nuclei 
and the free $n - \bar n$ transition rate seem to capture the dominant physics~\cite{Friedman08}, 
but various processes not previously considered are known to exist.  Since the isospin 
change for $n-\bar{n}$ in nuclei can be $\Delta I$ = 1, 2, or 3~\cite{Vain13} (for free neutrons only 
$\Delta I=1$ is possible) one can also get contributions to the nuclear conversion process from $n - \bar n$ oscillations in the nucleus and from dinucleon 
conversion into pions or kaons, which would give a very similar signal in the underground detectors. A qualitative discussion of the importance of
such processes relative to $n - \bar n$ in nuclei was presented in \cite{Dove83} and references therein. 

The constraint from underground detector data therefore corresponds to some combination 
of $n - \bar n$ oscillations and these other possibilities. Although the relative size of these 
different contributions can be constrained using QCD sum rule techniques, few such calculations have 
yet been performed. Such a calculation will help the scientific community to better judge the relative 
sensitivities of $\Delta\mathcal{B}$ = 2 search processes between free neutron and underground detectors.

%%%%%%%%%%%%%%%%%%%%%%%%%%%%%%%%%%%%%%%%%%%%%%%%%%%%%%%%%%%%%%%%%%%
\subsection{Improved Intranuclear Searches for $n - \bar n$ Oscillations}
\label{nnbar:subsec:improvedintra}
%%%%%%%%%%%%%%%%%%%%%%%%%%%%%%%%%%%%%%%%%%%%%%%%%%%%%%%%%%%%%%%%%%%

Future experiments with large underground detectors may improve the limits for $\Delta \mathcal{B} = 2$ processes.  The question of how much these limits might improve, along with the whole subject of the experimental prospects for improvement on the limits for proton 
decay, is beyond the scope of this work. However we mention some examples of projects 
under discussion or development which might be able to produce such improvements in the future. One example 
is a larger underground detector based on the well-established water-Cherenkov detector technology, such as Hyper-K~\cite{Abe11} and another is
the large liquid argon detector proposed for the long baseline neutrino experiment (LBNF/DUNE)~\cite{LBNE14}. For these detectors, one can make a 
reasonable set of extrapolations for the potential reach for $n - \bar n$ oscillations using the fact that the present versions of such detectors possess an 
irreducible background from atmospheric neutrinos that will be scaled up with detector mass and exposure time. In this case it will be possible to set a lower limit for $n - \bar n$, but will not be 
possible to claim a discovery of a new effect. For example, the new result of Super-K experiment on  $n - \bar n$ with the 
limit~\cite{Abe15} of $\tau_{m} > 1.9\times 10^{32}$ yr is based on the 24 selected candidate events with a background 
of 24.1 events. Here it cannot be excluded that a few events are due to the genuine $n - \bar n$ events. In the absence of background, one detected candidate in this data sample would correspond to the observation of a new $n - \bar n$ effect with a lifetime 
of $\sim 2.5\times 10^{34}$ yr.   Without improvements in water-Cherenkov 
technology the $n - \bar n$ lifetime limit obtained by proposed 500-kt Hyper-K~\cite{Abe11} detector for 10 years of exposure in our estimate would
increase only to $\sim 7.5\times 10^{32}$ yr.   
 
A newer technology which is under development and consideration for the LBNF/DUNE project~\cite{LBNE14} is based 
on liquid argon. In small volume prototypes these detectors have demonstrated impressive  spatial resolution 
and particle ID capability~\cite{ICARUS2011, Anderson2012, CAnderson2012, Acciarri2014}. The key question from the point of view of this work is whether these 
capabilities can be scaled up to very large detector volumes with the ability to eliminate the atmospheric neutrino background completely.  In the complete absence of 
atmospheric neutrino background, one candidate event in the 40-kt DUNE detector after 10 years of exposure time can correspond to a 
$n - \bar n$ discovery with a lifetime of $\sim 10^{35}$ yr.  If the complete elimination of background 
will not be possible, its significant suppression (comparing to the level of water-Cherenkov detectors) would
enable the exploration of $n - \bar n$ lifetimes in the range $10^{33}-10^{35}$ yr. These questions of potentially possible atmospheric neutrino background 
reduction in liquid argon detectors are under study~\cite{Kearns}.  The LBNF/DUNE detector will also allow exploration of other modes 
of nuclear instability related to $n - \bar n$ and to $\Delta\mathcal{B}$ = 2 and $\mathcal{B-L}$ violation, 
such as $NN\rightarrow K's$, $NN\rightarrow \pi's$, and others.
Modes of neutron decay with the poorest experimental limits, such as $n\rightarrow 3\nu$, $n\rightarrow 5\nu$ 
and $nn\rightarrow 2\nu$, are favored in some models~\cite{Mohapatra03}. If such processes were 
to occur in nuclei they would leave neutron holes whose de-excitation would leave characteristic 
signatures~\cite{Kamyshkov03}. Data from underground detectors built to observe neutrino oscillations can 
also place stronger constraints on such possible $n$ and $nn$ decay modes~\cite{Ahmed04, Back03, Araki05}. 

%%%%%%%%%%%%%%%%%%%%%%%%%%%%%%%%%%%%%%%%%%%%%%%%%%%%%%%%%%%%%%%%%%%%%%
\subsection{Complementarity of Intranuclear and Free Neutron Searches}
\label{nnbar:subsec:complementarity}
%%%%%%%%%%%%%%%%%%%%%%%%%%%%%%%%%%%%%%%%%%%%%%%%%%%%%%%%%%%%%%%%%%%%%%

In the free neutron oscillation phenomenology described in \S\ref{nnbar:sec:physics}, it was assumed that $m_{n}=m_{\bar n}$
as required by the $CPT$ theorem. However, this equality might be violated above the 
Planck scale~\cite{Abov84} with $\Delta m_{n} / m_{n} < m_{n} / m_{Planck}$ which is less than the established 
$\Delta m$ limit for neutral kaons~\cite{Beringer12}. 
We note that the current experimental upper bound on the mass difference 
for neutron and antineutron is quite poor: $\mid m_{n} - m_{\bar n}\mid/m_{n} = (9 \pm 6)\times 10^{-5}$
~\cite{Beringer12}.  More generally $\Delta M$ in Eq. (\ref{deltam}) might include some potentials different for neutron and 
antineutron, e.g. due to the gravity different for matter and antimatter ~\cite{Lamoreaux91} or the existence of a 
vector force with repulsive effect between baryons. In the recent paper~\cite{Babu2015} an interaction
different for neutron and antineutron was conjectured to come from Lorentz invariance violation. 

If $\Delta M \gg \delta m$ and $\Delta M \gtrsim 10^{-15}$ eV, vacuum oscillations of free neutrons 
will be strongly suppressed, while this suppression will not be essential for intranuclear $n - \bar n$ where many orders of 
magnitude stronger potential difference between neutron and antineutron is already present~\cite{Abov84}. It is possible that $n - \bar n$ transformation would occur for bound neutrons but will be suppressed for free neutron transformations. The following scenarios are possible: 

i) If $n - \bar n$ transformation is observed at rates in a free neutron experiment consistent with rates in large 
underground experiments with a corresponding suppression then this will yield a stronger limit on $\Delta m/m$ by a few orders of magnitude than the limit provided by the mass difference of neutral kaons.  
At the same time this observation will establish equivalence of gravitational interaction of neutron and antineutron
~\cite{Lamoreaux91} with the potential difference down to $\sim 10^{-15}$ eV.

ii) If $n - \bar n$ transformation will be observed only in large underground detectors in intranuclear transformations
and will not be seen with free neutrons due to $\Delta M \gg \delta m$, e.g. $\Delta M = 10^{-14}$ eV, 
then it will be possible to apply an external magnetic field that would compensate $\Delta M$ for free neutrons and thereby {\it unlock} the suppressed $n - \bar n$ free neutron transformation.

iii) If $n - \bar n$ will be observed only with free neutrons (e.g. if atmospheric neutrino background will be limiting 
the intranuclear measurements) then one could measure the appearance of antineutrons as a function of an 
externally controlled weak magnetic field. The field value which maximizes the $n - \bar n$ oscillation rate determines $\Delta M$.

A new mechanism of spontaneous baryon number violation under consideration~\cite{baryo-majoron} might have an interesting impact on intranuclear transformations: $n - \bar n$ inside nuclei can 
be significantly suppressed or amplified by this mechanism while free neutron transformations will remain 
unaffected. Thus, observation of different rates for free neutron and intranuclear $n - \bar n$ transformations might provide a hint for new physics.  In conclusion, we stress that $n - \bar n$ transformations in both types of experiments, with free neutrons and with neutrons bound inside nuclei, will be needed to extract the new physics.

%%%%%%%%%%%%%%%%%%%%%%%%%%%%%%%%%%%%%%%%%%%%%%%%%%%%%%%%%%%%
\section{Cold Neutron Sources and Optics for an Improved $n - \bar n$ Oscillations Search with Free Neutrons}
\label{nnbar:sec:coldneutrons}
%%%%%%%%%%%%%%%%%%%%%%%%%%%%%%%%%%%%%%%%%%%%%%%%%%%%%%%%%%%%

In this section we describe the major components of the spallation sources (targets and moderators) 
that fit the criteria for a modern $n - \bar n$ search experiment.  One needs a source 
that can provide maximum brightness from a large-area cold neutron moderator.  Although the coldest neutrons with 
velocities $v \leq$ 200 m/s cannot be efficiently used in the horizontal version of a $n - \bar n$ search 
experiment, these neutrons still constitute a small fraction of Maxwell-Boltzmann velocity distribution 
for cold neutron spectra that can be generated by practical cold moderators (i.e. cryogenic para-hydrogen,
or deuterium). Finding new mechanisms and materials for efficient neutron moderation to 
the lower spectrum of temperatures might enhance the sensitivity of a $n - \bar n$ search.  A very important 
factor for an advanced high-sensitivity $n - \bar n$ search is the ability to access a large solid angle of neutron emission from the moderator surface.  Most neutron scattering instruments use a typical acceptance angle 
of 1-3 degrees.  In order to take advantage of the focusing reflector option for a $n - \bar n$ search, the acceptance angle in the 
horizontal layout of experiment must be increased to 10-20 degrees. At the same time, the input aperture 
of the neutron transport system should allow one to view the whole area of the moderator. The actual acceptance angle
in the design of an experiment should be found from parameter optimization with overall construction cost 
included as one of the main parameters. In this section we also describe the concepts of the neutron transport 
system including the vacuum vessel, supermirror reflective optics, and active/passive magnetic shielding
that allow for creation of the ``quasi-free" conditions for neutrons flying between moderator and annihilation 
detector. The latter will be discussed in \S\ref{nnbar:sec:sensitivity}. 

%%%%%%%%%%%%%%%%%%%%%%%%%%%%%%%%%%%%%%%%%%%%%%%%%%%%%%%%%%%%
\subsection{Current and Future Spallation Sources}
\label{nnbar:subsec:current}
%%%%%%%%%%%%%%%%%%%%%%%%%%%%%%%%%%%%%%%%%%%%%%%%%%%%%%%%%%%%

Existing MW-class spallation sources include the SNS~\cite{Mason06} at ORNL,
SINQ~\cite{Blau09,Fischer97} at PSI, the Dubna IBR-2 pulsed reactor~\cite{Belushkin06}, and JSNS~\cite{Maekawa10} at J-PARC.  
The SNS uses a liquid mercury target running at 1 MW with a proton energy of 825 MeV and a
frequency of 60 Hz. The time-averaged flux of neutrons with kinetic energies below 5 meV at a
distance of 2 m from the surface of the coupled moderators is 1.4$\times 10^{9}$ ${\it n}/{\rm cm^{2}/s}$ at 1 MW~\cite{Iverson13}. The SINQ source is currently the strongest operating continuous mode
spallation neutron source in the world.  It receives an effectively continuous (51
MHz microstructure) 590 MeV proton beam at a current up to 2.3 mA.   Under normal
operation the beam current is typically 1.5 mA.  The SINQ source uses
a ${\it cannelloni}$ target made of an array of Zircaloy clad
lead cylinders.  The cold neutron beam contains a flux of 2.8$\times
10^{9}$ ${\it n}/{\rm cm^{2}/s}$ at 1 MW at a distance of 1.5 m from
the surface of the Target 8 coupled moderators~\cite{Wohlmuther11}.  The best potential for a $n-\bar n$ search among facilities now under construction will be provided by the European Spallation Source (ESS) that will be 
commissioned in Sweden near Lund in 2019~\cite{ESSTDR}.  The proton beam will be accelerated by a linac to an energy of 2 GeV and will be deposited on a rotating solid tungsten target with pulses widtha of $2.86$ ms at a repetition rate of 14 Hz. The time-average beam power on target will be 5 MW. The tungsten target will be arranged 
as a disk rotating in the horizontal plane with segments cooled by helium gas. High-density tungsten
will provide a very compact source of spallation neutrons.  A cold para-hydrogen moderator installed 
above the target disk and surrounded by a Be reflector will act as a cold source with brightness up to 
$7.4 \times 10^{13}$~$n/{\rm s/cm^{3}/sr}$.  The target design will allow for a beam 
port with large angle acceptance.  An option exists for a second moderator below the target~\cite{esben}.  We describe below some of the general issues associated with the design and operation of such sources. 

%%%%%%%%%%%%%%%%%%%%%%%%%%%%%%%%%%%%%%%%%%%%%%%%%%%%%%%%%%%%
\subsection{Spallation Target Assemblies and Remote Handling}
\label{nnbar:subsec:spalltgt}
%%%%%%%%%%%%%%%%%%%%%%%%%%%%%%%%%%%%%%%%%%%%%%%%%%%%%%%%%%%%

Spallation target assemblies are engineered to produce large numbers of neutrons that can later be moderated to very low energies.  Generally speaking spallation of ~GeV protons beans on high atomic number elements produce the highest brightness sources: at higher proton energies a growing fraction of the energy goes into pion production.  Possible liquid spallation neutron targets include Hg and Pb-Bi eutectic. Possible solid spallation targets include lead and tungsten alloys with a tantalum coating and either light or heavy water cooling. The biggest engineering problems associated with the spallation target bombarded by a 1-MW continuous proton beam are heat removal and radiation damage.  The proton beam spatial profile determines the largest energy density that can be cooled. Radiation damage in 316L alloy stainless steel has been researched extensively and is often employed as a key target component.  With some solid targets \lq\lq after-heat~\rq\rq, namely, the heat generated in the target by radioactive substances even after the proton beam is off, must be considered in the design.  Liquid targets such as mercury or lead bismuth flowing inside a 316L boundary are used at the SNS and JSNS~\cite{Maekawa10}. Solid target materials with internal water cooling channels have also been used in high intensity spallation target applications.  The target material is subject to high mass specific thermal loading, and both peak temperature and temperature gradients in the solid material can encounter structural design constraints.  Peak thermal fluxes at the solid to coolant interface in a high performance cooling circuit can approach 10$^{6}$ W/m$^{2}$.  The high thermal flux and typical small cooling channel cross sections naturally leads to rapid bulk temperature rise in the coolant and the need to limit coolant residence times.  These constraints in turn lead to high coolant flow velocities and rapid fluid pressure losses.  Aggressive cooling designs are required to reduce the volume of low density coolant in the target volume, which degrades the neutronic performance.   Solid rotating targets under study for the SNS second target station and for the ESS long-pulse spallation target can also be considered.

Spallation neutron targets rapidly become extremely radioactive. The primary purpose of the required remote handling system is to optimize the availability of the target systems while protecting personnel. This can be accomplished through the development of a remote handling system capable of safely performing the required maintenance operations and by designing the target system components in a modular fashion suitable for remote maintenance and/or replacement. Remote handling systems include methods of remote manipulation, lifting, viewing, and special tooling. Many facilities incorporate hot cells with shielded viewing windows and thru-wall mechanical manipulators. The SNS also incorporated a highly-dexterous, bridge-mounted servo-manipulator system as well as a separate bridge-mounted crane to provide remote maintenance capability throughout the entire cell; remote operations are augmented with a closed-circuit viewing system using in-cell cameras tolerant of high radiation doses. Special tools and lift fixtures are required for most component handling operations, and these are often unique to a particular task so that several are required for a given operation.

Remote maintenance facilities are also heavily influenced by the severity of radiation and contamination hazards; the use of liquid targets such as Hg or lead-bismuth eutectic (LBE) requires a dedicated hot cell to contain both the liquid and gaseous hazards. Shielded storage casks are required for any activated or contaminated components that are removed from the shielded target environment, and in some cases shielded containers are utilized within a hot cell to minimize radiation damage to the in-cell maintenance equipment. In most all cases, unique casks are needed for each component due to differing component geometry and radiation levels.

Waste handling is another aspect that must be included in the early phases of design of the remote handling systems, the target system components, and the target facility itself. For example, SNS utilizes the TN-RAM waste cask, which is the largest waste cask licensed for over-the-road transport of solid nuclear materials in the U.S.  At the moment there exists only a single cask of this type. The size constraints imposed by this cask have significantly influenced the design of SNS target components. The SNS hot cell incorporates a special waste port designed to allow docking of this cask to the hot cell for remote loading. If size reduction of components is required in order for them to fit within the TN-RAM cask, special remote tooling is used within the hot cell. Design for waste handling is required to ensure a cradle-to-grave path exists for each required target system component.

%%%%%%%%%%%%%%%%%%%%%%%%%%%%%%%%%%%%%%%%%%%%%%%%%%%%%%%%%%%%
\subsection{Neutron Moderator System}
\label{nnbar:subsec:nmod}
%%%%%%%%%%%%%%%%%%%%%%%%%%%%%%%%%%%%%%%%%%%%%%%%%%%%%%%%%%%%

The purpose of the moderator system is to reduce the energy of the spallation neutrons generated in the target to energies of a few meV. We discuss neutron moderation theory in~\ref{nnbar:apdx:nmod}. It is perhaps not surprising that the phase space compression (cooling) of neutral particles such as neutrons is especially challenging, but it is worth outlining the specific issues encountered in the slow neutron regime. Since the neutrons of interest lie in an energy range (meV) in which the neutron does not possess internal degrees of freedom accessible by the application of external fields (neutron spin flips in external magnetic fields are only in the $\mu$eV range), the only efficient way at present to increase neutron phase space density is through collisions in a cold material medium. In this case we must resolve a contradiction: when we cool the medium to very low temperatures, the degrees of freedom which can interact inelastically with slow neutrons whose wavelengths are larger than interatomic spacings of the material tend to freeze. It is known that the cross sections for neutron cooling through the universally-available mechanism of phonon creation suffer from a reduced accessible phase space for lower energy neutrons proportional to $\omega^{3}$ where $\omega$ is the phonon frequency. Furthermore the inelastic elementary excitations in condensed matter systems generally possess energy-momentum dispersion relations which intersect that for a free neutron at only a few specific energy and momenta, thereby further reducing the fraction of inelastic events in the medium. We also must take into account the increased importance of neutron absorption in nuclei for the slower neutrons, especially since the slowest neutrons are the most important ones for this experiment. Therefore one cannot simply stick slow neutrons in an arbitrary cold solid medium and expect the neutrons to efficiently cool to the temperature of the medium. 

Extensive research has been performed at neutron scattering facilities on slow neutron moderator materials which optimize various combinations of brightness and neutron energy spectrum shape. For a fixed transverse momentum acceptance of the neutron optical components downstream of the moderator, a slower spectrum allows neutron guides to accept and transport neutrons from a larger solid angle in the source and thereby increase the intensity in the experiment. The intense radiation field experienced by the moderator near a MW spallation target poses both a cryogenic engineering challenge and also a safety challenge which further restrict somewhat the choice of moderator materials. Both liquid hydrogen and liquid deuterium are known to be able to tolerate this environment. The neutron moderating properties are also strongly influenced by the spin states of the molecules as the spin flips cause transitions between rotational molecular levels with energy separations on the meV scale. The fact that there is still some room for optimization of the cold neutron brightness from liquid hydrogen moderators can be seen in a comparison between the brightness of the SNS liquid hydrogen moderators and those of the JSNS, which are brighter by about a factor of 2 because of their choice of liquid parahydrogen as the moderator medium combined with a beryllium reflector. Both materials possess a total neutron scattering cross section which falls rapidly for neutrons with energies below some critical energy, which allows the moderator to keep the neutrons in a relatively small spatial volume as they are slowing down from higher energies and release them as they fall below this critical energy. Recent investigations for the proposed moderators at the ESS have also exploited the special properties of liquid parahydrogen to demonstrate the possibility for increased neutron brightness.  

%%%%%%%%%%%%%%%%%%%%%%%%%%%%%%%%%%%%%%%%%%%%%%%%%%%%%%%%%%%%
\subsection{Neutron Optics}
\label{nnbar:subsec:noptics}
%%%%%%%%%%%%%%%%%%%%%%%%%%%%%%%%%%%%%%%%%%%%%%%%%%%%%%%%%%%%

Advances in slow neutron optics technology enable a more sensitive $n - \bar n$ oscillation experiment with cold neutrons. The ILL $n - \bar n$ oscillation search experiment employed what are now viewed as relatively low phase space acceptance nickel guides of nominally rectangular cross section for neutron propagation between the moderator and the free flight path of the experiment. The nickel presents an effectively repulsive average optical potential energy to the slow neutrons, and the reflectivity of a neutron with transverse momentum $p_{T}$ which satisfies $V_{opt}>{p_{T}^{2} \over 2m}$ is ideally unity. One can crudely estimate $V_{opt}$ as follows: if one assumes that the ~10 MeV potential energy of a neutron in the nucleus of radius ~1 fm is averaged over the volume of a material made of atoms whose size is larger than the nucleus by a factor of $~10^{5}$, then the average potential energy $V_{opt}\sim 100$ neV. For a non-relativistic neutron of kinetic energy of a few meV one can therefore get total external reflection for angles of incidence of a few mrad. Since this ILL experiment multilayer coatings of alternating materials with large neutron optical potential differences with a graded distribution of separations called \lq\lq supermirrors\rq\rq~\cite{Mezei76} have been developed which greatly increase the range of  transverse momenta which can be reflected with high probability and therefore the transverse phase space acceptance from the source by exploiting coherent interference of scattering from the different layers (diffraction) over the full spectrum of neutron wavelengths from the cold source.  Supermirror performance is roughly classified by a parameter $m$ defined by $p_{T}=mp_{T,Ni}$ which measures the increase of the critical angle of reflection relative to nickel of natural isotopic abundance where specular reflection still occurs. Commercially available supermirror materials constructed out of alternating layers of nickel and titanium now have $m$-values of up to seven, and since the transverse phase space acceptance of the guide system from the moderator scales as $m^{2}$ this represents a major increase in the number of neutrons that can be delivered to an antineutron annihilation target. In addition to this major extension in the range of transverse momenta of neutrons that can be reflected, supermirror neutron optics have also been successfully deployed on curved surfaces. This capability can also increases the transverse phase space acceptance of a neutron optical guide system. This becomes clear if one considers the extreme case of a point neutron source and a point antineutron annihilation target: in this case the optimum mirror arrangement consists of an elliptical mirror reflector with the source at one focus and the target at the other focus, which obviously delivers more neutrons than a guide system of rectangular cross section with the same $m$. Curved neutron guides can now be made commercially and are starting to find applications in neutron scattering facilities. Tests using a parabolically-shaped neutron guide coated with $m$=3 supermirrors showed excellent agreement between simulations and measurement. The neutron guide implemented for the WISH diffractometer~\cite{Schanzer04} at the ISIS second target station employs a horizontal 40 m elliptical neutron guide made of $m$=2 supermirrors enclosed in a large vacuum tank in a configuration not unlike that which could be adopted in a $n - \bar n$ oscillation experiment.

Note however that even high-m supermirror guides have relatively low angular acceptance (of order 10 mrad/${\rm \AA}$) for neutrons with typical energies from a cold neutron moderator. This makes the achievable increase in neutron phase space acceptance from the moderator heavily dependent on the geometry of the experiment. The proximity of the mirror reflectors to the neutron source and the maximum diameter of the reflection chamber should be chosen so that the phase space acceptance of the guides is fully illuminated. One also must take into account the susceptibility of the supermirrors to radiation damage. Helium gas cooled neutron supermirrors with $m=3.5$ deposited on polished flat aluminum substrates located are now known to withstand the SNS radiation environment about 1m from the cold neutron source over a few years of operation with no measurable degradation in performance. We are therefore optimistic that this issue will not become a show-stopper for the use of supermirror optics in a $n - \bar n$ oscillation experiment.

%{\bf Describe and include results from some of our previous simulations on this subject}

%%%%%%%%%%%%%%%%%%%%%%%%%%%%%%%%%%%%%%%%%%%%%%%%%%%%%%%%%%%%
\subsection{Vacuum and Magnetic Shielding}
\label{nnbar:subsec:vacuum}
%%%%%%%%%%%%%%%%%%%%%%%%%%%%%%%%%%%%%%%%%%%%%%%%%%%%%%%%%%%%

The magnetic moment of the antineutron has opposite sign to that of the neutron. An external magnetic field therefore splits the degeneracy of the neutron and antineutron energies and suppress the oscillation rate as mentioned above. To keep this effect negligible the neutrons must meet the quasi-free condition $\mu BT < 1$ where $\mu$ is the magnetic moment, {\it B} is the magnetic field, and {\it T} is the observation time. A rough estimate quickly shows that one requires a magnetic field in the $1-10$ nT regime to meet the quasi-free condition for the free observation time of a practical slow neutron oscillation experiment. This requires the suppression of the Earth's magnetic field by many orders of magnitude.

 The very large cylindrical volume of order 2 m radius and 100-200 m in length would be to our knowledge the largest magnetic shield ever constructed.  For the magnetic shielding geometries of both the previous ILL experiment and the proposed experiment, which use long shields with cylindrical symmetry, the dominant component of the residual magnetic field inside the shield is the component along the axis of the shield. The internal shield for the previous ILL experiment strongly suppressed transverse components of the magnetic field and rendered the longitudinal component sufficiently uniform that it could be largely compensated by a homogeneous external field generated by a coil wrapped on the outside of the shield~\cite{Bitter91,Kinkel92}. Once this major component to the residual field was removed, another set of coils were able to trim out the residual transverse fields. Current loops for shield demagnetization, an active compensation system for external magnetic field variations (including transverse fields), internal magnetometry, and removal of large external sources of magnetic field gradients were also required to ensure maintenance of the quasi-free condition. Since this experiment was performed a great deal has been learned about large volume magnetic shield technology in the course of R\&D performed for experiments which search for the neutron electric dipole moment~\cite{Altarev14}. We therefore do not foresee any issues of principle which would forbid us to reach this goal, but a significant research program to understand how to achieve this lower limit in a cost-effective manner, and to understand the possible reduction in sensitivity that might arise from residual field configurations would clearly be needed.

The maintenance of the quasi-free condition was verified experimentally in the ILL experiment~\cite{BaldoCeolin94} by polarizing the neutrons and measuring the rotation of their plane of polarization along the length of the zero field volume to bound the line integral of the longitudinal field along the neutron flight path~\cite{Schmidt92}. This was done by inserting a neutron polarizer and a neutron polarization analyzer before and after the magnetic field-free region. If the neutrons are polarized transverse to this axis, then the presence of a magnetic field along the axis of the shield will rotate the neutron polarization direction by an angle $\phi=\gamma B T$  where $\gamma$  is the neutron gyromagnetic ratio, $\mathcal{B}$ is the magnetic field, and $T$  is the time that the neutron spends in the field. For cold neutrons of $1$ second flight time and $B=1$ nT this angle is about $0.2$ radians. The neutron cold source produces a broad spectrum of neutron velocities and this will lead to a distribution of rotation angles. One can chop the beam mechanically if necessary, but a more elegant solution (used in the ILL experiment) is to employ a second magnetic coil before the polarization analyzer with a known internal field and length and operate the measurement in \lq\lq spin echo \rq\rq mode. In this case when the echo condition is met the final neutron polarization is high and insensitive to the velocity distribution in the beam, and the decrease of the polarization away from this resonance condition depends on the shape of the velocity distribution. Furthermore by changing the direction of the second magnetic coil in all 3 directions and using the spin echo detection method one can measure the line integral of all three magnetic field components in the shield as seen by the neutrons. It is therefore quite feasible if necessary to
use a polarized free neutron beam as a magnetometer to confirm the maintenance of the quasi-free condition.  

We foresee no issue of principle which would preclude a similar operation for an improved experiment. However the neutron polarizer and analyzer technology would need to be changed because the phase space acceptance of the optical system for an improved experiment will necessarily be larger than that of the neutron optical reflecting based devices usually used for slow neutron polarization and analysis. Fortunately since the ILL experiment neutron polarizers and analyzers with large phase space acceptance based on transmission through laser optically-pumped polarized $^{3}$He gas have been developed over the last decade for neutron scattering applications~\cite{Coulter90,Chupp07,Petoukhov06,Hussey05,Chen14}. These devices are now being adopted for routine operation at neutron scattering facilities and their present performance would suffice for the polarimetry needs of an oscillation experiment.

The vacuum requirement of 10$^{-5}$ Pa mentioned above is necessary to ensure that the difference between the neutron and antineutron optical potential in the residual gas does not cause a violation of the quasi-free condition. Fortunately there is extensive experience in the experimental physics community with a much larger vacuum chamber than needed for this experiment with a much more stringent vacuum requirement: namely, the 2 km long, 1.2 m diameter vacuum chambers for the LIGO gravitational wave observatory. The LIGO vacuum tubes are operated at the (much lower) pressure of order $10^{-7}$ Pa, achieved after 2 months of pumping with 9 pumps, each connected to a 25 cm diameter pumpout port. Once this ultimate pressure is reached they can be maintained at this pressure with only pumping at the ends of the chamber. In LIGO this relatively simple vacuum solution was achieved through the use of nonmagnetic 304L stainless steel which was baked at 150C by electrical heating of the tube (using the tube as a resistor). In addition the steel was heat-treated during manufacture to reduce the hydrogen outgassing rate from the usual 10$^{-11}$ torr liters/s/cm$^{2}$ to 5$\times$10$^{-14}$ torr liters/s/cm$^{2}$. The required purity of the internal surface of the stainless steel was maintained during on-site welding assembly through the development of a continuous spiral welding process~\cite{Weiss14}. For a vacuum chamber of diameter larger than LIGO that one would expect to use in a slow $n - \bar n$ oscillation experiment the gas dynamics in this pressure regime is faster. Because of the larger diameter of the tube and the less stringent vacuum demands for the $n - \bar n$ experiment even the outgassing rate from untreated stainless could be handled with a few hundred liters/s pumping speed. We therefore again see no issue of principle which would preclude a new experiment from meeting this condition. 

%%%%%%%%%%%%%%%%%%%%%%%%%%%%%%%%%%%%%%%%%%%%%%%%%%%%%%%%%%%%
\section{Design Considerations for an Improved $n - \bar n$ Oscillation Search Experiment with Free Neutrons}
\label{nnbar:sec:sensitivity}
%%%%%%%%%%%%%%%%%%%%%%%%%%%%%%%%%%%%%%%%%%%%%%%%%%%%%%%%%%%%

In considering the practicality of a future $n - \bar n$ oscillation experiment it is a useful to consider a specific design for evaluation. We present such a study in this section with a rough estimate for the sensitivity which can be achieved with the existing technology in neutron optics and
moderation discussed in the preceding section~\cite{Snow09}.
We do not include in our estimate the various methods by which the initial cold neutron brightness might be increased which are the subjects for ongoing research. 
Examples include neutronics techniques such as a reentrant moderator
designs~\cite{Ageron89}, strategic use of
neutron reflector/filters~\cite{Mocko13}, supermirror
reflectors~\cite{Mezei76}, and non-specular, high-albedo materials such as
diamond nanoparticle
composites~\cite{Nesvizhevsky08,Lychagin09el,Lychagin09em}.
Some of these techniques are not suitable for use at
multipurpose spallation sources serving a materials science user
community, where sharply defined neutron pulses in time may be
required. However they can be employed in a dedicated $n - \bar n$ experiment and we expect that they can be used to further improve the sensitivity of an experiment.

Supermirrors based on multilayer coatings greatly increase the
range of reflected transverse velocities relative to the nickel guides
used in the ILL experiment. Recall that ${\it m}$ denotes the increased critical angle above nickel for near-unity reflection. Supermirrors with $m = 4$ are now mass-produced and supermirrors with up to $m = 7$ can be produced~\cite{Swiss00}.
To enhance the sensitivity of the $n - \bar n$ search the supermirrors can be arranged in the shape of a truncated focusing
ellipsoid~\cite{Kamyshkov95}.  The focusing reflector with a large
acceptance aperture will intercept neutrons within a fixed solid angle
and direct them by single reflection to the target. The cold neutron
source and annihilation target will be located in the focal planes of
the ellipsoid. The geometry of the reflector and the parameter ${\it m}$ of the mirror material are chosen to maximize the sensitivity
$N_{n}\cdot t^{2}$ for a given brightness of the source and a given
size of the moderator and annihilation target.  Elliptical
concentrators of somewhat smaller scale have already been implemented
for a variety of cold neutron experiments~\cite{Boni10}.  

MCNPX~\cite{Mcnp08} simulation of the performance of the cold
source shown in Fig.~\ref{nnbarx:fig:horiz} produced a flux of
cold neutrons emitted from the face of cryogenic
moderator into the forward hemisphere with the spectrum shown in
Fig.~\ref{nnbarx:fig:source}.  Only a fraction of the integrated flux is
accepted by the focusing reflector. For sensitivity ($N_n\cdot t^{2}$) calculations, neutrons emitted from
the surface of neutron moderator were traced through  the detector
configuration shown in Fig.~3 with gravity
taken into account and with focusing reflector parameters that were
adjusted by a partial optimization procedure. The flux of cold
neutrons impinging on the annihilation detector target located at the
distance $L$ from the source was calculated after (mostly
single) reflection from the focusing mirror. The time of flight to the target
from the last reflection was also recorded. Each traced neutron contributed its $t^{2}$ to the total
sensitivity figure $N_n\cdot t^{2}$ that was finally normalized to the
initial neutron flux from the moderator. The sensitivity as a function of
distance between neutron source and target ($L$) is shown in
Fig.~\ref{nnbarx:fig:sensitivity}. 

The simulation has several parameters that affect the sensitivity: emission area of the moderator, distance
between moderator and annihilation target, diameter of the
annihilation target, starting and ending distance for truncated
focusing mirror reflector, minor semi-axis of the ellipsoid, and the
reflecting value ``$m$" of the mirror. Sensitivity is a complicated
function in the space of these parameters. Over a broad range of parameters we obtain a sensitivity which is a factor of 100
greater than the ILL experiment per year of operation assuming no background.  Configurations of
parameters that would correspond to even larger sensitivities are
achievable, but for the baseline simulation shown in 
Fig.~\ref{nnbarx:fig:sensitivity} we have chosen a set of parameters that
we believe will be reasonably achievable and economical after
inclusion of more engineering details than can be accommodated in our
simulations to date.

\begin{figure}
    \centering
    \includegraphics[width=5.3in]{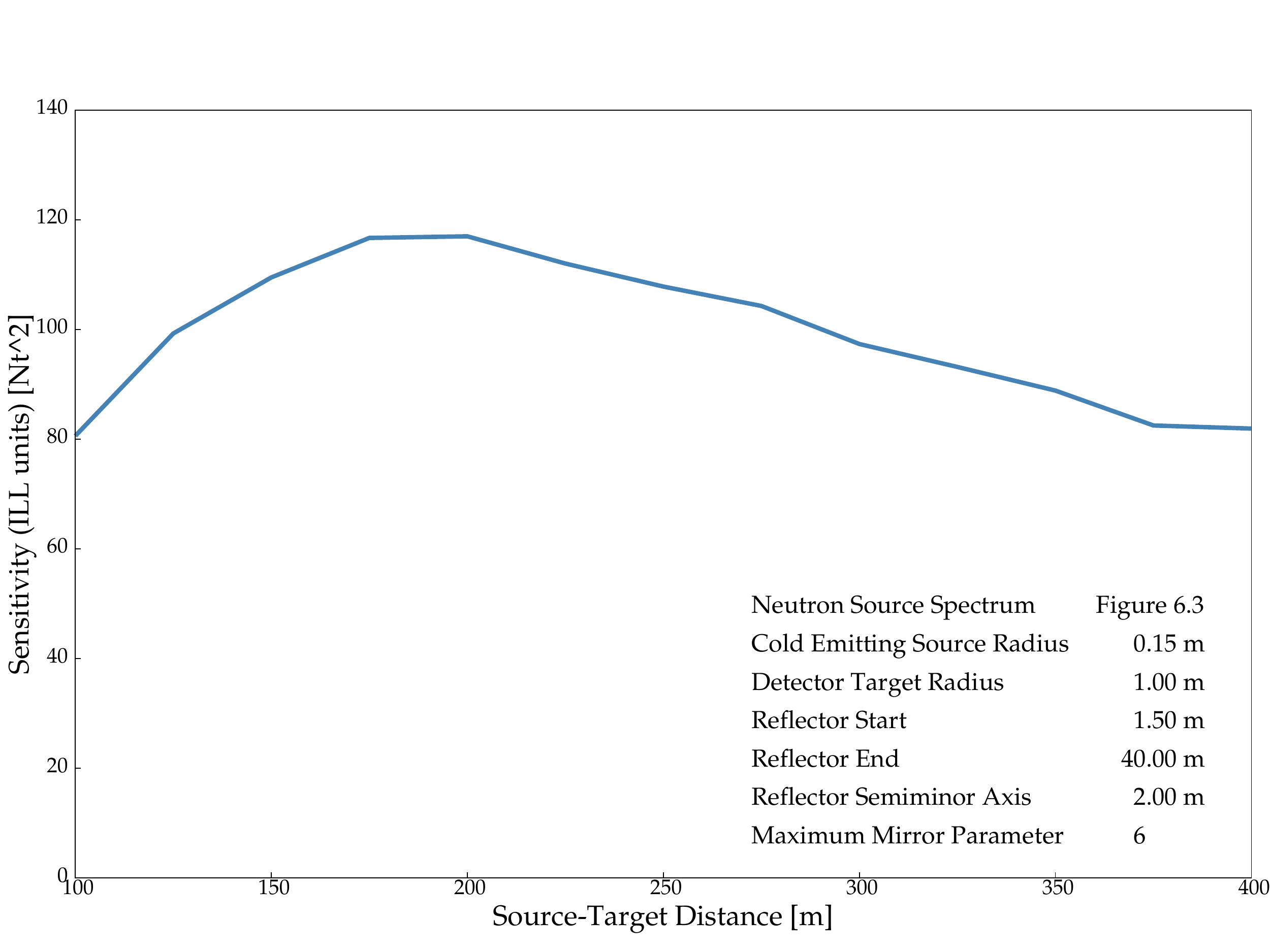}
    \caption{Result of a calculation of the $n - \bar n$ oscillation
sensitivity for a geometry similar to that in Fig.~\ref{nnbarx:fig:horiz}, where all parameters are fixed except for the source-target distance {\it L}. The semi-major axis of the elliptical reflector is equal to {\it L}/2, so one focus is at the source and the other is at the target.}
    \label{nnbarx:fig:sensitivity}
\end{figure}

The optimal optical configuration for a
$n - \bar n$ search is significantly different from anything that has
previously been built for materials research, so the full impact on the sensitivity of cost
and other engineering considerations is not straightforward to
predict. However representative parameters assumed for the optimization simulations (a source brightness of $3.5\times 10^{12} n/$s-cm$^{2}$-sec-MW and a moderator viewing area of $700$ cm$^{2}$) are reasonable~\cite{Maekawa10, Kai05}, the $200$m vacuum tube length and the $2$m $^{12}$C target diameter are about a factor of 2 larger than the ILL experiment, and the $0.2$sr solid angle of acceptance comes from the new high $m$ supermirror guides and the elliptical geometry. Our conclusion from this exercise is that a $n - \bar n$ oscillation experiment with a horizontal geometry using existing technology
can without difficulty improve the limit on the oscillation probability by at least two orders of magnitude per year of operation assuming no background in the 
antineutron detector.   

%\begin{table}[ht]
%\begin{threeparttable}
%\centering
%    	\caption{Comparison of parameters in simulations with
%   			existing practice.}
 %   	\label{edm:tab:lqcd}
%    	\begin{tabular}{cccc}
%        \hline\hline Parameter & Units & Used in & Existing MW \\ (References) & & Simulations & Facility Value \\ %\hline
%        Source brightness & $\frac{n}{{\rm s\cdot cm^{2} \cdot sr\cdot MW}}$ &
%        3.5$\times$10$^{12}$ & 4.5$\times$10$^{12}$ \\  ($E <$ 400 meV)~\cite{Maekawa10} &  &  &  \\
%        Moderator viewed area~\cite{Maekawa10} & cm$^{2}$ & 707 & 190 \\ Accepted solid
%        angle~\cite{Kai05}  & sr & 0.2 & 0.034 \\ Vacuum
%        tube length~\cite{BaldoCeolin94} & m & 200 & 100 \\ $^{12}$C
%        target diameter~\cite{BaldoCeolin94} & m & 2.0 & 1.1 \\
%        \hline\hline
%    \end{tabular}
%    \begin{tablenotes}
%    \end{tablenotes}
%    \end{threeparttable}
%\end{table}

%%%%%%%%%%%%%%%%%%%%%%%%%%%%%%%%%%%%%%%%%%%%%%%%%%%%%%%%%%%%
\subsection{Signal, Backgrounds, and Sensitivity in a Next-Generation $n - \bar n$ Oscillation Search Experiment with Free Neutrons}
\label{nnbar:subsec:backgrounds}
%%%%%%%%%%%%%%%%%%%%%%%%%%%%%%%%%%%%%%%%%%%%%%%%%%%%%%%%%%%%

We assign a figure of merit for the detector design based on the upper limit for a null result at the 90$\%$ confidence level for an ideal free neutron experiment, which can be written in terms of the oscillation time as~\cite{Prosper}:
\begin{equation}
\tau_{n\bar{n}} < \langle t \rangle_{rms} \sqrt{\frac{I_{t}T}{2.3}}.
\end{equation}
where $\langle t \rangle_{rms}$ is the average value of the free oscillation observation time ($t$ in Eq. 10), $I_t$ is the integrated neutron flux on target, and $T$ is the operation time of the experiment.  This expression assumes $100\%$ detection efficiency for the  integrated flux on target, and that the quasi-free condition is exactly satisfied.  For the 1994 ILL experiment, the sensitivity was written as~\cite{BaldoCeolin94}:

\begin{equation}
\tau_{n\bar{n}} < \langle t \rangle_{rms} \sqrt{\frac{I_{t}\langle\eta\rangle\epsilon(1-d)T}{2.3}},
\end{equation}
with $I_t = (1.25\pm 0.06)\times 10^{11}$ neutrons/s, $T=2.4\times 10^7$ s and an observation time $t=0.0109\pm 0.002$ s.  The correction for imperfectly satisfying the quasi-free condition results in the term $\langle \eta \rangle = 0.984\pm 0.003$, the annihilation detection efficiency is $\epsilon = 0.52\pm 0.02$ and the experimental deadtime results in a value of $d=0.068$.  No valid signal or background events were recorded, indicating that the essentially ideal, background free condition was met.

This makes it clear that it is reasonable to design experiments with a total oscillation produced $\bar{n}$ detection efficiency greater than 50$\%$ and which satisfy the background free criterion.  However, for the experiment we consider, sited at a spallation source with greatly increased integrated neutron flux on target and increased running time relative to the ILL experiment, it is worthwhile to provide a definite target for the background rate.  If we assume that no events are detected at the 90$\%$ confidence level over a running time T, we require a background which would produce on average 0.11 counts, or a background rate less that 0.11/T.  For three years of running, this gives a required background rate below 1.2$\times 10^{-9}$ s$^{-1}$ in the signal window.  We consider the sources of background with this goal in mind.

%%%%%%%%%%%%%%%%%%%%%%%%%%%%%%%%%%%%%%%%%%%%%%%%%%%%%%%%%%%%
\subsubsection{Backgrounds in a $n - \bar n$ Oscillation Search Experiment at a Spallation Source}
\label{nnbar:subsubsec:spallbkgd}
%%%%%%%%%%%%%%%%%%%%%%%%%%%%%%%%%%%%%%%%%%%%%%%%%%%%%%%%%%%%

An $n - \bar{n}$ experiment at a spallation source must contend with backgrounds from three primary sources: (1) high energy products from the spallation process, (2) cold neutron beam-generated backgrounds and (3) cosmic rays.  Neutron spallation sources can be pulsed 
or continuous. 
Protons with energy 1-2 GeV interact with a spallation target producing high-energy particles (including 
protons, pions, muons, gammas) and most essentially neutrons with an energy 
range from MeV to GeV as well as the slower neutrons of interest. These high energy particles will produce a new background 
that was not present in previous reactor experiments~\cite{BaldoCeolin94, Fidecaro, Bressi90}.  At a pulsed 
spallation source the fast particle backgrounds can be excluded by vetoing data from the small fraction of time when the proton beam strikes the spallation 
target. For a continuous spallation source this background requires more careful consideration.  

The issue is illustrated in Fig.~\ref{nnbar:fig:striganov}, where we show the simulated flux at 90 degrees from a 1 GeV proton beam incident on a lead spallation target.  Of course, the shielding placed around the spallation target will greatly reduce these fluxes, but the requirement of a low attenuation path between the cold neutron source and the target means that some of this fast flux will escape.  There are roughly 20 high energy neutrons produced in the spallation target per proton.  These neutrons are extremely difficult to shield and can produce tracks in the calorimeter.  High energy neutrons, protons and gammas can scatter in the target, creating spurious events which appear to originate from the target.  Because the background rate in the detector depends heavily on the configuration and shielding of the source, beamline and detector, a realistic assessment of these backgrounds is challenging. We will return to this issue after we review other background sources.  

\begin{figure}
    \centering
    \includegraphics[width=4.7in]{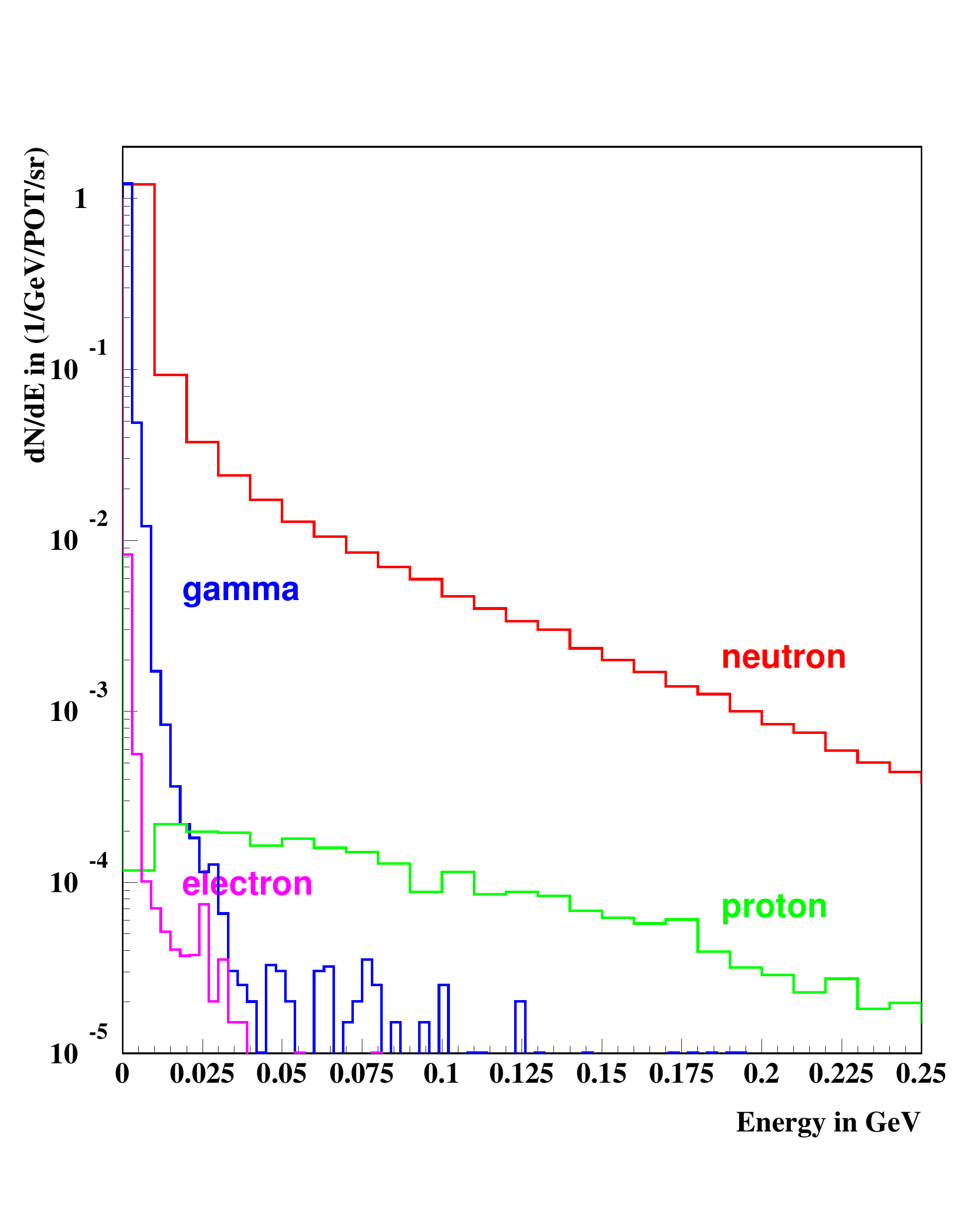}
    \caption{Spectra from S. Striganov MARS simulations~\cite{MARS} for 1 GeV proton beam incident on a lead spallation target viewed normal to the incident beam.}
    \label{nnbar:fig:striganov}
\end{figure}

Cold neutron beam-generated backgrounds and cosmic ray backgrounds were present in all previous experiments.  The cold neutron beam background contains an ``irreducible" component in the form of MeV gammas from neutron capture in the annihilation target.  In the ILL experiment only 5.2$\%$ of the cold neutron beam was lost in the beam optics and in the journey to the target. The beam halo was absorbed by boron-loaded glass collimators, and the beam dump was constructed of $^{6}$Li-loaded tiles (the neutron absorption reaction in $^{6}$Li emits no gammas, but it possesses a very small $\mathcal{O}$(10$^{-5}$) branch for fast neutron production)~\cite{Lone80}.   These backgrounds do not produce track-like background but they greatly influence the trigger requirements, as the instantaneous energy deposition from these backgrounds is anticipated to be well over the desired energy threshold for annihilation events.  For the envisioned next-generation experiment, there will be an enormous increase in the incident flux (on the order of factor of 100 to 1000) and an even larger increase in the beam which does not reach the target, resulting in more stringent requirements on the beam line shielding, detector granularity, tracking resolution, and trigger cuts.  In the ILL experiment, spurious events above threshold produced by multiple gamma-ray hits during the 150 ns trigger timing window account for about 32$\%$ (1 Hz) of the total trigger rate of $~4$Hz.

Cosmic rays (CR) were the dominant backgrounds for all previous experiments.  For the ILL experiment, they accounted for the remaining 3 Hz of the trigger rate, with 2.7 Hz coming from CR muons which evaded the veto (efficiency $\sim$99.5$\%$) and 0.3 Hz due to neutral cosmic rays.  The neutral cosmic rays were of particular concern in earlier $n - \bar{n}$ experiments~\cite{Bressi90}.  Neutral CRs evade the CR veto and can produce events which originate from the target and beam tube, which were assessed to be the leading contributors to possible background in the signal window.  Given the larger annihilation target area and detector volume of a next generation experiment, these events are expected to potentially contribute to backgrounds and a corresponding improvement in vertex reconstruction and event identification will almost certainly be required.  

%%%%%%%%%%%%%%%%%%%%%%%%%%%%%%%%%%%%%%%%%%%%%%%%%%%%%%%%%%%%
\subsubsection{Background Reduction and Measurement Strategy}
\label{nnbar:subsubsec:bkgdmeas}
%%%%%%%%%%%%%%%%%%%%%%%%%%%%%%%%%%%%%%%%%%%%%%%%%%%%%%%%%%%%

Experiments which use free neutrons provide several unique tools to evaluate the very small background signals anticipated for a properly designed measurement.  In particular there are two methods which identify events that must come from background:

\begin{itemize}
\item
(1)	By ``switching off" the magnetic shielding, one can suppress the $n - \bar{n}$ oscillation effect, and produce a population of background-only events while essentially leaving the entire experiment unperturbed.
\item
(2)	By adding one or more targets downstream of the annihilation target but within the sensitive volume of the detector, one can produce additional ``sources" for background events without an annihilation signal, because the $\bar{n}$ content in the cold neutron beam is removed by the primary annihilation target.  This method to characterize backgrounds was explored in the $n - \bar n$ search at 
Pavia University's Triga Mark II reactor, which employed a second downstream target for this purpose~\cite{Bressi90}.  In a modern detector with improved tracking reconstruction, one might be able to accommodate a larger number of ``background" targets as well.   
\end{itemize}  

These features of the apparatus can provide a robust method to directly test a false ``positive" signal, and provide data to characterize aspects of the detector response and allow a more robust evaluation of the background rejection scheme.  When taken together with the more conventional approach of parameterizing the background cuts to extrapolate to the expected backgrounds in the signal window, free neutron experiments are well equipped to provide robust evidence for $n - \bar{n}$ oscillations in the event of a positive signal. 

The critical tool for background rejection is the tracking capability of the annihilation detector.  Given the advances in the past 20 years of tracking calorimetry, existing technology (which we review in \S\ref{nnbar:subsec:detector}) seems to be adequate for a next generation experiment.  Specific strategies can also address particular components of the background.  The high energy backgrounds associated with the proton beam can be eliminated using a pulsed beam.  Since the high energy backgrounds reach the target in times less than about 10 $\mu$s, one can safely ``veto" a time interval during and after the proton beam arrival, with essentially no impact to the ``live time" of the cold neutron beam, which takes times on the order of 0.1 s to travel from the moderator to the target.  Modulating the proton beam intensity does not help with possible beta-delayed neutron background from the spallation target.  For non-fissile spallation targets, these neutrons have typical energies below 10 MeV and come primarily from $\gamma$-n processes resulting from activation in the target materials.  Although they are not expected to pose a large challenge for backgrounds, their intensity is very dependent on the target composition and shielding configuration, and must be modeled and measured to constrain their contribution to the background budget~\cite{Ridikas07}.

Given the copious production of high energy neutrons, a low-mass inner-tracking detector to help differentiate neutral from charged tracks emanating from the target seems advantageous.  Recessing the detector to eliminate direct line-of-sight from the moderator geometry to the detector should also help.

The general strategy for eliminating CR events is in principle the same as in earlier experiments: one can design a CR veto or (with a modern fast-timing calorimeter) use the entire calorimeter as a CR veto to reduce muon events.  Coupled with the anticipated improved vertex reconstruction capability, these events should be adequately controlled.  The neutral CR events remain a potential source of background, but only a modest improvement in the expected rejection of these events is required since the mass of the target in the next generation experiment is envisioned to only be a factor of 4 greater than the target for the ILL experiment.

Capture gamma fluxes are expected to be many orders of magnitude larger in the scaled up experiment, putting some stress on the trigger criterion.  Since these events do not create tracks per se, demanding ``track-like" cuts on the detector are likely to provide adequate control for these backgrounds.  We now present some considerations on the antineutron detector.

%%%%%%%%%%%%%%%%%%%%%%%%%%%%%%%%%%%%%%%%%%%%%%%%%%%%%%%%%%%%
\subsection{Requirements for an Annihilation Detector}
\label{nnbar:subsec:detector}
%%%%%%%%%%%%%%%%%%%%%%%%%%%%%%%%%%%%%%%%%%%%%%%%%%%%%%%%%%%%

As mentioned in \S\ref{nnbar:subsec:vacuum}, a free $n - \bar n$
transformation search could require a vacuum of
10$^{-5}$ Pa and a magnetic field of $|{\vec B}| <$ 1 nT along the flight path
of the neutrons. The target vacuum is achievable with standard vacuum
technology, and the magnetic fields could be achieved with an
incremental improvement on the ILL experiment through passive
shielding and active field
compensation~\cite{BaldoCeolin94,Kronfeld13}.

Identification of antineutrons in the beam proceeds through the detection 
of the (on average five) pions produced in antineutron annihilation with a neutron or proton
in the annihilation target.  Background is rejected by reconstructing the 
annihilation vertex and measuring the annihilation products' invariant mass.

\begin{figure}[hbtp]
  \centering
    \scalebox{0.6}{\includegraphics{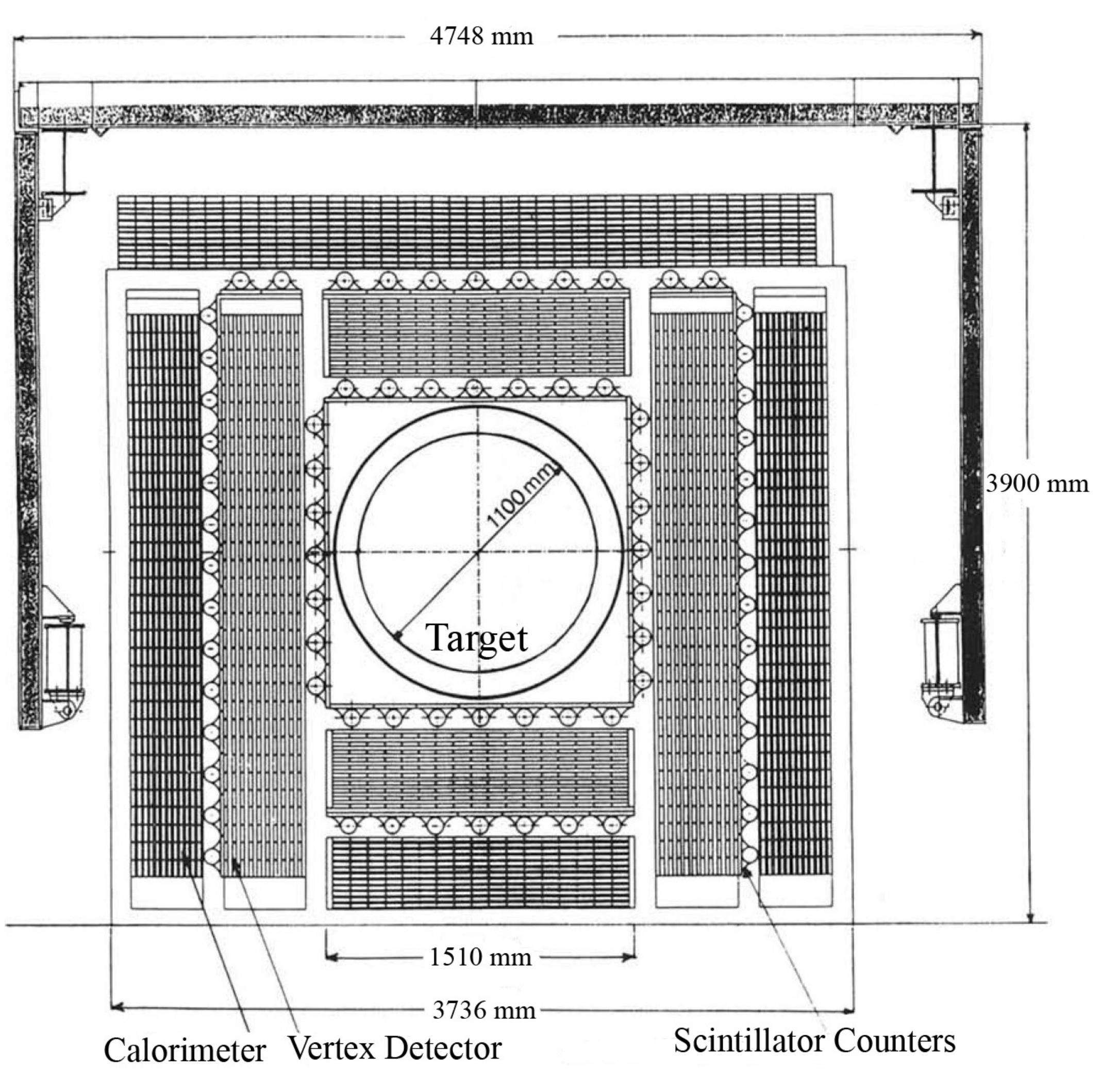}}
    \caption{A cross sectional view, perpendicular to the cold neutron beam axis, of the Institut Laue Langevin annihilation detector~\cite{BaldoCeolin90}. }
  \label{nnbarx:fig:illdetector}
\end{figure}

Our strategy in the design of the annihilation detector is to develop
an updated realization of the design concept used successfully in the ILL
experiment~\cite{BaldoCeolin94} (see
Fig.~\ref{nnbarx:fig:illdetector}).  Major subsystems of the
annihilation detector ordered radially outward include:  (i) the annihilation target and detector vacuum
region; (ii) the tracker; (iii) the time of flight systems before and
after the tracker; (iv) the calorimeter; and (v) the cosmic veto
system. Requirements for these subsystems are formulated below. In comparison with detectors 
of comparable scale and function in other nuclear and particle physics experiments, the $n - \bar n$ detector 
doesn't require premium performance, but needs rather careful optimization of the
cost due to its size. The detector should be built along the detector vacuum region
with several layered detection subsystems (subsystems (ii) - (v)) and
should cover a significant solid angle (coverage in the $\theta$-projection from
$\sim$20$^{\circ}$ to 160$^{\circ}$ corresponds to a solid angle
acceptance of $\sim$94$\%$). In the $\phi$-projection, the detector
configuration can be cylindrical, octagonal, hexagonal, or square
(similar to the ILL experiment~\cite{BaldoCeolin94}).

The spallation target introduces a new
consideration in the annihilation detector design, because of the
possible presence of fast neutron and proton backgrounds.  These
backgrounds were absent from the ILL
experiment, which produced fewer high energy particles in the reactor
source and eliminated the residual fast backgrounds using a curved
guide system to couple the cold source to the $n - \bar n$ guide. 
We propose to integrate our shielding scheme for
fast particles into the design of the source and beamline.  The residual fast backgrounds
at the detector are a strong function of the guide tube length,
detector threshold, and pulse structure for the proton beam.  
In particular the natural pulsed structure of neutron spallation sources built to 
measure slow neutron energies using neutron time-of-flight is very useful as the source is off when
the slow neutrons arrive at the annihilation target.

%%%%%%%%%%%%%%%%%
\subsubsection{Annihilation Target}
\label{nnbar:subsubsec:target}
%%%%%%%%%%%%%%%%%

A uniform carbon disc with a thickness of $\sim$ 100 $\mu$m and
diameter $\sim$ 2 m could serve as the antineutron annihilation target. It would be
stretched on a low-${\it Z}$ material ring and installed in the center
of the detector vacuum region. The choice of carbon is guided by its low
capture cross section for thermal neutrons, $\sigma \sim$ 4 mb, and high antineutron
annihilation cross section, $\sigma \sim$ 4 kb.  The fraction of hydrogen in
the carbon film should be maintained below $\sim$ 0.1$\%$ to reduce
generation of capture $\gamma$'s. If it is too difficult to extract enough hydrogen from the 
carbon foil a different material choice for the foil might well be preferable.

In order to clearly establish background rates one can imagine placing several identical target 
foils downstream of the first foil as mentioned above. The additional foils would be spaced 
with a separation greater than the spatial resolution of the tracking detector.  Given the ratio of annihilation and capture cross sections, only 
the first foil would be sensitive to $\bar{n}$ annihilation events 
but all would be sensitive to background.  
Since the background rate would be enhanced by having several foils, 
the (lack of) observation of candidate events 
in the downstream foils would provide an accurate, data-based estimate of the 
expected background in the ``signal foil.''

%%%%%%%%%%%%%%%%%%%
\subsubsection{Detector Vacuum Region}
\label{nnbar:subsubsec:vacuum}
%%%%%%%%%%%%%%%%%%%

The detector vacuum region should be a thin tube made of
low-${\it Z}$ material (Al) to reduce multiple scattering for tracking
and provide a low (${\it n}$,$\gamma$) cross section, but thick enough to maintain structural integrity.  Simulation studies on how the vacuum tube thickness affects the $n - \bar n$ signal mode acceptance are underway and are described in \S\ref{nnbar:subsec:simulation}.  Additional lining of the inner surface of the vacuum region with $^{6}$LiF pads
will reduce the generation of $\gamma$'s by captured neutrons.  The
detector vacuum region is expected to be the source of $\sim$ 10$^{8}$
$\gamma$'s per second originating from neutron capture.  Unlike in the
neutron beam flight vacuum region, no magnetic shielding is required
inside the detector vacuum region. As mentioned before, the vacuum
level should be better than 10$^{-4}$ Pa via connection with the neutron beam
vacuum region.  We plan to have a section of the vacuum tube in the
detector recessed.  This area will have no support or detector
elements in the neutron beam, which will reduce the rate of neutron
captures.

%%%%%%%%%%%%%%%
\subsubsection{Tracker}
\label{nnbar:subsubsec:tracker}
%%%%%%%%%%%%%%%

The tracker's primary purpose is to identify the position of the annihilation
vertex in the foil with high accuracy.  It should have the largest possible solid angle
coverage,  $\sim$20$^{\circ} < \theta <$ 160$^{\circ}$ would yield 94\% 
acceptance.  
It should provide
annihilation vertex reconstruction accuracy with rms $\leq$ 1 cm in all directions (compared to 4-12 cm in
ILL experiment). 
In addition to ensuring candidate pions originate from a single vertex,
accurate vertex position determination is an important factor in 
measuring overall momentum balance and achieving precise invariant mass reconstruction.
Relevant tracker
technologies can include straw tubes, proportional and drift
detectors. Limited Streamer Tubes (LST), as used in the ILL
experiment, are presumed to be worse than proportional mode detectors
due to better discrimination of the latter for low-energy capture
$\gamma$'s.

The ATLAS transition radiation tracker (TRT) is an example of a large 
straw tube detector with good accuracy.
The ATLAS TRT
covers a pseudorapidity range less than 2 and has a measured position
resolution in the direction transverse to the straws of about 125 $\mu$m.  
It has a 3 ns time resolution.  The
ATLAS TRT provides tracking for charged particles down
to a transverse momentum of $p_{T} =$ 0.25 GeV with an efficiency
above 90$\%$.  Tracks with lower transverse momentum loop in ATLAS'
solenoidal magnetic field and are not reconstructible in the TRT.
For
tracks that have at least 15 TRT hits, a transverse momentum $p_{T} >$
1.00 GeV, and are within 1.3 mm of the anode, the efficiency was found
to be 94.4$\%$ for the 7 TeV ATLAS data with similar results for the
0.9 TeV ATLAS dataset~\cite{Boldyrev12}. 
For a cut of $p_{T} >$ 0.25 GeV, the efficiency drops down to
93.6$\%$.  For higher momentum tracks (e.g. $p_{T} >$ 15 GeV), the
efficiency increases to 97$\%$, and this is more indicative of the
single-straw efficiency~\cite{Vankooten13} as such tracks are
essentially straight.  The efficiency drops
at the edges of the straw due to geometric and reconstruction effects.  Fig.~\ref{nnbarx:fig:atlastrt} shows a schematic of the prototype ATLAS TRT straw module that has been studied and will continue to be studied in upcoming fast neutron beam tests (see \S\ref{nnbar:subsec:fastneutrons}).    

The straw tubes in the TRT have a diameter of 4 mm and are made from
wound Kapton reinforced with thin carbon fibers. 
The anode at the
center of each straw is gold plated tungsten wire with a diameter of
31 $\mu$m.  In ATLAS the cathodes are kept at -1.5 kV, while the anodes are
kept at ground.  The tubes are filled with a gas mixture of 70$\%$ Xe,
27$\%$ CO$_{2}$, and 3$\%$ O$_{2}$, however we will have to optimize
our gas mixture for a different set of backgrounds in this experiment,
particularly fast ${\it n}$-backgrounds and proton backgrounds.  
Furthermore, transition radiation detection is likely to be of limited value.
If the tracker must be moved inside the
detector vacuum region for better accuracy (also giving rise to the
problem of gas and electrical vacuum feedthroughs), then the
requirements on the detector tube material and thickness should be
revisited.

\begin{figure}[hbtp]
  \begin{center}
    \scalebox{0.53}{\includegraphics{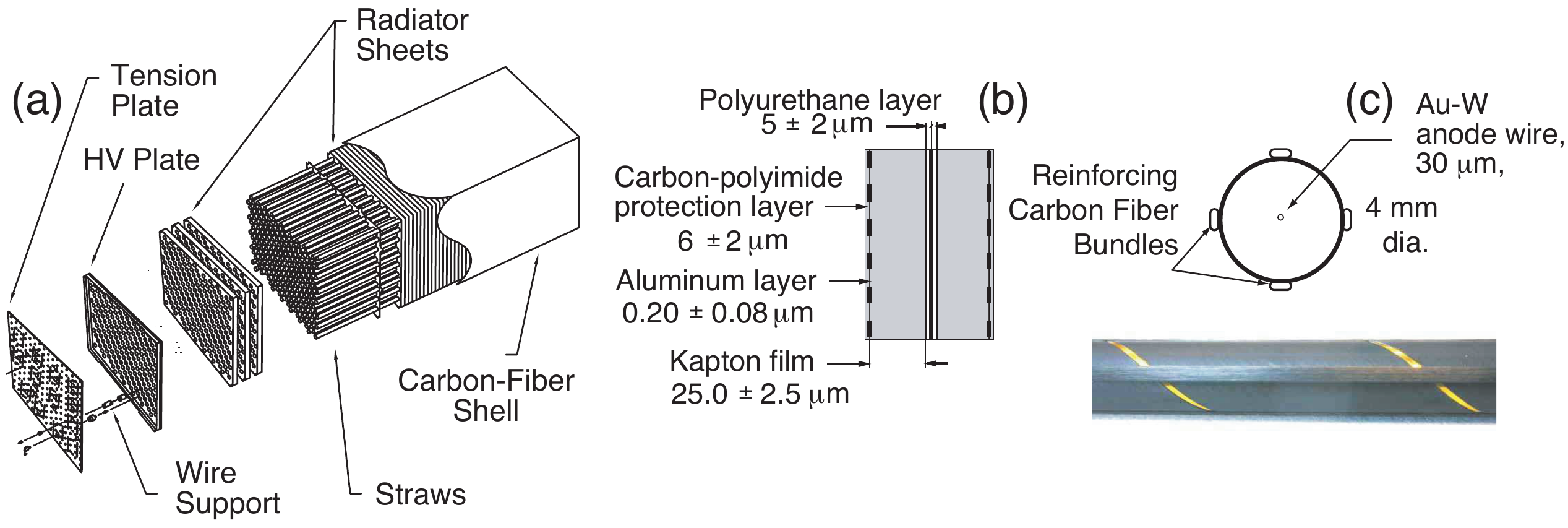}}
    \caption{(a) Prototype ATLAS TRT straw module; (b) ATLAS TRT straw layers; (c) photo and cross section of ATLAS TRT straw.}
     \label{nnbarx:fig:atlastrt}
  \end{center}
\end{figure}

%%%%%%%%%%%%%%%%%%
\subsubsection{Time of Flight System}
\label{nnbar:subsubsec:tof}
%%%%%%%%%%%%%%%%%%

The primary role of the time of flight (TOF) system is to ascertain that 
tracks originate from the annihilation target, and it should therefore 
have sufficient timing accuracy to discriminate the annihilation-like tracks from the
cosmic ray background originating outside the detector volume.
It would consist of two layers of fast
detectors (e.g.~plastic scintillation slabs or tiles) before and after
the tracker with solid angle coverage matching the tracker's
and sufficient segmentation to allow matching of TOF signals to 
reconstructed tracks.  The CDF TOF system~\cite{Cabrera02}, built of 4$\times$4$\times$279 cm$^{3}$ scintillator bars, achieved 130 ps timing resolution for 
tracks passing close to the photomultiplier tubes.

%%%%%%%%%%%%%%%
\subsubsection{Calorimeter}
\label{nnbar:subsubsec:calorimeter}
%%%%%%%%%%%%%%%

The calorimeter will measure the pion energies and should
provide trigger signal and energy measurements in the solid angle
$\sim$20$^{\circ}$ to 160$^{\circ}$.  The average multiplicity of
pions in annihilation at rest equals 5, so an average charged pion can be
stopped in $\sim$20 cm of dense material (like lead or iron). For low
multiplicity (but small probability) annihilation modes, the amount of
material needs to be larger. 
Detailed performance requirements for the measurement of total
energy of annihilation events and momentum balance in $\theta$- and
$\phi$-projections should be determined from simulations. 

Technology options include lead-glass, as used for example at Jefferson
Lab~\cite{Mkrtchyan13}, and reaching a resolution of $\sim$5\%$/\sqrt{E}$
for electrons, or scintillating fibers embedded in lead as used by 
the CHORUS experiment~\cite{Eskut97}, which reached $\sim$13\%$/\sqrt{E}$
for electrons and $\sim$33\%$/\sqrt{E}$ for hadrons.
An approach using MINER$\nu$A-like wavelength shifting fibers
coupled to scintillating bars is also being
considered~\cite{Mcfarland06}. Another example of a calorimeter technology which could be used for nnbar is the lead/plastic scintillator technology used in the DAPHNE calorimeter at KLOE~\cite{Babura1993}. The calorimeter configuration used in the ILL
experiment consisted of 12 layers of Al/Pb interspersed with
layers of streamer tubes.  For this option, the
proportional mode of calorimeter detector operation possibly can be
less affected by copious low-energy $\gamma$-background than the LST
mode.

%%%%%%%%%%%%%%%%%%%
\subsubsection{Cosmic Veto System}
\label{nnbar:subsubsec:veto}
%%%%%%%%%%%%%%%%%%%

The cosmic veto system (CVS) should have excellent coverage to identify all cosmic ray
background. 
%All annihilation products should be totally stopped in the
%calorimeter. %pion DIF??
Large area detectors similar to MINOS scintillator
supermodules~\cite{Michael08} might be a good approach to the
configuration of the CVS. The use of timing information should be
studied in connection with the TOF system, and might obviate the need for 
at least one TOF layer. 

%%%%%%%%%%%%%%%%%%%
\subsubsection{Trigger and Data Acquisition System}
\label{nnbar:subsubsec:tdaq}
%%%%%%%%%%%%%%%%%%%

The ILL experiment already observed high single hit rates, and the environment at a
spallation source is likely to produce higher background trigger rates.  Solid trigger logic,
with the ability to reconstruct track segments and discriminate single photons 
from $\pi^{0}$ decays is likely to be required.  The detector's channel count will
be quite manageable by modern FPGA standards, and the algorithms' complexity should
be relatively low, so that the main requirement is that the detectors themselves are fast enough.  
For lead-glass 
or scintillator-based calorimetry this should not be a problem, and ATLAS has 
demonstrated that straw tubes can be used for asynchronous triggering on cosmic
muons~\cite{Fratina09}.

The data volume should also be very moderate by modern standards: assuming 100 kB raw 
data per event and a 100 Hz trigger rate, the ``rate-to-tape'' would be 10 MB/s and the yearly data 
volume of order 100 TB.  This is comparable to the data storage and processing 
capabilities of university-based (``Tier-3'') LHC computing clusters.

%%%%%%%%%%%%%%%%%%%%%%%%%%%%%%%%%%%%%%%%%%%%%%%%%%%%%%%%%%%%
\subsection{Modeling of Annihilation Events}
\label{nnbar:subsec:annihmodel}
%%%%%%%%%%%%%%%%%%%%%%%%%%%%%%%%%%%%%%%%%%%%%%%%%%%%%%%%%%%%

The model for low-energy $\bar{n}$ annihilation on the nuclear targets is based on the intranuclear cascade (INC) approach.  This method uses the main ideas of the optical-cascade model which was earlier applied for the analysis of the annihilation of stopped antiprotons on nuclei~\cite{Iljinov82}. Using the optical-cascade model the experimental data on stopped antiproton annihilation on nuclei obtained at LEAR~\cite{Minor90,Hofmann90} were successfully described.

In the optical-cascade model the annihilation of a slow antinucleon on a nuclear target is considered as a multistage process:
\begin{enumerate}
\item
absorption of the antinucleon by the nucleus;
\item
annihilation of the antinucleon with a nucleon inside the nucleus;
\item
development of the intranuclear cascade initiated by the annihilation products; 
\item
de-excitation of the residual nucleus.
\end{enumerate}
As in the case of stopped antiproton annihilation~\cite{Iljinov82}, the 1st stage of the process is described by the optical model. Unlike antiprotons, slow antineutrons are absorbed by nucleus from the $S$-wave. The main result from the optical approach is the radial distribution of the $\bar{n}$ absorption probability, which can be written as:

\begin{equation}
P_{abs}(r) \sim 4\pi r^{2}\rho(r)|\Phi(r)|^{2}, 
\end{equation}

where $\rho(r)$ is a nuclear density, $\Phi(r)$ is a $\bar{n}$ wavefunction found from the numerical solution of the wave equation for $^{12}$C~\cite{Golubeva96}.  Carbon is preferred as a target material because of its low neutron capture cross section ($\sim$ 12 mb) and high antineutron annihilation cross section ($\sim$ 5 kb).  The radial distribution of absorption probability $P_{abs}(r)$ for $\bar{n}$ and $\bar{p}$ calculated for  $^{12}$C is shown in Fig. 7 of reference~\cite{Golubeva96}.  It is seen that the radial dependence of absorption probability for antiproton ($\bar{p}$) and antineutron ($\bar{n}$), calculated with the optical model for the $^{12}C$, is not much different and annihilation happens mostly on the periphery of the nucleus.  The $\bar{n}$ annihilation point inside the nucleus is determined by Monte Carlo in accordance with this distribution function.

Elementary antinucleon-nucleon $\bar{N} - N$ annihilation in the second stage can proceed through a large number ($\sim$ 10$^{2}$) of open channels.  Experimental information is available only for a small fraction of all possible annihilation channels.  In the optical-cascade approach the description of the elementary $\bar{N} - N$ annihilation is based on the statistical model with SU(3) symmetry which allows the production of two to six intermediate particles.  The intermediate particles can be $\pi$, $\eta$, $\omega$, $\rho$, $K$ and $K^{*}$ mesons.  Since neither the statistical model nor experiment can give precise and complete exclusive information about elementary $\bar{N} - N$ annihilation, it seems reasonable to use the results of both together.  Mean values of meson multiplicities and spectra of $\pi$- and $\eta$-mesons calculated from the Monte Carlo based on this semi-empirical approach are in good agreement with experiment~\cite{Golubeva92}.  In our $\bar{n} - ^{12}$C  annihilation model we consider separately the $\bar{n}$ annihilation on intranuclear protons and neutrons as those induce different reactions with
a different number of possible final states. This is why two semi-empirical tables of intermediate channel probabilities are used in the simulation: for $\bar{n} - n$ we use the same table as for $\bar{p} - p$ and for $\bar{n} - p$ the same table as for $\bar{p} - n$ (with changed charge) because no data on $\bar{n} - n$ or $\bar{n} - p$ exist.        

The primary annihilation of a $\bar{n}$ on one of the nucleons from the nucleus, in the process of which the large released energy is transformed into mesons, serves as a boundary between the initial and the final stages of the reactions. The following stages, where the energy released in the primary annihilation event is dissipated in the nucleus (stage 3) and the residual nucleus deexcites (stage 4), are described (respectively) by the INC approach~\cite{Iljinov94} and by the evaporation model~\cite{Botvina90}.  The INC + evaporation model has been successfully used in analysis of the inelastic interactions of intermediate-energy mesons and nucleons with nuclei and also for antiproton annihilation at rest~\cite{Golubeva92}.  Each $\bar{n}$-nucleus annihilation event is simulated by Monte Carlo.  

There is currently no experimental data on the annihilation of slow
antineutrons, but the experimental data on stopped $\bar{p}$-annihilation on light nuclei ($^{12}$C, $^{14}$N) obtained at LEAR~\cite{Minor90,Hofmann90}, including the multiplicity and charge distribution of pions, and momentum spectrum of pions and protons, was described successfully within the framework of the optical-cascade model~\cite{Golubeva96}.  Due to the surface character of a $\bar{p}$-nucleus annihilation at rest, most of the annihilation pions escape from the nucleus. In the case of light nuclei, the effects of rescattering and nuclear absorption of pions are not large and, as a result, the average number of emitted pions and average pion energy are close to the values corresponding to a $\bar{p} - p$ annihilation in vacuum (Table~\ref{nnbar:tab:avgne}).  

\begin{table}[ht]
\begin{threeparttable}
\centering
    	\caption{Average multiplicity, $\bar{\mathfrak{M}}$, and energy, $\bar{E}$, of pions after $\bar{p} - p$ and $\bar{p} - ^{12}$C annihilation at rest. Experimental measurements are given with references.}
    	\label{nnbar:tab:avgne}
    	\begin{tabular}{cccccc}
        \hline\hline Annihilation & $\bar{\mathfrak{M}}_{\pi}$ & $\bar{\mathfrak{M}}_{\pi^{+}}$ & $\bar{\mathfrak{M}}_{\pi^{-}}$ & $\bar{\mathfrak{M}}_{\pi^{0}}$ & $\bar{E}_{\pi}$(MeV) \\ \hline
        $\bar{p} - p$~\cite{Ghesquiere74}\tnote{**} & 5.01$\pm$0.23 & & & 1.96$\pm$0.23 & 370 \\
        $\bar{p} - p$~\cite{Minor89}\tnote{**} & 4.94$\pm$0.14 & 1.52$\pm$0.06 & 1.52$\pm$0.06 & 1.90$\pm$0.12 & \\
        $\bar{p} - p$ Calc. & 5.04 & 1.525 & 1.525 & 1.99 & 371\\
        $\bar{p} - ^{12}$C~\cite{Minor90} & 4.57$\pm$0.09 & 1.25$\pm$0.06 & 1.59$\pm$0.08 & 1.73$\pm$0.01 & 380$\pm$2 \\
        $\bar{p} - ^{12}$C Calc. & 4.69 & 1.17 & 1.61 & 1.91 & 367 \\
        \hline\hline
    \end{tabular}
    \begin{tablenotes}
    \item [**] This data may also be found in Ref.~\cite{Golubeva92}.
    \end{tablenotes}
    \end{threeparttable}
\end{table}

As the developed optical-cascade model provides a good description of the available experimental data on annihilation of antiprotons at rest on light nuclei, we expect that it can predict rather well the characteristics of nuclear absorption of slow antineutrons.  In the framework of the proposed model, the following characteristics of slow $\bar{n}$ absorption by $^{12}$C nuclei were calculated:  

\begin{itemize}
\item
the multiplicity distributions of $\pi^{+}$, $\pi^{-}$, and $\pi^{0}$,
\item
the multiplicity distribution of protons and neutrons,
\item
the energy spectra of protons and pions,
\item
the energy carried by pions and the energy carried by charged particles~\cite{Golubeva96}.
\end{itemize}

Due to the surface character of $\bar{n} - ^{12}$C annihilation, these characteristics are very similar to the ones for $\bar{p} - ^{12}$C absorption with the exception of the charged pion multiplicity distribution, which is different owing to the different charges of the initial states (Table~\ref{nnbar:tab:avgmult}).  This model can be used in the $\bar{n} - ^{12}$C event generator for the detector design and optimization in a $n - \bar n$ oscillation search experiment using a beam of cold neutrons.  The model can be extended to other target nuclei for analysis of $n - \bar n$ transitions of bound neutrons in the nucleus.  The most important step in this model is to correctly set the initial conditions (stage 1) via the $\bar{n}$ absorption probability function.  

\begin{table}[H]
\centering
    	\caption{Average multiplicities, $\bar{\mathfrak{M}}$, of various particles after $\bar{p} - ^{12}$C or $\bar{n} - ^{12}$C annihilation at rest.}
    	\label{nnbar:tab:avgmult}
    	\begin{tabular}{ccccccc}
        \hline\hline Annihilation & $\bar{\mathfrak{M}}_{ch}$ & $\bar{\mathfrak{M}}_{\pi}$ & $\bar{\mathfrak{M}}_{\pi^{+}}$ & $\bar{\mathfrak{M}}_{\pi^{-}}$ & $\bar{\mathfrak{M}}_{\pi^{0}}$ & $\bar{\mathfrak{M}}_{pr}$ \\ \hline
        $\bar{n} - ^{12}$C & 4.00 & 4.73 & 1.63 & 1.19 & 1.92 & 1.19 \\
        $\bar{p} - ^{12}$C & 3.99 & 4.69 & 1.17 & 1.61 & 1.91 & 1.22 \\
        \hline\hline
    \end{tabular}
\end{table}

A detailed treatment of $\bar{n} - n$ annihilation modes in $^{12}$C is under development. Here we present a list of $\bar{n} - n$ annihilation modes in $^{16}$O~\cite{Abe15} (see Table~\ref{edm:tab:sim}), which we expect to be similar (but not identical) to those for $^{12}$C.  We are currently refining an event generator for $\bar{n}$- and $\bar{p}$-annihilation on $^{12}$C nuclei using the GENIE Monte Carlo generator~\cite{Andreopoulos10} (version 2.8.0).  GENIE models the intranuclear propagation of mesons from the initial $\bar{n} \to n$ or $\bar{n}\to p$ vertex.  Annihilation vertices are distributed in the $^{12}$C nucleus according to a Gaussian nuclear density model with particle content given by the branching ratios in Table~\ref{edm:tab:sim}.  GENIE uses the INTRANUKE INC simulation package, originally developed by the Soudan-2 Collaboration, to model the hadron transport inside the nucleus. Final state interactions are calculated using the \textit{hA} mode of INTRANUKE, which simulates the intranuclear transport using the measured cross sections of each possible process instead of performing a full INC calculation.  Stable final state particles including charged and neutral mesons, gammas, and proton and neutrons from nuclear fragmentation are recorded for input into the detector simulation.  We are comparing the GENIE approach with the benchmark approach detailed at the beginning of this subsection in order to assess the effectiveness of the \textit{hA} mode of INTRANUKE for the simulation of for $\bar{n}$- and $\bar{p}$-annihilation on $^{12}$C nuclei.

\begin{table}[H]
\centering
    	\caption{List of $\bar{n} - n$ annihilation modes and branching ratios from the Super-Kamiokande simulation study.~\cite{Abe15}}
    	\label{edm:tab:sim}
    	\begin{tabular}{cc}
        \hline\hline $n - \bar n$ Annihilation Mode & Branching Ratio \\ \hline
        $\pi^{+}\pi^{-}3\pi^{0}$ & 28$\%$ \\
        $2\pi^{+}2\pi^{-}\pi^{0}$ & 24$\%$ \\
        $\pi^{+}\pi^{-}2\pi^{0}$ & 11$\%$ \\
        $2\pi^{+}2\pi^{-}2\pi^{0}$ & 10$\%$ \\
        $\pi^{+}\pi^{-}\omega$ & 10$\%$ \\
        $2\pi^{+}2\pi^{-}$ & 7$\%$ \\
        $\pi^{+}\pi^{-}\pi^{0}$ & 6.5$\%$ \\
        $\pi^{+}\pi^{-}$ & 2$\%$ \\
        $2\pi^{0}$ & 1.5$\%$ \\
        \hline\hline
    \end{tabular}
\end{table}

%%%%%%%%%%%%%%%%%%%%%%%%%%%%%%%%%%%%%%%%%%%%%%%%%%%%%%%%%%%%
\subsection{Detector Simulation}
\label{nnbar:subsec:simulation}
%%%%%%%%%%%%%%%%%%%%%%%%%%%%%%%%%%%%%%%%%%%%%%%%%%%%%%%%%%%%

Developing a detector model that allows us to reach our goal of zero background and optimum signal event detection efficiency is the primary goal of our simulation.  We use Geant 4.9.6~\cite{Geant13} to simulate the passage of annihilation event products through the annihilation detector geometry. This simulation consists of tracking a random sampling of 5000 of the 50000 events available from the GENIE event generator through the annihilation detector geometry.  The primary particles escaping the nucleus after annihilation can be different than the annihilation modes given in Table~\ref{edm:tab:sim} due to nuclear interactions.  A list of the most prevalent final state primary modes that escape the nucleus after a $n - \bar n$ or $p - \bar{n}$ annihilation is given for the GENIE event generator in Table~\ref{genie:tab:sim}.  Each event contains 3.84 $\pm$ 1.08$_{RMS}$ pionic primaries as displayed in the distribution on the left-hand side of Fig.~\ref{nnbarx:fig:primaries}, where a pionic primary is defined as either a pion or an $\omega$(782).  Fig.~\ref{nnbarx:fig:primaries} also shows the kinetic energy per final state primary.    

\begin{table}[H]
\centering
    	\caption{List of the most prevalent final state primary modes that escape the nucleus after a $n - \bar n$ or $p - \bar{n}$ annihilation from the GENIE event generator (see \S\ref{nnbar:subsec:annihmodel}).}
    	\label{genie:tab:sim}
    	\begin{tabular}{ccc}
        \hline\hline Final State Primary Mode & N$_{events}$ & Fraction of Total \\ \hline
        $\pi^{+}\pi^{-}2\pi^{0}$ & 530 & 10.60$\%$ \\
        $2\pi^{+}\pi^{-}\pi^{0}$ & 486 & 9.72$\%$ \\
        $\pi^{+}\pi^{-}\pi^{0}$ & 417 & 8.34$\%$ \\
        $2\pi^{+}\pi^{-}2\pi^{0}$ & 409 & 8.18$\%$ \\
        $\pi^{+}\pi^{-}3\pi^{0}$ & 329 & 6.58$\%$ \\
        $2\pi^{+}2\pi^{-}\pi^{0}$ & 315 & 6.30$\%$ \\
        $\pi^{+}2\pi^{0}$ & 290 & 5.80$\%$ \\
        $\pi^{+}3\pi^{0}$ & 219 & 4.38$\%$ \\
        $\pi^{+}\pi^{-}\omega$ & 145 & 2.90$\%$ \\
        $\pi^{+}\pi^{0}$ & 137 & 2.74$\%$ \\
        \hline\hline
    \end{tabular}
\end{table}

\begin{figure}
    \centering
    \includegraphics[width=5.58in]{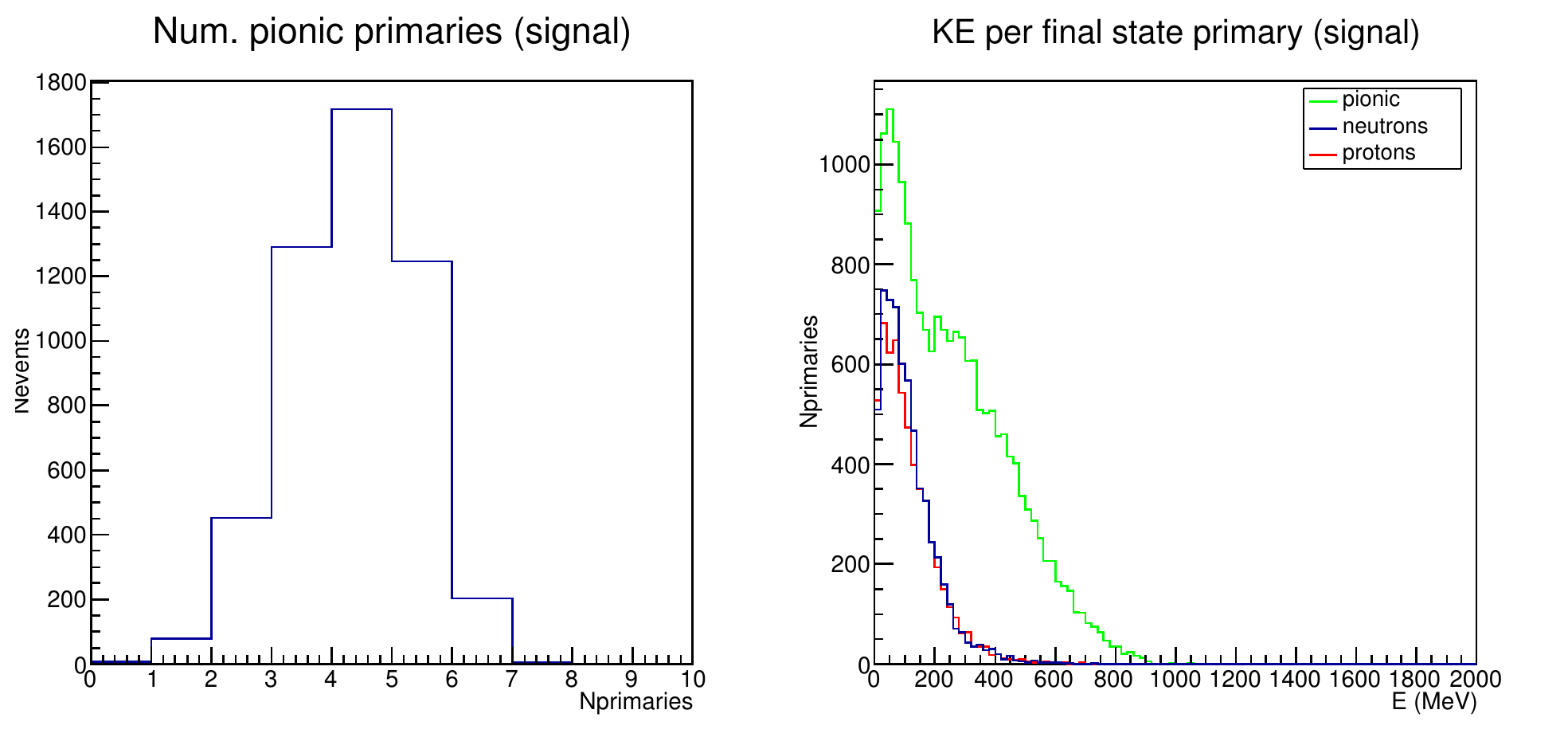}
    \caption{Histograms of the number of pionic primaries ($\pi^{+,-,0}$ or $\omega$(782)) on the left and the kinetic energy per primary particle on the right.}
    \label{nnbarx:fig:primaries}
\end{figure}

The base model for the annihilation detector geometry consists of a target, vacuum tube, straw tube tracker, and polystyrene/Pb calorimeter.  The target is a 2 m diameter, 100 $\mu$m thick $^{12}$C disk while the vacuum tube is 2 cm thick Al.  The tracker in the simulation is 0.5 m thick and consists of 5 mm diameter Kapton straws  organized into 50 XY planes with a fill gas of 70$\%$ Ar and 30$\%$ CO$_{2}$.  The simulation calorimeter geometry is divided into 20 alternating layers of 4 cm thick polystyrene and 0.2 cm thick Pb.  The length of the proto-detector in this simulation geometry is 15 m.  Fig. \ref{nnbarx:fig:nnbarxdetector} shows an event display from our preliminary Geant4 simulation of a $\pi^{+}\pi^{-}2\pi^{0}$ annihilation event in our proto-detector geometry.  

\begin{figure}
    \centering \includegraphics[width=1.08\textwidth]{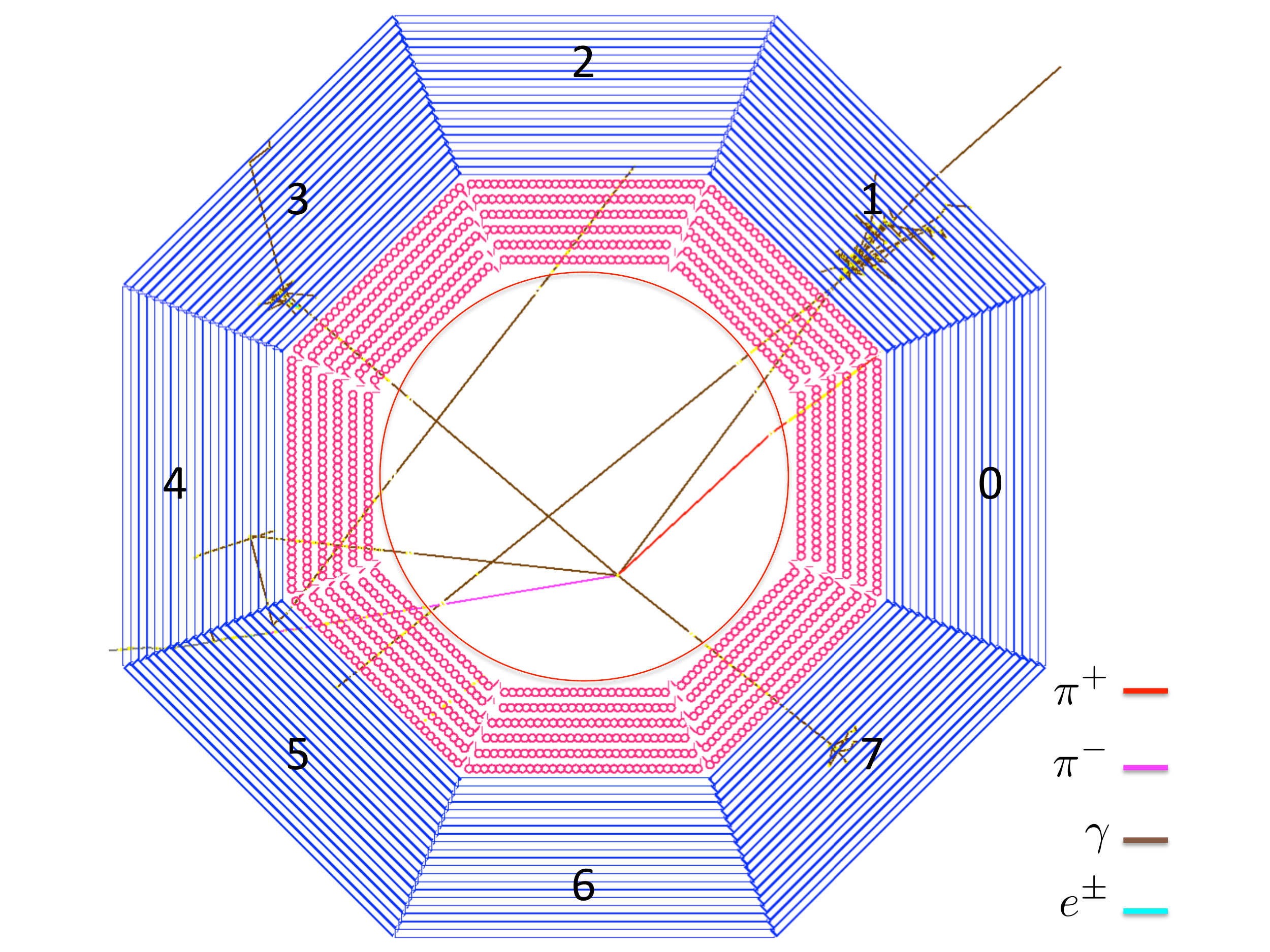}
\caption{Event display generated in our preliminary Geant4~\cite{Geant13} simulation for an annihilation event in a generalized 8-sided NNbar detector geometry.  A depiction of the numbering scheme for the detector towers in our simulation is shown.  The distance from the center of the apparatus to the edge of the active calorimeter (shown in blue) is 2.5 m, while the target (shown in red) had a radius of 1.0 m.  Between the target and the active calorimeter was the tracker, which contained straw tubes with a radius of 2.5 mm.  In this figure the straw tubes had a radius of 2.0 cm for visualization purposes.}
    \label{nnbarx:fig:nnbarxdetector}
\end{figure}

Our analysis of the simulations includes studies on energy deposition, timing and hit positioning.  Energy deposition studies have been carried out on all available deposition regions of the annihilation detector as displayed in Fig.~\ref{nnbarx:fig:eneana}.  A key point of study for future analysis, specifically comparisons of signal to background, is the energy deposition in the polystyrene scintillator, which we shall heretofore define as the active calorimeter where events typically deposit (776 $\pm$ 159$_{RMS}$) MeV as shown in the distribution on the left-hand side of Fig.~\ref{nnbarx:fig:eneana}.  The right-hand side of Fig.~\ref{nnbarx:fig:eneana} shows a study of how the signal mode acceptance in the simulation varies with the amount of energy deposited in the active calorimeter.  Applying the ILL trigger condition~\cite{BaldoCeolin94} of retaining events that deposit $\ge$ 450 MeV in the active calorimeter yields a signal mode acceptance of 97.42$\%$ in our simulation.  

Assessing the structural integrity of the vacuum tube will be a key issue down the line when the design of the annihilation detector apparatus is being finalized.  We have studied the signal mode acceptance for vacuum tube thicknesses from 2 cm (the thickness throughout the body of this article) to 5 cm.  The mean energy absorbed per signal mode event for a vacuum tube thickness of 2 cm was found to be 113 MeV, which increased to 254 MeV for a thickness of 5 cm.  The signal mode acceptance for events that deposit $\ge$ 450 MeV in the active calorimeter would drop from 97.42$\%$ to 93.40$\%$ when changing the thickness from 2 cm to 5 cm (see Fig.~\ref{nnbarx:fig:enevtana}).       

\begin{figure}
    \centering
    \includegraphics[width=5.6in]{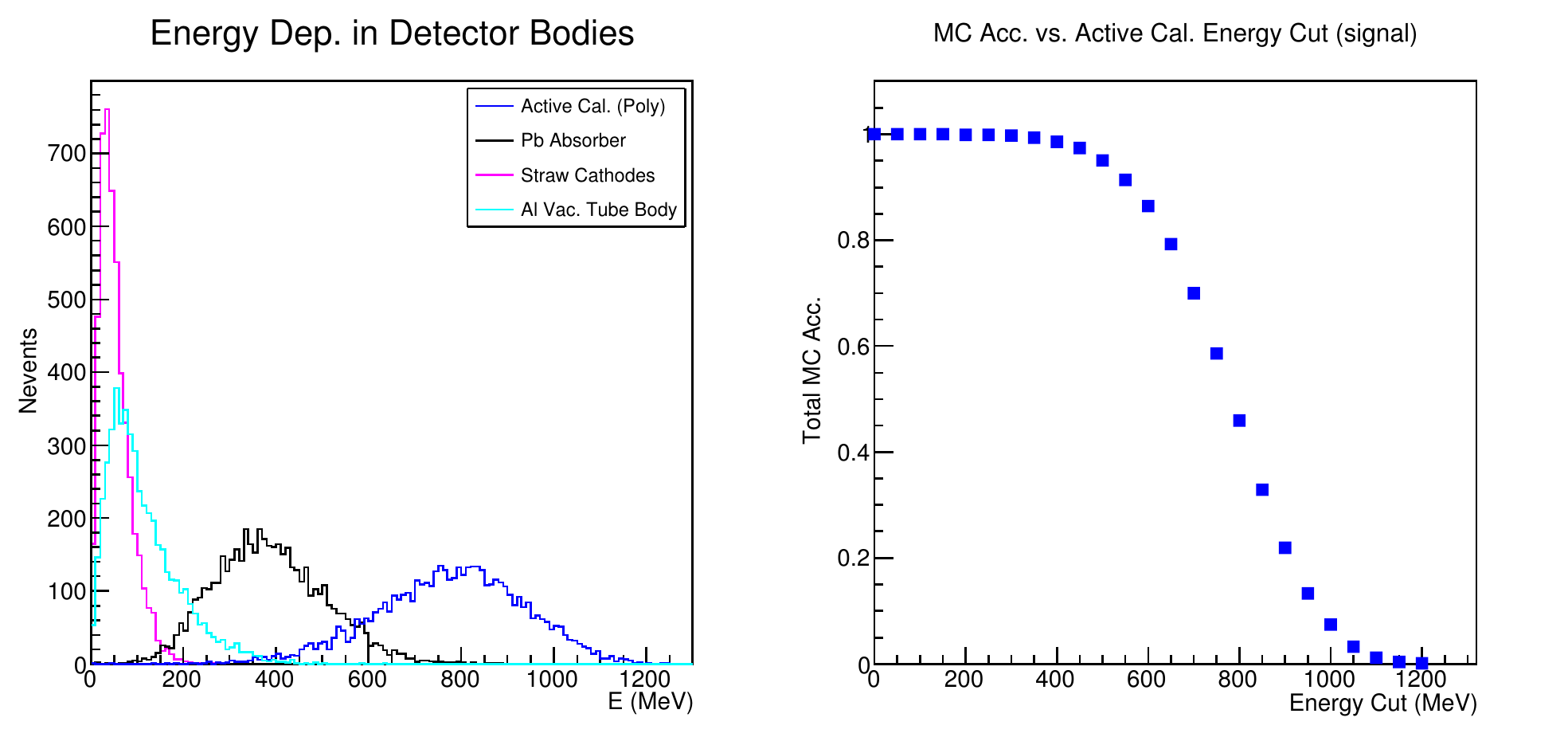}
\caption{Total energy deposited in the active calorimeter per event (left) and the fraction of accepted events in the simulation (MC Acc.) after a cut on the total energy in the active calorimeter.  The thickness of the Al vacuum tube in this study was 2 cm.}
    \label{nnbarx:fig:eneana}
\end{figure}

\begin{figure}
    \centering
    \includegraphics[width=5.6in]{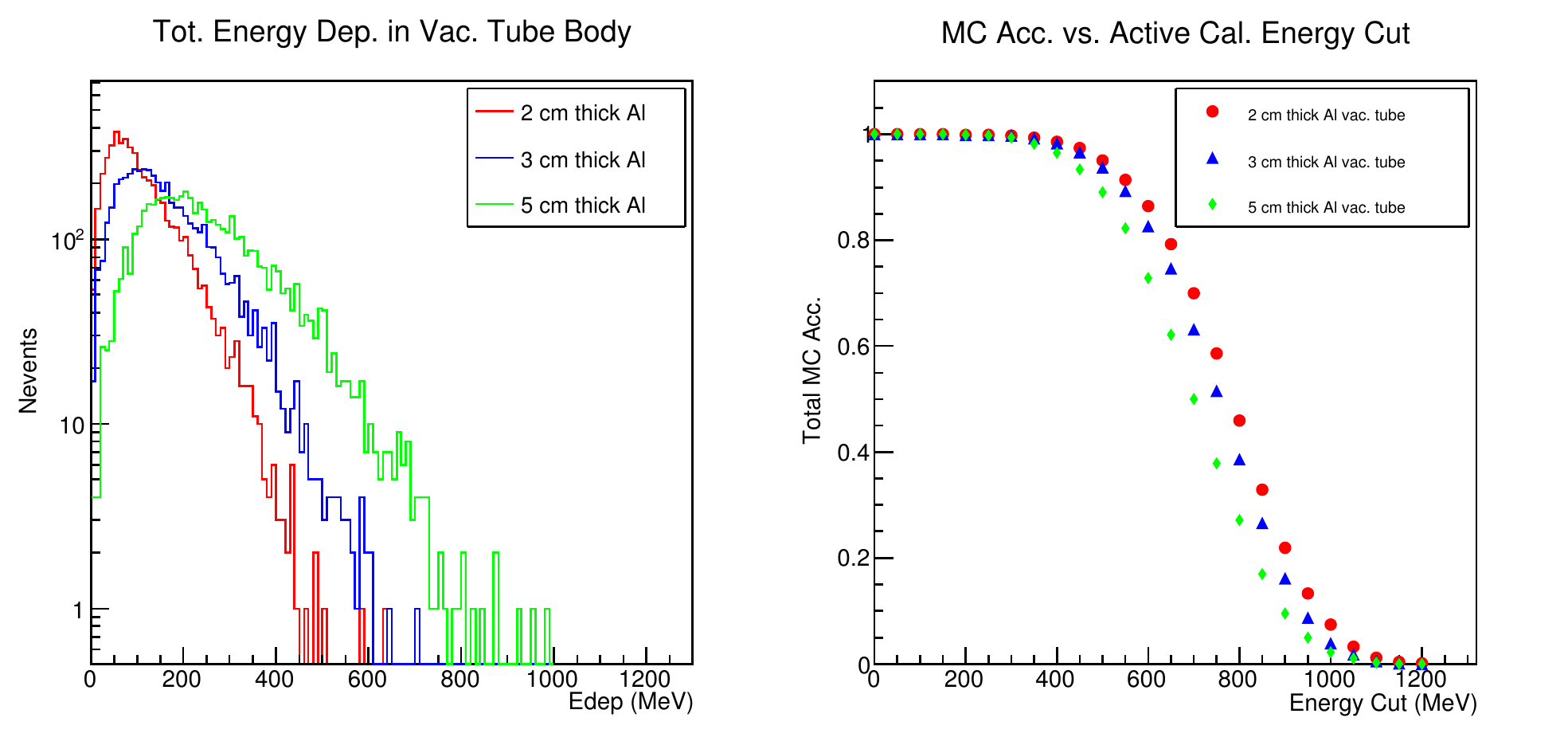}
    \caption{Total energy deposited in the vacuum tube for various vacuum tube thicknesses (left) and the fraction of accepted events in the simulation (MC Acc.) for each vacuum tube thickness after a cut on the total energy in the active calorimeter.}
    \label{nnbarx:fig:enevtana}
\end{figure}

Another useful tool for comparing signal to background in a future $n - \bar n$ experiment is the angular distribution of hits in the active calorimeter and tracker, where a hit in the tracker and active calorimeter is currently defined as an energy deposition point in any one of the straw tube cathodes and polystyrene scintillator panels respectively.  As displayed in the left-hand side of Fig.~\ref{nnbarx:fig:towerana}, most of the signal mode events create hits in over half of the tracker and active calorimeter towers, where the tower numbering scheme is shown in Fig.~\ref{nnbarx:fig:nnbarxdetector}.  Requiring hits in at least two tracker towers, two active calorimeter towers and a minimum energy deposition of 450 MeV in the active calorimeter gives a signal mode acceptance of 96.18$\%$.   

\begin{figure}
    \centering
    \includegraphics[width=5.6in]{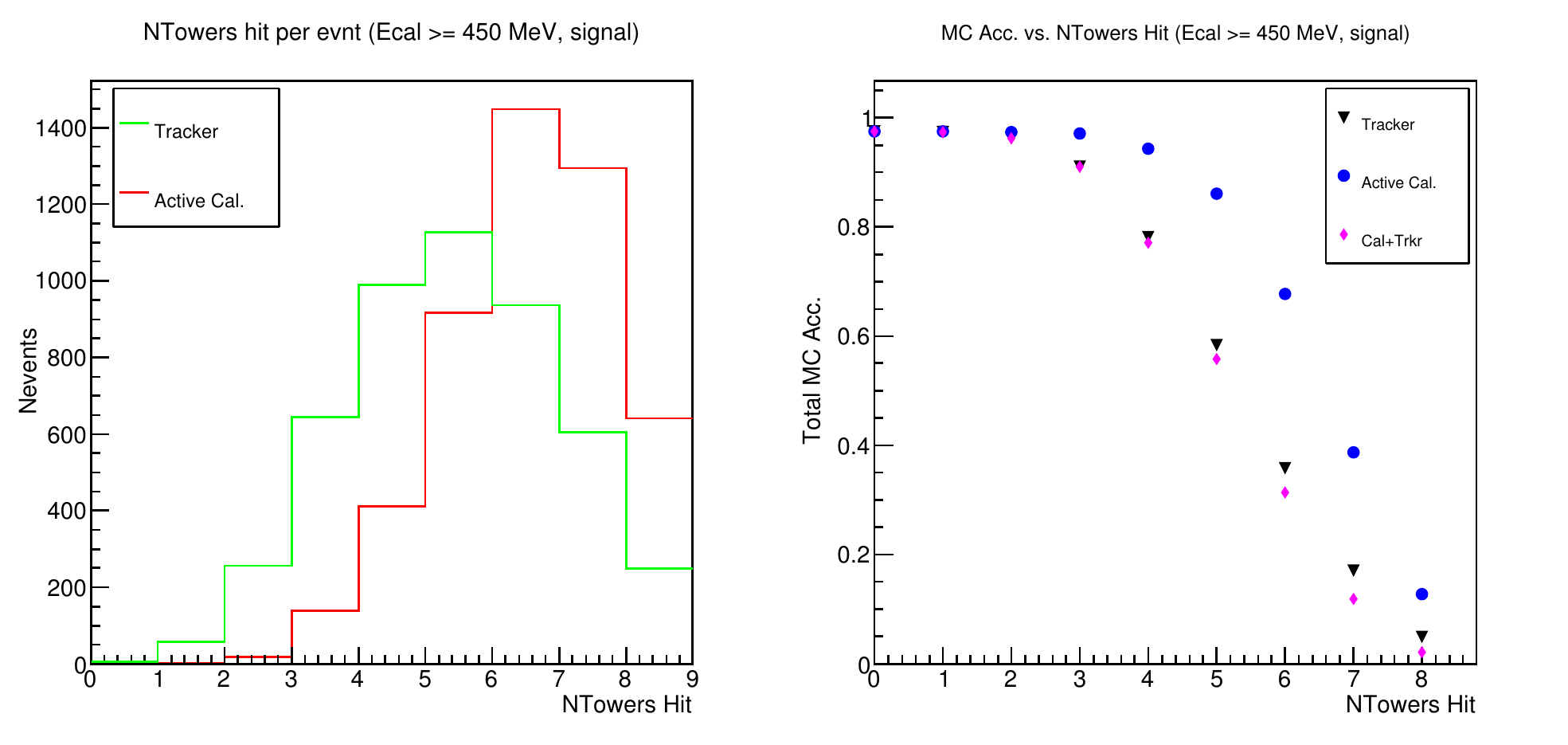}
    \caption{Number of tower hits per event on the left for the tracker (green) and active calorimeter (red).  The graph on the right side shows the fraction of accepted events (MC Acc.) vs. the number of towers hit.  Both plots require an energy of at least 450 MeV to be deposited in the active calorimeter for each event.}
    \label{nnbarx:fig:towerana}
\end{figure}

Trigger timing will play a key role in the comparison of signal to background.  The timing in the active calorimeter has been studied for signal mode events, where the active calorimeter time of energy deposition in our studies has been calculated in reference to the first active calorimeter hit.  The trigger timing window in the ILL experiment was 150 ns~\cite{BaldoCeolin91}.  The majority of all signal mode events in our simulations deposit a large fraction of their energy within 50 ns after the first active calorimeter hit (see Fig.~\ref{nnbarx:fig:timeana}).  Implementing the active calorimeter energy deposition and tower hit conditions described in the previous two paragraphs along with the requirement that all events deposit at least 500 MeV in the active calorimeter within 50 ns results in a signal mode acceptance of 91.74$\%$ for our simulations.  Changing the vacuum tube thickness from 2 cm to 5 cm would lower this acceptance to 84.6$\%$. 

\begin{figure}
    \centering
    \includegraphics[width=5.45in]{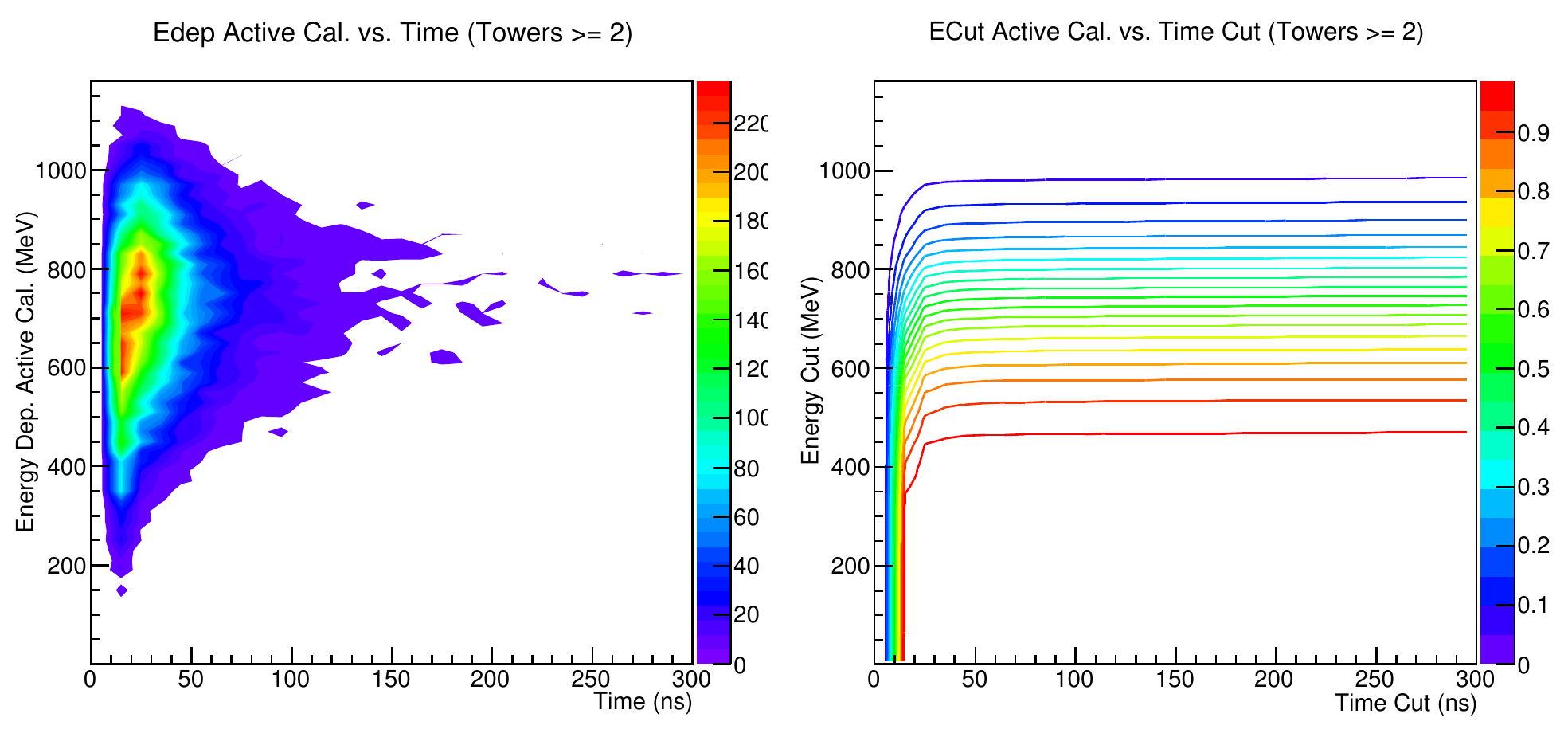}
    \caption{Number of events for energy deposited in the active calorimeter vs. time (left) and a contour plot of the fraction of accepted events for the energy deposited in the active calorimeter vs. size of the time window (right).  Each plot includes at least 2 calorimeter towers and 2 tracker towers hit per event and a bin size of 10 ns.}
    \label{nnbarx:fig:timeana}
\end{figure}

We are analyzing backgrounds (see \S\ref{nnbar:subsec:backgrounds}) through creation of a cosmic ray background event generator in addition to the generator for beam-related backgrounds.   
Development of a track finding and vertex position routines, creation a more detailed tracker geometry, implementation of a hit timing routine for the tracker, and a detailed analysis of beam products scattered off the target into the detector are in progress.  A summary of the current state of our analysis is given in Table~\ref{mc:tab:analysis} and the number of missing events per final state primary mode after all analysis steps is given in Table~\ref{mc:tab:missing}.

\begin{table}[H]
\centering
    	\caption{Simulation signal mode analysis summary.}
    	\label{mc:tab:analysis}
    	\begin{tabular}{ccc}
        \hline\hline $\#$ & Description & Acceptance \\ \hline
        1 & E$_{cal} \ge$ 450 MeV & 97.42$\%$ \\
        2 & 1) + N$_{towerhits, trkr} \ge$ 2 + N$_{towerhits, cal} \ge$ 2 & 96.18$\%$ \\
        3 & 2) + E$_{cal} \ge$ 500 MeV ($\it {t}_{window} \le$ 50 ns) & 91.74$\%$ \\
        \hline\hline
    \end{tabular}
\end{table}

\begin{table}[H]
\centering
    	\caption{Number of missing events per final state primary mode after step 3) in the MC signal analysis (see Table~\ref{mc:tab:missing}).}
    	\label{mc:tab:missing}
    	\begin{tabular}{ccc}
        \hline\hline Final State Primary Mode & N$_{missing}$ & N$_{missing}$/N$_{events,mode}$ \\ \hline
        $\pi^{+}\pi^{-}\pi^{0}$ & 40 & 9.59$\%$ \\
        $\pi^{+}\pi^{-}$ & 34 & 36.96$\%$ \\
        $2\pi^{+}\pi^{-}\pi^{0}$ & 25 & 5.14$\%$ \\
        $\pi^{+}2\pi^{0}$ & 23 & 7.93$\%$ \\
        $2\pi^{+}2\pi^{-}\pi^{0}$ & 23 & 7.30$\%$ \\
        $2\pi^{+}2\pi^{-}$ & 22 & 17.74$\%$ \\
        $\pi^{+}\pi^{-}2\pi^{0}$ & 22 & 4.15$\%$ \\
        $2\pi^{+}\pi^{-}$ & 21 & 19.44$\%$ \\
        $\pi^{+}2\pi^{-}\pi^{0}$ & 19 & 14.39$\%$ \\
        $\pi^{+}2\pi^{-}$ & 16 & 2.74$\%$ \\
        \hline\hline
    \end{tabular}
\end{table}

%%%%%%%%%%%%%%%%%%%%%%%%%%%%%%%%%%%%%%%%%%%%%%%%%%%%%%%%%%%%
\section{Ongoing Research and Development for Future $n - \bar n$ Oscillation Searches}
\label{nnbar:sec:randd}
%%%%%%%%%%%%%%%%%%%%%%%%%%%%%%%%%%%%%%%%%%%%%%%%%%%%%%%%%%%%

The preceding section indicates that significant improvements over current limits on
$\tau_{n-\bar n}$ would be achievable with a horizontal experiment based on a MW-class spallation source with relatively straightforward extensions of the general concepts and approach of the ILL experiment.
In this section we outline some strategies that could provide further improvements in sensitivity and research activities
that will help to define the cost and sensitivity of such an experiment. 
We also outline the considerations that should go into the
design of an annihilation detector suitable for use with a spallation source. In many cases these activities
will be of benefit not only to $n - \bar n$ searches, but to other fundamental physics investigations as well.

%%%%%%%%%%%%%%%%%%%%%%%%%%%%%%%%%%%%%%%%%%%%%%%%%%%%%%%%%%%%
\subsection{Novel Neutron Moderator Concepts}
\label{nnbar:subsec:noveln}
%%%%%%%%%%%%%%%%%%%%%%%%%%%%%%%%%%%%%%%%%%%%%%%%%%%%%%%%%%%%

The figure of merit for the statistical accuracy of the proposed $n - \bar n$
oscillation experiment is the number of neutrons observed times the square of the mean
free flight time (FOM={\it N}$_{n}T^{2}$). Both the spectral temperature of the moderated
neutrons and the source brightness directly impact the FOM. The optimization of the
free $n - \bar n$ oscillation experiment will rely on accurate simulation of
not only the total number of neutrons but also their energy spectrum. Unlike
the moderators at accelerator-based scattering facilities like the SNS, the emission
time and spatial distribution of the neutrons from the moderator are largely irrelevant for this experiment. This allows one to exploit well-understood concepts for increasing
moderator brightness (such as reentrant or grooved moderators) that are problematic
at pulsed sources due to their complicated emission time distributions and beam intensity non-uniformities. 
Novel ideas such as the \lq\lq convoluted~\rq\rq moderator, in which single crystals are placed within the moderator~\cite{Iverson2014}
or diamond nanoparticles as a slow neutron reflector around the moderator may provide significant improvements
to the low-energy neutron flux from a moderator suitable for a $n - \bar n$
oscillation search. 

Before an experiment can be designed to exploit these ideas,
however, reliable simulation tools need to be developed for accurately treating
the interactions of very cold neutrons with materials.  In the cold neutron regime of most interest to optimizing the
$n - \bar n$ oscillation experiment, available scattering kernels for
materials used in neutron moderation either do not model the relevant processes
accurately or do not even exist. The reason for this perhaps surprising circumstance is that in the slow neutron regime
there are strong interference effects in neutron scattering from condensed matter, and collective elementary excitations of the medium and their dispersion relations
play a major role in determining the elastic and inelastic cross sections. As a result there is no universally-applicable treatment of the slow neutron moderation problem.
This interaction is typically described in simulation codes through scattering kernels that
attempt to model the relevant double differential cross section ${d^{2}\sigma \over d\Omega dE}$ as a 
function of energy and or incident direction in an efficient manner. 

The flux from a neutron moderator depends not only on the primary flux from the target
and the design of the moderator, but also on the neutron reflector system that typically
surrounds both and whose task is to return the neutrons to the cold moderator to
increase its brightness. The reflector material must also withstand the radiation damage
and thermal load from the spallation target. The usual choice for a neutron reflector
for a cold neutron moderator is either graphite or beryllium. However the low energy
neutrons in the cold moderator possess de Broglie wavelengths which are large compared
to the atom spacings in materials. In this case one can imagine neutron reflectors of
enhanced reflectivity based on coherent scattering, and the longer time of flight of
such low energy neutrons in the free flight region of the experiment can greatly
improve the figure of merit. Recent experiments on mm-thick slabs of diamond
nanoparticles show a large albedo due to this coherent scattering effect~\cite{Cubitt10,Nesvishevsky082,Nesvishevsky10}, and such nanoparticle
composites should be highly immune to radiation damage. It should also be possible
to combine this concept with other design choices such as the convoluted moderator
or large parahydrogen moderator to realize compound gains in neutron flux.  The
optimal configuration for using these materials in the vicinity of the moderator
is not yet known: it depends on many details such as the size and polydispersity
of the nanoparticles and the absorption and incoherent scattering from contamination
with residual hydrogen, which has already been identified as a limiting factor in existing 
measurements~\cite{Nesvishevsky10} and might be addressed through displacement with
deuterium. Perhaps other materials (glassy carbon, alumina, ...) could provide even
greater albedo for very cold neutrons. We are therefore very optimistic that new reflector
materials can increase the cold neutron brightness from a moderator optimized for a
$n - \bar n$ oscillation experiment, but caution that experimental verification
of these gains must be demonstrated before they could be considered for a role in an
experimental design.

%%%%%%%%%%%%%%%%%%%%%%%%%%%%%%%%%%%%%%%%%%%%%%%%%%%%%%%%%%%%
\subsection{Additional Advantages of a Pulsed Neutron Source for $n - \bar n$ Oscillation Searches}
\label{nnbar:subsec:pulsedn}
%%%%%%%%%%%%%%%%%%%%%%%%%%%%%%%%%%%%%%%%%%%%%%%%%%%%%%%%%%%%

Most high-power spallation sources constructed for neutron scattering facilities are designed to operate in pulsed mode at relatively low frequencies (10-60 Hz) to enable neutron energy measurement in neutron scattering spectrometers using neutron time of flight. It is therefore useful to emphasize the additional advantages a pulsed neutron source have for this experiment. We can think of many aspects of the experiment which could benefit from this mode of operation.

One obvious advantage of pulsed operation is the possibility of improved background rejection for the antineutron detector.  As protons strike the spallation target, the slow neutrons take enough time to get from the source to the antineutron detector that they will arrive at the annihilation target well after most of the high-energy backgrounds consisting of fast neutrons, gammas, and charged particles.  Vetoing the short time interval of proton pulses on the target can effectively remove this fast background together with reducing cold neutron flux by a few percent.  The knowledge of the neutron energy spectrum from the moderator allows one to easily determine the neutron beam intensity on the annihilation target as a function of time and therefore also improves background discrimination and would also provide a consistency check should a set of several candidate events be found. Such improved background rejection features might be sufficient to allow for a straight path from the neutron source to the annihilation target which would improve the beam intensity (it is anticipated that a neutron bender would be needed in a CW source to get the antineutron target away from direct line-of-sight of the moderator).

A pulsed source would also allow one in principle to actively form the slow neutron beam as it moves between the moderator to the detector. The static optical elements which we described above conserve the phase space density of the neutron beam, and at a continuous neutron source there is no easy way to act separately on different neutron velocity classes using reflection devices. However the speed and approximate angle of the neutron as it moves through the experiment is known at a pulsed neutron source from geometry and neutron time-of flight. One can therefore imagine introducing a time-dependent neutron optical element which acts on different neutron velocity classes with a time-dependent interaction which can in principle increase the phase space density of the neutron beam.

As one example one can imagine reducing the transverse size of the beam (and therefore the cost of the experiment) using a phased supermirror reflector array. Although lengthening the apparatus can increase the free observation time, gravity leads to dispersion in the focusing of the neutron trajectories and eventually this poses a practical problem for the transverse size of the vacuum chamber in a horizontal geometry. Since the speed and approximate angle of the neutron as it strikes a location on the reflector array is known at a pulsed neutron source from geometry and neutron time-of flight, one can imagine a neutron reflector tiled with individually adjustable elements whose orientations can be changed in phase with the source frequency. One could imagine this function being effected by piezoelectric transducers and optimized to reduce the transverse size of the neutron beam passing thought the magnetically-shielded vacuum chamber. Since the cost of the long vacuum chamber and magnetic shield for the experiment is roughly proportional to its cross sectional area, such an active array could be used to reduce the cost of the experiment. 

%%%%%%%%%%%%%%%%%%%%%%%%%%%%%%%%%%%%%%%%%%%%%%%%%%%%%%%%%%%%
\subsection{Advantages of a $n - \bar n$ Oscillation Search using Ultracold Neutrons}
\label{nnbar:subsec:advultracold}
%%%%%%%%%%%%%%%%%%%%%%%%%%%%%%%%%%%%%%%%%%%%%%%%%%%%%%%%%%%%

Ultracold neutrons (UCN) are neutrons which are so slow that they can reflect from a material surface at all angles of incidence and can therefore be trapped in a fixed volume for times on the order of the neutron beta decay lifetime. A UCN based experiment envisions an intense source of UCN's filling a large UCN bottle surrounded by magnetic shielding and an antineutron detector. It has the advantage of forming a much more compact and therefore less expensive experiment than the long focusing arrangement needed for cold neutrons. Figure~\ref{fig:UCN} indicates a possible realization. The hope is that the increased observation time for UCN can counteract the lower brightness of UCN sources relative to cold neutron sources, but again gravity introduces some limitations. Here the relevant advance lies mainly in the development of \lq\lq superthermal\rq\rq UCN sources, which promise higher brightness than achieved at the present ILL turbine facility. A UCN source based on solid D2 now operates at LANSCE and PSI, other solid D2-based UCN sources are under construction at FRM, and NC State, and superfluid helium-based UCN sources are under consideration at PNPI and under construction at TRIUMF. The brightest source now in operation (PSI) after optimization could foresee making enough UCN to improve the current limits in neutron antineutron oscillation time by about a factor of $7$, corresponding to an improvement in the oscillation probability of a factor of $50$. 

\begin{figure}

\centering

\includegraphics[scale=0.6]{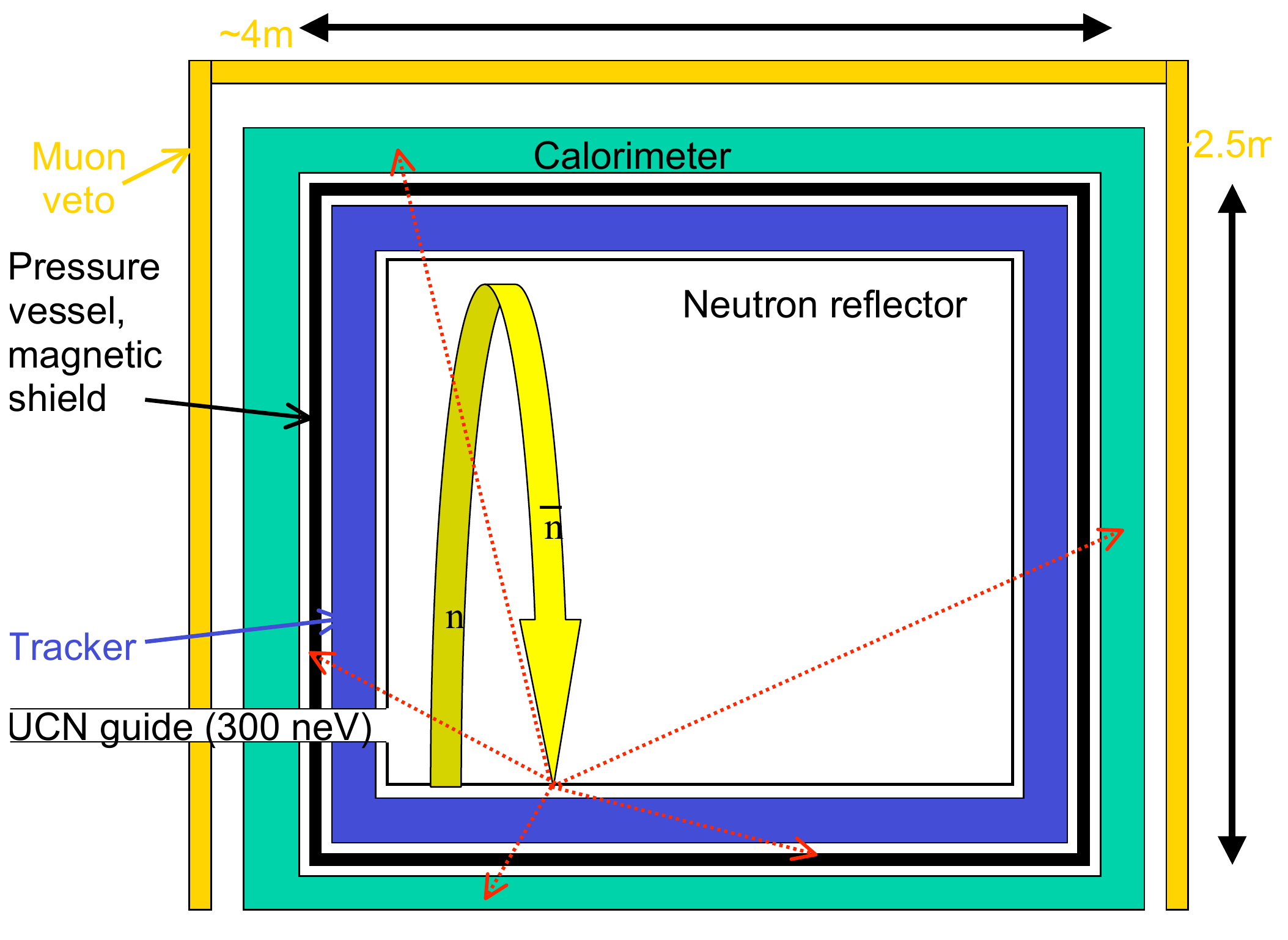}

\caption{Schematic design for an improved $n - \bar n$ oscillation search using ultracold neutrons. A large bottle filled from one of the superthermal UCN sources under construction could be used to improve the present limits.}

\label{fig:UCN}

\end{figure}

Neither of the above estimates takes any credit for the possibility of making use of those antineutrons in the system which coherently reflect from matter. Despite the rather large antineutron annihilation cross section in matter (about $1000$ barns at thermal neutron energies), there is a high degree of specularity (and therefore quantum mechanical coherence) in the reflection of the superposition of neutron and antineutron states from the mirrors. However this absorption probability eventually takes its toll after many reflections. Recent work~\cite{Friedman08, Kerbikov03, Kerbikov04}, building on earlier considerations~\cite{Yoshiki89, Golub92}, has improved the precision with which low energy antineutron scattering amplitudes can be estimated and has allowed a sharpened analysis of this problem.

%%%%%%%%%%%%%%%%%%%%%%%%%%%%%%%%%%%%%%%%%%%%%%%%%%%%%%%%%%%%
\subsection{Fast Neutron Response of Candidate Detectors}
\label{nnbar:subsec:fastneutrons}
%%%%%%%%%%%%%%%%%%%%%%%%%%%%%%%%%%%%%%%%%%%%%%%%%%%%%%%%%%%%

High energy neutron backgrounds produced in the spallation target can be mitigated through timing cuts, shielding, and selecting detectors with low sensitivity to the spallation products. Initial MCNP simulations of a 1 MW, 1 GeV target station project an average flux of neutrons above 100 MeV to be roughly 10$^{10}$ $n/{\rm s}$, requiring a $10^{18}$ reduction of possible backgrounds via hardware and on/offline software cuts to achieve $<$ 1 spurious event per year. We are conducting an evaluation of various detectors at the Weapons Neutron Research (WNR) facility at the Los Alamos Neutron Science Center (LANSCE) which provides a pulsed neutron flux up to 800 MeV.  The pulsed structure of WNR enables neutron detection efficiency measurement as a function of neutron energy using neutron time-of-flight.  Initial measurements have employed carbon fiber drift tubes produced by the Los Alamos muon radiography group~\cite{Wang09}, polystyrene scintillator bars produced at Fermilab, and the ATLAS TRT straw tubes. The ability of detectors to track charged particles in a high neutron flux environment is a common issue for several high energy physics experiments~\cite{Engl2010}.   

Large diameter (4.2 cm) Ar/ethane filled carbon fiber drift tubes are a robust and cost effective method for building large area arrays for charge particle tracking and represent an interesting technology for the interior vertex tracker. The integrated neutron detector efficiency for $E_n > 100$ MeV was measured to be $\sim 10^{-6}$, a factor of 10 lower than Monte Carlo prediction in GEANT4. This discrepancy is partially because of gain reduction in the detector over the duration of a 625 $\mu$s beam pulse train. The measured event rate at the end of the beam pulse train was suppressed by a factor of 10 - 10000 depending on the neutron flux intensity.  This can be attributed to a build up of charge in the gas suppressing the gain and thereby reducing the trigger rate. Tubes filled with an Ar/ethane/CF$_4$ (50/7.25/42.75\%) gas mixture, which is known to have a faster drift speed, suffered significantly less gain reduction.  The results of these studies and future studies at WNR will be used to benchmark the GEANT4 neutron scattering physics lists included in simulation of the background for the full detection system.

%%%%%%%%%%%%%%%%%%%%%%%%%%%%%%%%%%%%%%%%%%%%%%%%%%%%%%%%%%%%
\section{Conclusion and Avenues for Future Work}
\label{nnbar:sec:conclusions}
%%%%%%%%%%%%%%%%%%%%%%%%%%%%%%%%%%%%%%%%%%%%%%%%%%%%%%%%%%%%

In this paper we have  summarize{d}  our theoretical understanding of  $n - \bar n$ oscillations. There is no question that the discovery of such oscillations would
 constitute a result of fundamental importance for physics. {The implications of a lower bound on the oscillation time depend on theoretical context.}  {A} null result for oscillations{,} in combination with
 other data{,} might be sufficient to eliminate one subclass of logically-possible
 scenarios for  {generating} the baryon asymmetry of the universe below
 the electroweak phase transition{.} {W}e suggest that further investigation
 of this possibility is a worthwhile goal for particle phenomenology. 

{A} dedicated beamline at a MW-class spallation source could enable an experiment using free neutrons  to improve the sensitivity to the oscillation
 probability by a few orders of magnitude. As was recently demonstrated in ESS design work~\cite{Kamyshkov95,Boni10} the environment of spallation 
source allows optimized configurations of cold moderators where brightness of neutrons 
exceeds those obtained in the environment of research reactors.  We have identified a number of
 areas where additional R\&D could improve the ultimate reach of such an experiment. 

Examples of additional work needed to optimize the sensitivity of a new experiment to search for $n - \bar n$ oscillations include the following:
\begin{enumerate}
\item Continued R\&D on the technology for efficient, low-background antineutron detection. Extensive developments conducted for detector technologies in high energy physics since the last free neutron oscillation experiment can greatly extend the radiation environments in which the experiment
can be operated in the ``background-free'' mode required to conduct a sensitive search.
\item Continued R\&D on slow neutron moderation and optics technology.
In our view these components in combination with a properly-designed source
hold the most promise for improving the sensitivity of a free neutron oscillation experiment.  Several new technologies were 
described in this paper.  Although some of the required technical developments will
be conducted by scientists who apply slow neutrons to materials research
studies, other possibilities must be pursued with this specific experiment in
mind. Research facilities exist where this work can be carried out effectively.  Two of these facilities are ILL and the Indiana University Low Energy Neutron Source (LENS)~\cite{Leuschner07}.
\item Studies of possible improved capabilities of new underground detector
technologies{,} {such as those based
on liquid argon technology}{,} for $n - \bar n$ oscillations in nuclei. If these or any other large underground detector
technology can reduce backgrounds enough so that one enters an essentially
``zero background'' regime and at the same time can contain a large enough
volume of nuclei it would constitute a very interesting {complementary} approach to neutron oscillation physics.
\item Development of improved models of the antineutron annihilation process and of the propagation of the annihilation products through the nuclear
medium. This study is important both for free neutron oscillations searches
and for searches using underground detectors.
\end{enumerate}

Recent progress in studies of n-nbar processes raised new theoretical issues
that can be addressed by the following additional theoretical work:
\begin{enumerate}
\item  An effective field theory analysis of possible $\Delta\mathcal{B} = 2$ operators involving Standard-Model fields. To our knowledge a complete analysis has not
 been conducted, although an initial analysis for a certain subclass of operators has been performed [129]. Such an analysis could be useful in conjunction
 with specific models for $n - \bar n$ oscillations and improved constraints on new processes from the LHC and other rare decay searches and
 symmetry tests to judge whether or not the interesting possibility of post
sphaleron baryogenesis can be eliminated by laboratory experiments and (if
 so) what is the required experimental sensitivity.  {The prospect of eliminating} an entire
 subclass of possible baryogenesis mechanisms through a null
 search for $n - \bar n$ oscillations would be an attractive goal for
an experiment, and establishing the experimental sensitivity
required to do so would be of great interest.
\item Better calculations of the matrix elements of the 6-quark operators relevant for $n - \bar n$ oscillations. These matrix elements are needed
to relate the $n - \bar n$ transition amplitude to the mass scale
probed in the oscillation process. Existing calculations of the matrix elements of the 6-quark operators relevant for $n - \bar n$ oscillations
are quite old and use bag model wave functions  {that} could give large errors. QCD based calculations are now becoming possible on the lattice as a
byproduct of techniques developed to calculate slow neutron-nucleon parity violation amplitudes. We strongly encourage researchers to pursue these calculations.
\item A more thorough and quantitative analysis of the relationship between free neutron oscillations and $n - \bar n$ oscillations in nuclei,
including uncertainties due to the strong interaction.  {Among} studies
 {that} could help illuminate this relationship {we mention} estimates of the relative sizes of the different modes of nucleon disappearance which can exist in
nuclei in addition to direct $n - \bar n$ oscillation and subsequent
annihilation, and a QCD sum rule calculation of the conversion factor $R$,
which would serve as a cross-check on existing calculations and have the
benefit of a more direct connection to QCD.  {The simplicity} of the
deuteron {and} our  {detailed} knowledge of its properties  might
well enable a more precise calculation of $R$ {that} would  {inform the extraction of} oscillation constraints from SNO.
\end{enumerate}
  
%%%%%%%%%%%%%%%%%%%%%%%%%%%%%%%%%%%%%%%%%%%%%%%%%%%%%%%%%%%%  
\section{Acknowledgements}
%%%%%%%%%%%%%%%%%%%%%%%%%%%%%%%%%%%%%%%%%%%%%%%%%%%%%%%%%%%%
 
The work of W. M. Snow,  C-Y. Liu, and R. Van Kooten was supported in part by the Indiana University Center for Spacetime Symmetries. The University of Tennessee group was supported in part by the ORU Funding Program of the Office of Research of the University of Tennessee.  Fermilab is operated by Fermi Research Alliance, LLC, under contract DE-AC02-07CH11359 with the United States Department of Energy Office of High Energy Physics.

\appendix

%%%%%%%%%%%%%%%%%%%%%%%%%%%%%%%%%%%%%%%%%%%%%%%%%%%%%%%%%%%%
\section{Neutron Sources}
\label{nnbar:apdx:nsources}
%%%%%%%%%%%%%%%%%%%%%%%%%%%%%%%%%%%%%%%%%%%%%%%%%%%%%%%%%%%%

Intense sources of free neutrons are usually created through either fission reactions in a nuclear reactor or through spallation in high $Z$ targets struck by GeV proton beams.  We briefly examine these neutron sources and the process by which slow neutrons are produced starting from neutrons with energies several orders of magnitude greater. Neutrons are produced from fission in a research reactor at an average energy of  2\,MeV. They are slowed to thermal energy in a moderator such as heavy or light water, graphite, or beryllium, surrounding the fuel. The peak core fluence rate of research reactors is typically in the range $10^{14} n/{\rm cm^2/s}$ to $10^{15} n/{\rm cm^2/s}$.  To maximize the neutron density it is necessary to increase the fission rate per unit volume, but the power density is ultimately limited by heat transfer and material properties. In the spallation process, protons (typically) are accelerated to  energies in the  GeV range and strike a high $Z$ target, producing approximately 20 neutrons per proton with energies in the fast and epithermal region~\cite{Windsor81}. This is an order of magnitude more neutrons per nuclear reaction than from fission. Existing spallation sources yield neutron rates of $10^{16}$\,s$^{-1}$ and $10^{17}$\,s$^{-1}$. Although the time-averaged fluence from spallation neutron sources is presently about an order of magnitude lower than for fission reactors, there is potentially more room for technical improvements in the near-term future due in part to the relaxation of the constraints needed to maintain a nuclear chain reaction~\cite{Difilippo91} and from the potential for future developments in GeV proton accelerator technology. It is worth remembering that this paper has demonstrated that even in the absence of these improvements, a modern spallation source can host a free neutron oscillation experiment that could greatly improve upon the existing limits. 

The main feature that differentiates spallation sources from reactors is the  convenient operation in a pulsed mode. At most reactors one obtains continuous beams with a thermalized Maxwellian energy spectrum. In a spallation source, neutrons arrive at the experiment while the production source is off, and the frequency of the pulsed source can be chosen so that slow neutron energies can be determined by time-of-flight methods.  The lower radiation background and convenient neutron energy information can be advantageous for certain experiments. The frequency is typically chosen to lie in the 10\,Hz to 60\,Hz range, so that the subsequent neutron pulses from the moderator do not overlap for typical neutron spectra and distances to the experimental area.

Fast neutrons reach the thermal regime most efficiently through a logarithmic energy cascade of roughly 20 to 30 collisions with matter rich in hydrogen or deuterium. Cold neutrons are produced by a cryogenic neutron moderator adjacent to the reactor core or spallation target held at a temperature of $\approx 20$\,K.  One generally wants the moderator as cold as possible to increase the phase space density of the neutrons.   As the neutron wavelengths become large compared to the atomic spacings, the total scattering cross sections in matter are dominated by elastic or quasielastic processes, and it becomes more difficult for the neutrons to thermalize. The development of new types of cold neutron moderators is therefore an important area for research.

%%%%%%%%%%%%%%%%%%%%%%%%%%%%%%%%%%%%%%%%%%%%%%%%%%%%%%%%%%%%
\section{Neutron Moderation}
\label{nnbar:apdx:nmod}
%%%%%%%%%%%%%%%%%%%%%%%%%%%%%%%%%%%%%%%%%%%%%%%%%%%%%%%%%%%%

The phase space compression of neutron ensembles, usually referred to as neutron moderation, is concerned with the energy, space and time distribution of neutrons in a moderating medium. The fundamental object of interest in neutron moderation is $f({\vec r}, {\vec v}, t)$, the phase space distribution function, where $ f({\vec r}, {\vec v}, t) d{\vec r}d{\vec v}$ is the average number of neutrons in the phase space element $d{\vec r}d{\vec v}$ at time $t$.  Since the density of neutrons is orders of magnitude lower than the density of scatterers in a medium, neutron-neutron scattering can be ignored, and since the mean free path of neutrons in matter is much larger than the separation  between atoms in the medium, the neutron motion between collisions can be approximated as classical motion with velocity ${\vec v}$. This leads to a linear integral equation for the neutron phase space density of the form:

\begin{eqnarray*}\label{eq:1-10}
( {{\partial } \over {\partial t}} +{\vec v}\cdot {\vec \Delta}) f({\vec r}, {\vec v}, t)&=&-\rho v \sigma_{a}(v)f({\vec r}, {\vec v}, t)\\\nonumber
&+& \int{[\rho v_{0}\sigma_{s}({\vec v_{0}}, {\vec v}) f({\vec r}, {\vec v_{0}}, t) -\rho v \sigma_{s}({\vec v}, {\vec v_{0}}) f({\vec r}, {\vec v}, t)]}d{\vec v_{0}}
\end{eqnarray*}

where $v=|{\vec v}|$ is the neutron speed, $\rho$ is the density of the medium, $\sigma_{a}(v)$ is the neutron absorption cross section, and $\sigma_{s}({\vec v_{0}}, {\vec v})$ is the scattering kernel. The first term in the integral can be understood to represent \lq\lq downscattering" of the neutron of velocity ${\vec v_{0}}$ to velocity ${\vec v}$ and the second term can be understood to represent \lq\lq upscattering" of the neutron of velocity ${\vec v}$ to velocity ${\vec v_{0}}$. In terms of the double differential scattering cross section $d^{2}\sigma$, defined as the average number of incident particles with velocity ${\vec v_{0}}$ scattered into the volume element $d{\vec v}$ per atom per unit time per unit incident flux, the scattering kernel is defined by
\begin {equation}\label{eq:1-11}
d^{2} \sigma=\sigma_{s}({\vec v_{0}}, {\vec v})d{\vec v},
\end{equation}
which is expressed in spherical coordinates in terms of the scattered neutron energy $E=\frac{1}{2}{mv^{2}}$. If ($\theta,\phi$) are the polar coordinate of $\vec{v}$ with respect to $\vec{v}_{0}$ as polar axis, then $d\vec{v}=v^{2}dv\textrm{sin}\theta d\theta d\phi={v\over m} d\Omega dE$. The double differential scattering cross section becomes
\begin {equation}\label{eq:1-12}
{{d^{2} \sigma} \over {d\Omega dE}}= {v\over m} \sigma_{s}({\vec v_{0}}, {\vec v}).
\end{equation}

On the other hand, the usual expression for the double differential scattering cross section is
\begin{equation}\label{eq:1-13}
{{d^{2} \sigma} \over {d\Omega dE}}=\left({\sigma_{s} \over 4\pi\hbar}\right){v\over v_{0}} S({\vec Q}, \omega),
\end{equation}
where $S(\vec{Q}, \omega)$ is the dynamic structure factor~\cite{Lovesey86}, $\sigma_s$ is the bound scattering cross section per atom,  and ${\vec{Q}}$ and $\omega$ are the momentum and energy transfers from the neutron to the medium. Comparing Eq. \ref{eq:1-12} and Eq. \ref{eq:1-13}, one can see that the scattering kernel is directly proportional to the dynamic structure factor,
\begin{equation}\label{eq:1-14}
\sigma_{s}({\vec {v_{0}}}, {\vec{ v}})=({{m \sigma_{s}} \over {4\pi\hbar v_{0}}})S({\vec{Q}}, \omega).
\end{equation}

Neutron moderation theory has been developed for nuclear engineering and related applications in the energy regime between the typical energy which neutrons possess upon liberation from nuclei (MeV) to the kinetic energies of atoms in matter at room temperature ($\sim$25 meV), since this is the regime of greatest interest to nuclear energy technology.  The physics is simple.  In this energy regime the neutron-matter interaction is dominated by the strong neutron-nucleus interaction, and since the scattering is predominantly s-wave in this energy regime it can be parameterized by (accurately-measured) neutron scattering lengths. Also, since these energies are large compared to the kinetic energy and the binding energy of the atoms in the moderating medium, the total cross section is given to a good approximation by the incoherent sum of the scattering cross sections from the individual atoms in the medium. In this limit, therefore, the theory of neutron moderation only needs to apply energy and momentum conservation to a sequence of non-relativistic collisions of neutrons of mass $m$ and initial energy $E$ with target nuclei of mass $M$ and zero kinetic energy, with probabilities given by the known cross sections. Such an analysis forms the core of neutron moderation theory in nuclear reactors~\cite{Duderstadt76,Davison57,Foderaro71}.  For neutron moderation in hydrogen, the results are especially simple: the average energy after $n$ collisions is simply $<E_{n}>={E_{0}/2^{n}}$ and the average number of collisions of a neutron with initial energy $E_{0}$ and final energy $E$ is $<n(E_{0}, E)>={\ln(E_{0}/E) \over \ln(2)}$. For a neutron with initial energy 1 MeV and final energy $\sim 1$ eV,  this number is about $15$.

For neutron energies at or below about 1 eV, however, the situation is qualitatively different. In this case, the neutron wavelength becomes comparable to the separation between atoms in the medium and the neutron energy is below the binding energy of the atoms in the medium. In this case the dynamic structure factor $S({\vec {Q}}, \omega)$  exhibits a large coherent elastic component and strong interference effects which depend on the details of the structure and modes of motion of the medium, thereby precluding any universal solution to the problem of neutron moderation. For the coherent elastic scattering contribution to the scattering cross section, the state of the medium is unchanged. Therefore the neutron motion can be described through motion in an effective potential (optical potential), which means that Liouville's theorem applies and the phase space density for this subset of processes remains unchanged. Only the contribution to the total cross section from inelastic processes is responsible for phase space compression. This fraction typically decreases as the neutron energy drops below the binding energy of atoms in the material and the $meV$ excitation energies of collective modes in the medium. The amount of phase space compression per collision for slow neutrons in a moderating medium, therefore, tends to decrease as the energy decreases.

The obvious optimal solution for neutron cooling would be to match the elementary excitation spectrum to the neutron phase space to be cooled and reduce the neutron energy to zero by the creation of one or more elementary excitations, with the refrigerator that maintains the moderating medium at $T=0$ K providing the dissipation required for phase space compression.This is precisely what happens for fast neutron motion in media with hydrogen mentioned above: the equal neutron-proton mass causes the energy-momentum dispersion relation for the neutron and the proton to fall on top of each other and thereby maximize the fractional energy loss per collision and the neutrons are cooled rapidly. As mentioned above, this process works well for collision energies above the binding energy of the protons in the medium.  With the exception of free protons, however, whose dispersion relation is essentially the same as that for a free neutron and therefore possesses the optimum overlap needed for rapid phase space compression through collisions, at lower energies the dispersion relation for the elementary excitations in the medium which can cool the neutrons typically overlap the neutron dispersion relation $E={p^{2} \over 2m}$ only in some localized regions of  $(Q,\omega)$ space.

In so-called superthermal neutron moderators optimized for the creation of ultracold neutrons~\cite{1991GolubBook} one concentrates on that small subset of processes in which the initial neutron phase space intersects the energy-momentum dispersion relation of the medium at one or a few points: for these cases the neutrons can lose virtually all their
kinetic energy and momentum in one collision and come close to rest. A neutron moderator can increase the phase space density starting from an initial state without forcing the neutrons to come into thermal equilibrium with the moderator. Thermal equilibrium of the moderating medium allows one to use detailed balance to relate the scattering kernels for the forward and reverse directions as $vw({\vec v})\sigma_{s}({\vec v}, {\vec v_{0}})=v_{0} w({\vec v_{0}})\sigma_{s}({\vec v_{0}}, {\vec v})$ or, in terms of initial and final energies, $E_{0}\sigma(E_{0}\to E, \Omega_{0}\to \Omega)=\exp{((E_{0}-E)/kT)} E\sigma(E \to E_{0}, \Omega \to \Omega_{0})$. If we consider the extreme case of a cold moderating medium at temperature $T\to 0$ with only two energy states separated by $\Delta\gg kT \to 0$, the scattering kernel for the second (upscattering) term in the integral equation for the phase space density becomes negligible and disappears at $T=0$ due to conservation of energy. If the initial phase space distribution is characterized by a Maxwellian of temperature $T_{n}\gg\Delta$, the phase space density will increase to a value proportional to $(E/\Delta)\exp({\Delta/kT})$, thereby increasing exponentially as a function of temperature. This mechanism is known as the \lq\lq superthermal \rq\rq\ principle in the field of ultracold neutron production~\cite{1991GolubBook}, which concentrates on that subset of the neutron phase space distribution with final energies $E$ near zero. It emphasizes especially those scattering events in which the neutron loses essentially all of its energy in one collision. These events are often dominated by the coherent elementary excitations in the medium, whose dispersion relations define which subclasses of neutrons from the initial distribution can conserve energy and momentum in the collision and be brought to rest. Liouville's theorem is evaded in this case by the avoidance of thermal equilibrium: the entropy production in the neutron moderation process is removed by the refrigerator that maintains the $\Delta\gg T$ condition.

One can in principle increase the phase space density in a neutron moderator indefinitely by cooling the medium all the way to $T=0$ and allowing the neutrons to come into thermal equilibrium. Since the phase space density in thermal equilibrium in a uniform medium is proportional to the Maxwell velocity distribution $w({\vec{ v}})\propto (m/kT)^{3/2} \exp({-mv^{2}/2kT})$, the phase space density increases as $1/T^{3/2}$ as the distribution is cooled.  In practice the finite neutron absorption cross section of all media (other than those composed of nuclei like $^{4}He$, $^{2}H$, $^{16}O$, $^{12}C$, and a few other nuclei which possess absorption cross sections of millibarns and below) places an upper bound on the number of collisions that can be tolerated, and then once again the amount of phase space compression per collision depends on the microscopic properties of the medium even at $T=0$. Unfortunately the inelastic modes available in a condensed medium tend to freeze as $T \to 0$, thereby reducing the efficiency of the moderating medium even further~\cite{Dingenen62}. This phenomenon is clearly observed in the neutron energy distributions measured from cold neutron sources. If one characterizes the energy distribution from a finite moderator medium at a temperature $T$ approximately as a Maxwellian with an effective temperature $T_\mathrm{eff}$, not only is $T_\mathrm{eff}>T$ but also the fractional deviation ${(T_\mathrm{eff}-T)}\over T$ increases as $T \to 0$ in the cold neutron regime~\cite{Butterworth57}. This is not difficult to understand for a simple solid: the energies of molecular vibrations lie above the cold neutron energy range, and the inelastic cross section in the limit as the moderator temperature $T \to 0$, which is dominated by one-phonon creation, becomes $\sigma_{10}\propto (E_{0}/\omega_{D})^{3}$ where $\omega_{D}$ is the Debye frequency and therefore decreases as $E^{3}$ for small $E$. In fact, one can construct a simple theory for the case of ice which fits the measured dependence of $T_\mathrm{eff}$ on the moderator temperature rather well~\cite{Utsuro73}.

In the case of cold neutron moderation, in which we are trying to shift a broad distribution of neutron velocities to lower values, the width of the distribution of neutrons in phase space that we wish to cool is typically large compared to the range in $(Q,\omega)$ space over which $S(Q,\omega)$ is large. Typically $S(Q,\omega)$ is large only over narrow ridges in $(Q,\omega)$ space due to the presence of well-defined elementary excitations of the medium such as phonons, magnons, librons, etc. For cooling a broader phase space distribution of neutrons to energies in the $100 \mu {\rm eV} \sim {\rm meV}$ range, therefore, it is expected that in addition to processes involving the creation of single elementary excitation that multi-excitation processes may also be important. It is therefore important for $S(Q,\omega)$ to possess some strength near $\omega \rightarrow 0$ so that the largest number of neutrons in the distribution have energy losses $\omega_{i}$ in some sequence of collisions $S(Q, \omega_{i})$ that can allow them to approach $E_{f} \rightarrow 0$.

These considerations guide the search for improved neutron moderating media into
some obvious directions. As mentioned above, it is clear that normal $3D$ phonon excitation becomes inefficient at low energies, since the density of states and therefore $S(Q,\omega)$ vanishes as  $\omega \to 0$. The abnormally large scattering cross section of hydrogen, more an order of magnitude larger than for other nuclei, makes hydrogenous materials the obvious choice if the number of collisions for phase space cooling is not limited by the $0.3$ barn neutron absorption cross section of hydrogen at 25 meV and its $1/\sqrt{E}$ increase at lower energies. The typical reduction of $S(Q,\omega)$ with $\omega$ can, in some circumstances, be mitigated by employing materials with atypical dispersion in their excitation spectra. In CH$_{4}$, for example, the partially-free rotational motion of the molecules leads to considerable spectral weight at low energies since these excitations are local rather than collective~\cite{Shin10}.  Other possible examples include disordered materials, which often possess an excess of low energy inelastic modes relative to phonons. Confinement of the moderating medium in porous media can greatly shift both its thermodynamic properties and its excitation spectra in addition to generating new local modes of atoms bound on the surface.  On the other hand our poor understanding of disordered media, as demonstrated for example in the still-active controversy surrounding the physical explanation of the inelastic boson peak excitations around 1 meV observed in many disordered media, seems to preclude a general theoretical analysis.

Although the inelastic cross sections for neutrons are the main concern in neutron phase space compression, the elastic processes are also important, since they typically dominate the total cross section and the mean free path, and therefore strongly influence the spatial distributions of the neutrons and determine the brightness of the moderator. If the moderator is large enough that the number of scattering collisions is sufficiently large before absorption, neutrons may come close to thermal equilibrium within the moderator. The thermal neutron flux spectrum is approximately a Maxwellian distribution,
\begin{equation}
\phi_{M}(E)={1\over {(k_{B} T)^{2}}}E ~\textrm{exp} \left(- {E\over {k_{B} T} }\right),
\end{equation}
where $k_{B}$ is Boltzmann's constant and $T$ is the absolute temperature of the moderator. In the case of neutron moderation it is also known that the under-moderated component possesses a $1/E$ spectrum, and a real moderated neutron spectrum is typically a weighted sum of these two distributions. 
 
The optimization of a neutron moderator for a particular application is therefore a complicated task, in which suitable compromises must be struck between the often competing processes of elastic scattering, inelastic scattering, absorption, and neutron leakage out of the moderator (after all, it does no good to cool the neutrons if they do not end up in your experiment).  In addition to this multidimensional optimization task, the designer must face engineering realities of heat removal, radiation tolerance and system service. The selection of materials and the physical geometry of both the moderator itself and its surrounding systems can play significant roles in determining the flux of neutrons available to the experimenter.  Moderator design typically involves intensive computer simulations of proton, neutron, and photon transport through multiple materials in complex geometries. Unfortunately, today we have adequate kernels for modeling slow neutron transport through only very few of the relevant materials that could be used in constructing the next generation of cold neutron sources. Our inability to model slow neutron transport through real materials accurately can also have an impact on the ability to model accurately neutron-related backgrounds in rare decay and other low background environment experiments. There is, therefore, a need within the fundamental physics community for improving the available low-energy neutron data and for facilities that could be used to validate models that might be developed from those data. The recent development of small scale accelerator-driven neutron sources~\cite{Baxter05, Lavelle08} provides suitable infrastructure for this work. There is a need for this community to develop consensus on the materials and configurations that need to be pursued for maximum impact on the field. 

%%%%%%%%%%%%%%%%%%%%%%%%%%%%%%%%%%%%%%%%%%%%%%%%%%%%%%%%%%%%
\section{Neutron Optics}
\label{nnbar:apdx:noptics}
%%%%%%%%%%%%%%%%%%%%%%%%%%%%%%%%%%%%%%%%%%%%%%%%%%%%%%%%%%%%

A basic summary of neutron optics is very useful for understanding how neutron guides work. For a rigorous development of neutron optics theory see Sears~\cite{Sears89}.
An optical description for particle motion in a medium is possible for that subset of amplitudes in which the quantum mechanical state of the medium is unchanged.  For these events the wave function of the particle is determined by a one-body Schr\"{o}dinger equation
\begin{equation}
    \label{eq:coherent}
    \left[\frac{-\hbar^{2}}{ 2m}\Delta +v(r)\right]\psi(r)=E\psi(r),
\end{equation}
where $\psi(r)$ is the coherent wave and $v(r)$ is the optical potential of the medium. The coherent wave and the optical potential satisfy the usual coupled system of equations (Lippmann-Schwinger) of non-relativistic scattering theory. To an excellent approximation the optical potential for slow neutrons is
\begin{equation}
    \label{eq:opticalpotential}
    v_{opt}(r)~=~(2\pi \hbar^2/m)\sum_{l} N_{l} b_{l},
\end{equation}
where $N_{l}$ is the number density of scatterers and $b_{l}$ is the coherent scattering length for element $l$. All neutron-matter interactions contribute to the scattering length: the dominant contributions come from the neutron-nucleus strong interaction and the interaction of the neutron magnetic moment with magnetic fields. Both interactions are spin-dependent and therefore possess two channels with $J=I\pm 1/2$. The coherent scattering length $b$ is the sum of the scattering lengths in both scattering channels weighted by the number of spin states in each channel.  From a quantum mechanical point of view, this is simply the total amplitude for a neutron to scatter without a change in the internal state of the target. The effect of the optical potential for a non-absorbing uniform medium with positive coherent scattering length is to slow down the neutrons as they encounter the potential step due to the matter, thereby decreasing the neutron wave vector, \textit{K}, within the medium. A neutron index of refraction can be defined by this relative change in the magnitude of the wavevector $n=K/k$. Conservation of energy at the boundary determines the relation to the optical potential
\begin{equation}
    \label{eq:refractionindex}
    n^{2}=1-{v_{0}/E}.
\end{equation}
Incoherent effects due to neutron absorption and incoherent scattering with the medium remove probability density from the coherent wave and can be modeled by adding an imaginary part to the optical potential whose magnitude is determined by the optical theorem of non-relativistic scattering theory. The scattering lengths therefore become complex and in general are also spin-dependent.

It may seem counterintuitive that the attractive neutron-nucleus interaction can give rise to a repulsive neutron optical potential. For a weak potential there is indeed a one-to-one relation between the sign of the zero-energy scattering phase shift and the sign of the potential. The neutron-nucleus interaction is strong enough to form bound states, and therefore this perturbative intuition becomes inadequate. If one recalls Levinson's theorem from non-relativistic scattering theory, which says that the zero energy phase shift increases by $\pi$ for each bound state, one can see that it is possible for an attractive potential to produce a negative phase shift. Simple models~\cite{Peshkin71} and more detailed treatments~\cite{Aleksejev98} of the neutron-nucleus interaction show that for most nuclei the scattering lengths are indeed positive, which corresponds to a repulsive optical potential. A summary of all neutron scattering lengths up to 1991~\cite{Koester91} and a set of recommended values~\cite{Rauch00b} exist in the literature.

The fact that the neutron optical potential is repulsive for most materials makes neutron guides possible. Since the neutron index of refraction is less than unity, the well-known light optical phenomenon of total {\it internal} reflection for rays at an interface becomes for neutrons total {\it external} reflection. Neutron guides~\cite{Maier63}, the neutron equivalent of optical fibers, can be used to conduct neutron beams far from the neutron source by total external reflection from mirrors with negligible loss in intensity within their phase-space acceptance. The phase-space acceptance can be increased over that of a uniform medium by multilayer coatings called supermirrors, which expand the effective critical angle for total external reflection through constructive interference of scattering from different layers.

The expression for the low energy neutron scattering length of an atom away from nuclear resonances for unpolarized atoms and neutrons and nonmagnetic materials is
\begin{equation}
    \label{eq:alllengths}
    b=b_{nuc}+Z(b_{ne}+b_{s})[1-f(q)]+b_{s}+b_{pol},
\end{equation}
where $b_{nuc}$ is the scattering amplitude due to the neutron-nucleus strong force, $b_{ne}$ is the neutron-electron scattering amplitude due to the internal charge distribution of the neutron, $b_{s}$ is the Mott-Schwinger scattering due to the interaction of the magnetic moment of the neutron with the ${\bf
v} \times {\bf E}$ magnetic  field seen in the neutron rest frame from electric fields, $b_{pol}$  is the scattering amplitude due to the electric polarizability of the neutron in the intense electric field of the nucleus, and $f(q)$ is the charge form factor (the Fourier transform of the electric charge distribution of the atom). The electromagnetic contribution to the scattering lengths from $b_{ne}$ and $b_{s}$ are exactly zero for forward scattering due to the neutrality of the atoms, which forces the charge form factor $f(q) \to 1$ as $q \to 0$~\cite{Sears89}. The weak interaction also makes a contribution to the scattering length which possess a parity-odd component proportional to ${\vec{s_{n}} \cdot {\vec{p}}}$, where ${\vec{s_{n}}}$ is the neutron spin and ${\vec{p}}$ is the neutron momentum. The relative sizes of these contributions to the scattering length from the strong, electromagnetic, and weak interactions are roughly in the ratio $1:10^{-3}:10^{-7}$.



%% References


\bibliographystyle{model1b-num-names}

\begin{thebibliography}{00}
\bibitem{Bressi11} G. Bressi {\it et al.}, Phys. Rev. A {\bf 83}, 052101 (2011).
\bibitem{Baumann88} J. Baumann {\it et al.}, Phys. Rev. D {\bf 37}, 3107 (1988).
\bibitem{Stueckelberg38} E. C. G. Stueckelberg, Helv. Phys. Acta {\bf 11}, 312 (1938).
\bibitem{Schlamminger08} S. Schlamminger {\it et al.}, Phys. Rev. Lett. {\bf 100}, 041101 (2008).
\bibitem{Fields12} B.~D.~Fields, P.~Molaro, and S.~Sarkar, in K. A. Olive {\it et al}. (Particle Data Group), Chin. Phys. C {\bf 38}, 090001 (2014), http://pdg.lbl.gov/2014/reviews/rpp2014-rev-bbang-nucleosynthesis.pdf.
\bibitem{Babu2015} K. S. Babu and R. N. Mohapatra, Phys. Rev. D {\bf 91}, 096009 (2015).
\bibitem{Dolgov92} A. D. Dolgov, Phys. Rep. {\bf 222}, 309 (1992).
\bibitem{Dolgov98} A. D. Dolgov, Surv. High Energy Phys. {\bf 13}, 83 (1998).
\bibitem{Sak67} A. D. Sakharov, JETP Lett. {\bf 5}, 24 (1967); Sov. Phys. Usp. {\bf 34}, 392 (1991).
\bibitem{Zeldovich81} Ya. B. Zeldovich and A. Dolgov, Rev. Mod. Phys. {\bf 53}, 1 (1981). 
\bibitem{Dolgov10} A. Dolgov, Phys. Atom. Nucl. {\bf 73}, 588 (2010).
\bibitem{Dolgov14} A. Dolgov and V. Novikov, Phys. Lett. B{\bf 732}, 244 (2014). 
\bibitem{Bennett03} C. L. Bennett {\it et al.},  Ap. J  {\bf 148},1 (2003).
\bibitem{Hooft76} G. 't Hooft, Phys. Rev. Lett. {\bf 37}, 8 (1976).
\bibitem{Hooft78} G. 't Hooft, Phys. Rev. D {\bf 14}, 3432 (1978).
\bibitem{Kuzmin85} V. A. Kuzmin, V. A. Rubakov, and M. E.  Shaposhnikov, Phys. Lett. B {\bf 155}, 36 (1985).
\bibitem{Ramsey_Musolf12} D. J. E. Morrissey and M. Ramsey-Musolf, New J. 
Phys. {\bf 14}, 125003 (2012).
\bibitem{Shaposhnikov87} M. E.  Shaposhnikov, Nucl. Phys. B {\bf 287}, 757 (1987).
\bibitem{Fukugita86} M. Fukugita and T. Yanagida, Phys. Lett. {\bf 174} , 45 (1986).
\bibitem{Georgi74} H. Georgi and S. Glashow, Phys. Rev. Lett. {\bf 32}, 438 (1974).
\bibitem{Raby08} S. Raby {\it et al.}, arXiv:0810.4551 [hep-ph].
\bibitem{Beringer12} K. A. Olive {\it et al.} (Particle Data Group), Chin. Phys. C {\bf 38}, 090001 (2014).
\bibitem{seesaw} P.~Minkowski, Phys. Lett. B {\bf 67}, 421 (1977);
T.~Yanagida in {\em Workshop on Unified Theories, KEK Report
79-18}, p.~95 (1979);
M.~Gell-Mann, P.~Ramond and R.~Slansky, {\em Supergravity},
p.~315; Amsterdam: North Holland (1979);
S.~L. Glashow, {\em 1979 Cargese Summer Institute on Quarks and
Leptons}, p.~687; New York: Plenum (1980);
R.~N. Mohapatra and G.~Senjanovic, Phys. Rev. Lett. {\bf 44}, 912 (1980).
\bibitem{marshak80} R. N. Mohapatra and R. E. Marshak, Phys. Rev. Lett. {\bf 44}, 1316 (1980).
\bibitem{Elliott12} S. R. Elliott, Mod. Phys. Lett. A {\bf 27}, 1230009 (2012).
\bibitem{Duerr13} R. Foot, G. C. Joshi, and H. Lew, Phys. Rev. D {\bf 40}, 2487 (1989); M. Duerr, P. Filevitz Perez and M. B. Wise, Phys. Rev. Lett. {\bf 110}, 231801 (2013).
\bibitem{Abov84} Yu. G. Abov, F. S. Dzheparov, and L. B. Okun, JETP Lett. {\bf 39}, 493 (1984).
\bibitem{Lamoreaux91} S. K. Lamoreaux, R. Golub, and J. M. Pendlebury, Europhys. Lett. {\bf 14}, 503 (1991).
\bibitem{Yao06} W.M. Yao {\it et al.} (Particle Data Group), J. Phys. G. Nucl. Part. Phys. {\bf 33}, 1 (2006).
%\bibitem{Kostelecky04} V. A. Kostelecky, Phys. Rev. D {\bf 69}, 105009 (2004).
\bibitem{Pati74} J. Pati and A. Salam, Phys. Rev. D {\bf 10}, 275 (1974).
\bibitem{Mohapatra75rm} R. N. Mohapatra and J. Pati, Phys. Rev. D {\bf 11}, 566 (1975).
\bibitem{Mohapatra75jp} R. N. Mohapatra and J. Pati, Phys. Rev. D {\bf 11}, 2558 (1975); G. Senjanovi\'c and R. N. Mohapatra, Phys. Rev. D {\bf 12}, 1502 (1975).
\bibitem{Kuzmin70} V. A. Kuzmin, JETP Lett. {\bf 12}, 228 (1970).
\bibitem{Okun2013} L. B. Okun, arXiv:1306.5052 [hep-ph].
\bibitem{Glashow79} S. L. Glashow, Proc. Neutrino '79, Bergen, Vol. {\bf 1}, 518 (1979).
\bibitem{Kuo80} T. K. Kuo and S. T. Love, Phys. Rev. Lett. {\bf 45}, 93 (1980).
\bibitem{Chang80} L.-N. Chang and N.-P. Chang, Phys. Lett. B {\bf 92}, 103 (1980).
\bibitem{Mohapatra80rn} R. N. Mohapatra and R. E. Marshak, Phys. Lett. B {\bf 94}, 183 (1980).
\bibitem{Cowsik81} R. Cowsik and S. Nussinov, Phys. Lett. B {\bf 101}, 237 (1981).  
\bibitem{Rao82} S. Rao and R. Shrock, Phys. Lett. B {\bf 116}, 238 (1982).  
\bibitem{Misra83} S. P. Misra and U. Sarkar, Phys. Rev. D {\bf 28}, 249 (1983).
\bibitem{Rao84} S. Rao and R. Shrock, Nucl. Phys. B {\bf 232}, 143 (1984).  
\bibitem{Huber01} S. J. Huber and Q. Shafi, Phys. Lett. B {\bf 512}, 365 (2001). 
\bibitem{Babu01} K. S. Babu and R. N. Mohapatra, Phys. Lett. B {\bf 518}, 269 (2001).
\bibitem{Nussinov02} S. Nussinov and R. Shrock, Phys. Rev. Lett. {\bf 88}, 171601 (2002).
\bibitem{Mohapatra05} R. N. Mohapatra, S. Nasri, and S. Nussinov, Phys. Lett. B {\bf 627}, 124 (2005).
\bibitem{Babu06} K. S. Babu, R. N. Mohapatra, and S. Nasri, Phys. Rev. Lett. {\bf 97}, 131301 (2006).
\bibitem{Dutta06} B. Dutta, Y. Mimura, and R. N. Mohapatra,  Phys. Rev. Lett. {\bf 96}, 061801 (2006).
\bibitem{Berezhiani06} Z. Berezhiani and L. Bento, Phys. Rev. Lett. {\bf 96}, 081801 (2006).
\bibitem{Babu09} K. S. Babu, P. S. Bhupal Dev, and R. N. Mohapatra, Phys. Rev. D {\bf 79}, 015017 (2009).
\bibitem{Mohapatra09} R. N. Mohapatra, J. Phys. G {\bf 36}, 104006 (2009).
\bibitem{Gu11} P. Gu and U. Sarkar, Phys. Lett. B {\bf 705}, 170 (2011).
\bibitem{Babu13} K. S. Babu, P. S. B. Dev, E. C. F. S. Fortes, and R. N. Mohapatra, Phys. Rev. D {\bf 87}, 115019 (2013). 
\bibitem{Arnold13} J. M. Arnold, B. Fornal, and M. B. Wise, Phys. Rev. D {\bf 87}, 075004 (2013).
%\bibitem{Morrissey2012} D. E. Morrissey and M. J. Ramsey-Musolf, New
J. Phys. {\bf 14}, 125003 (2012).
\bibitem{Canetti2013} L. Canetti, M. Drewes, T. Frossard, and M. Shaposhnikov, 
Phys. Rev. D {\bf 87}, 093006 (2013).
\bibitem{Degouvea14} A. de Gouv${\rm \hat{e}}$a, J. Herrero-Garc{\rm \'{i}}a, and A. Kobach, Phys. Rev. D {\bf 90}, 016011 (2014). 
\bibitem{zeld-B} Ya. B. Zeldovich, Phys. Lett. A {\bf 59}, 254 (1976).
\bibitem{Bambi07} C. Bambi, A. Dolgov, and K. Freese, Nucl. Phys. B {\bf 763}, 91 (2007).
\bibitem{cms} The CMS Collaboration, Eur. Phys. J. C {\bf 74}, 3149 (2014).
\bibitem{deppisch} F. F. Deppisch, T. E. Gonzalo, S. Patra, N. Sahu and U. Sarkar, Phys. Rev. D {\bf 90}, 053014 (2014); M.~Heikinheimo, M.~Raidal and C.~Spethmann, arXiv:1407.6908 [hep-ph]; J.~A.~Aguilar-Saavedra and F.~R.~Joaquim, Phys. Rev. D {\bf 90}, 115010 (2014).
\bibitem{baryo-majoron} R.~Barbieri and R.~N.~Mohapatra, Z. Phys. C {\bf 11}, 175 (1981); Z. Berezhiani, arXiv:1507.05478 (2015).
\bibitem{Wietfeldt:2009} F. Wietfeldt (Tulane University), {\it private communication} (2009).
\bibitem{BaldoCeolin94} M. Baldo-Ceolin {\it et al.}, Z. Phys. C {\bf 63}, 409 (1994).
\bibitem{Gardner15} S. Gardner and E. Jafari, Phys. Rev. D {\bf 91}, 096010 (2015).  
\bibitem{Dove83}C. B. Dover, A. Gal, and J. M. Richard, Phys. Rev. D
{\bf 27}, 1090 (1983); Phys. Rev. C {\bf 31}, 1423 (1985); Nucl. Instrum. Methods
Phys. Res., Sect. A {\bf 284}, 13 (1989).
\bibitem{Friedman08} E. Friedman and A. Gal, Phys. Rev. D {\bf 78}, 016002 (2008).
\bibitem{Albe82}W. M. Alberico, A. Bottino, and A. Molinari,
Phys. Lett. {\bf 114B}, 266 (1982); W. M. Alberico, J. Bernabeu, A. Bottino,
and A. Molinari, Nucl. Phys. A {\bf 429}, 445 (1984).
\bibitem{Kang2010} Xian-Wei Kang, Hai-Bo Li, and Gong-Ru Lu, Phys. Rev. D {\bf 81}, 051901 (2010).
\bibitem{Sokoloff2014} M. D. Sokoloff, {\it private communication} (2014). 
\bibitem{Buchoff12} M. I. Buchoff, C. Schroeder, and J. Wasem, LLNL-PROC-563736, arXiv:1207.3832 [hep-lat].
\bibitem{Berezhiani2015} Z. Berezhiani and A. Vainshtein, arXiv:1506.05096.
\bibitem{Babu12} K.~S.~Babu and R.~N.~Mohapatra, Phys. Lett. B {\bf 715}, 328 (2012).
\bibitem{Babu07} K. S. Babu, R. N. Mohapatra and S. Nasri, Phys. Rev. Lett. {\bf 98}, 161301 (2007).
\bibitem{Dolgov06} A. D. Dolgov and F. R. Urban, Nucl. Phys. B {\bf 752}, 297 (2006).
\bibitem{EPJ-More} Z. Berezhiani, Eur. Phys. J. C {\bf 64}, 421 (2009).
\bibitem{Ban} G. Ban {\it et al.}, Phys. Rev. Lett. {\bf 99}, 161603 (2007).
\bibitem{Serebrov} A. Serebrov {\it et al.}, Phys. Lett.  B {\bf 663}, 181 (2008).  
\bibitem{Altarev} I. Altarev {\it et al.}, Phys. Rev. D {\bf 80}, 032003 (2009).     
\bibitem{Bodek} K. Bodek {\it et al.}, Nucl. Instrum. Meth. A {\bf 611}, 141 (2009).
\bibitem{Serebrov2} A. Serebrov {\it et al.}, Nucl. Instrum. Meth. A {\bf 611}, 137 (2009).
\bibitem{Nesti} Z. Berezhiani and F. Nesti, Eur. Phys. J. C {\bf 72}, 1974 (2012).
\bibitem{Foot} R. Foot, Phys. Rev. D {\bf 82}, 095001 (2010).
\bibitem{magnetic} Z. Berezhiani, A. D. Dolgov, and I. I. Tkachev, Eur. Phys. J. C {\bf 73}, 2620 (2013). 
\bibitem{Mohapatra_Nussinov_Teplitz2002}
R. N. Mohapatra, S. Nussinov, and V. Teplitz, Phys. Rev. D {\bf 66}, 063002 (2002).
\bibitem{Fidecaro} G. Fidecaro {\it et al.}, Phys. Lett. B {\bf 156}, 122 (1985).
\bibitem{Bressi90} G. Bressi {\it et al.}, Il Nuovo Cimento {\bf 103 A}, 731 (1990).
\bibitem{Bitter91} T. Bitter {\it et al}., Nucl. Inst. Meth. A {\bf 309}, 521 (1991).
\bibitem{Kinkel92} U. Kinkel, Z. Phys. C {\bf 54}, 573 (1992).
\bibitem{Schmidt92} U. Schmidt, T. Bitter, P. El-Muzeini, O. Scharpf, and D. Dubbers, Nucl. Inst. Meth. A {\bf 320}, 569 (1992).
\bibitem{BaldoCeolin90} M. Baldo-Ceolin {\it et al}., Phys. Lett. B {\bf 236}, 95 (1990).
\bibitem{Kamyshkov95} Y. Kamyshkov et al., {\it Proceedings of the ICANS-XIII meeting of the
International Collaboration on Advanced Neutron Sources}, Paul Scherrer Institute, Villigen,
Switzerland, October 11-14, p. 843 (1995).
\bibitem{Chung02} J. Chung {\it et al.}, Phys. Rev. D {\bf 66}, 032004 (2002).
\bibitem{Abe15} K. Abe {\it et al.}, Phys. Rev. D {\bf 91}, 072006 (2015).
\bibitem{Bergevin10} M. Bergevin,  {\it Search for Neutron-Antineutron Oscillations at the Sudbury Neutrino Observatory}, PhD Thesis University of Guelph (2010).
\bibitem{Huef98} J. Huefner and B. Z. Kopeliovich, Mod. Phys. Lett. A
{\bf 13}, 2385 (1998).
\bibitem{Jones84} T. W. Jones {\it et al.}, Phys. Rev. Lett. {\bf 52}, 720 (1984).
\bibitem{Takita86} M. Takita {\it et al.}, Phys. Rev. D {\bf 34}, 902 (1986).
\bibitem{Berger90} C. Berger {\it et al.}, Phys. Lett. B {\bf 240}, 237 (1990).
\bibitem{Kopel11} V. Kopeliovich and I. Potashnikova, JETP Lett. {\bf 95}, 1 (2012) [Pisma Zh. Eksp. Teor. Fiz. {\bf 95}, 3 (2012)].
\bibitem{Abe11} K. Abe {\it et al.}, arXiv:1109.3262 [hep-ex].
\bibitem{LBNE14} C. Adams {\it et al}. (The LBNE Collaboration), BNL-101354-2014-JA, FERMILAB-PUB-14-022, LA-UR-14-20881, arXiv:1307.7335 [hep-ex].
\bibitem{GLACIER} A. Rubbia, J. Phys. Conf. Ser. {\bf 375}, 042058 (2012).
\bibitem{Vain13} A. Vainshtein (University of Minnesota), {\it private communication} (2012).
\bibitem{ICARUS2011} C. Rubbia {\it et al}., JINST {\bf 6} P07011 (2011).
\bibitem{Anderson2012} C. Anderson {\it et al}., Phys. Rev. Lett. {\bf 108}, 161802 (2012).
\bibitem{CAnderson2012} C. Anderson {\it et al}., JINST {\bf 7} P10019 (2012).
\bibitem{Acciarri2014} R. Acciarri {\it et al}., Phys. Rev. D {\bf 89}, 112003 (2014).
\bibitem{Kearns} E. Kearns (Boston University), {\it private communication} (2014).
\bibitem{Mohapatra03} R. N. Mohapatra and A.  Perez-Lorenzana, Phys. Rev. D {\bf 67}, 075015 (2003).
\bibitem{Kamyshkov03} Y. Kamyshkov and E. Kolbe, Phys. Rev. D {\bf 67}, 076007 (2003).
\bibitem{Ahmed04} S. N. Ahmed {\it et al.} (SNO Collaboration), Phys. Rev. Lett. {\bf 92}, 102004 (2004).
\bibitem{Back03} H. O. Back {\it et al.} (Borexino Collaboration), Phys. Lett. B {\bf 563}, 23 (2003).
\bibitem{Araki05} T. Araki {\it et al.} (KamLAND Collaboration), Phys. Rev. Lett. {\bf 96}, 101802 (2006).
\bibitem{Mason06} T. E. Mason {\it et al.}, Physica B {\bf 385}, 955 (2006).
\bibitem{Blau09} B. Blau {\it et al.}, Neutron News {\bf 20 (3)}, 5 (2009).
\bibitem{Fischer97} W. Fischer {\it et al.}, Physica B {\bf 234}, 1202 (1997).
\bibitem{Belushkin06} A. V. Beluskin, Neutron News {\bf 2}, 14 (2006);  E. P. Shabalin, {\it Fast Pulsed and Burst Reactors}, Pergamon Press, Oxford (1979).
\bibitem{Maekawa10} F. Maekawa {\it et al.}, Nucl. Inst. Meth. A {\bf 620}, 159 (2010). 
\bibitem{Iverson13} E. Iverson (Oak Ridge National Laboratory), {\it Private Communication} (2013).
\bibitem{Wohlmuther11} M. Wohlmuther {\it et al.}, {\it The Improved SINQ Target}, Proceedings of the 10th International Topical Meeting on Nuclear Application of Accelerators (AccApp '11), ISBN: 978-0-89448-706-4 (2011).
\bibitem{ESSTDR} ESS Technical Design Report, ESS-doc-274, ISBN: 978-91-980173-2-8, http://eval.esss.lu.se/DocDB/0002/000274/015/TDR\_online\_ver\_all.pdf (2013).
\bibitem{esben} E. Klinkby {\it et al.}, arXiv:1401.6003 [physics.ins-det]. 
\bibitem{Mezei76} F. Mezei, Comm. Phys. {\bf 1}, 81 (1976).
\bibitem{Schanzer04} C. Schanzer, P. Boni, U. Filges, and T. Hils, Nucl. Inst. Meth. A {\bf 529},  63 (2004);  N. Kardjilov, P. Boni, A. Hilger, M. Strobl, and W. Treimer, Nucl. Inst. Meth. A {\bf 542}, 248 (2005); R. White and M. Bull, Neutron News {\bf 17}, 36 (2006). 
\bibitem{Altarev14} I. Altarev {\it et al.}, Rev. Sci. Instrum. {\bf 85}, 075106 (2014).
\bibitem{Coulter90} K. P. Coulter {\it et al.}, Nucl. Inst. Meth. A {\bf 288}, 463 (1990).
\bibitem{Chupp07} T. E. Chupp {\it et al.}, Nucl. Inst. Meth. A {\bf 574}, 500 (2007).
\bibitem{Petoukhov06} A. Petoukhov {\it et al.}, Physica B {\bf 385-386}, 1146 (2006).
\bibitem{Hussey05} D. R. Hussey {\it et al.}, Rev. Sci. Inst. {\bf 76}, 053503 (2005).
\bibitem{Chen14} W. C. Chen {\it et al.}, J. Phys. Conf. Ser. {\bf 528}, 012014 (2014). 
\bibitem{Weiss14} R. Weiss (MIT), {\it Private Communication} (2014).
\bibitem{Snow09} W. M. Snow, Nucl. Inst. Meth. A {\bf 611}, 144 (2009).
\bibitem{Ageron89} P. Ageron {\it et al.}, Nucl. Inst. Meth. A {\bf 284}, 197 (1989).
\bibitem{Mocko13} M. Mocko and G. Muhrer, Nucl. Inst. Meth. A {\bf 704}, 27 (2013).
\bibitem{Nesvizhevsky08} V. Nesvizhevsky {\it et al}., Nucl. Inst. Meth. A {\bf 595}, 631 (2008).
\bibitem{Lychagin09el} E. Lychagin {\it et al}., Nucl. Inst. Meth. A {\bf 611}, 302 (2009). 
\bibitem{Lychagin09em} E. Lychagin {\it et al.}, Phys. Lett. B {\bf 679}, 186 (2009).
\bibitem{Swiss00} R. Maruyama {\it et al.}, Thin Solid Films {\bf 515}, 5704 (2007); T. Krist {\it et al.}, {\it Neutron Supermirror Development}, Modern Developments in X-Ray and Neutron Optics, 355, ISBN: 978-3-540-74560-0, Springer Berlin Heidelberg (2008); D. Yamazaki {\it et al.}, J. Phys. Conf. Ser. {\bf 251}, 012076 (2010); http://www.swissneutronics.ch/
\bibitem{Boni10} P. Boni, F. Grunauer and C. Schanzer, Nucl. Inst. Meth. A {\bf 624}, 162 (2010). 
\bibitem{Mcnp08} {\it MCNP-X Users Manual Version 2.6.0}, report LA-CP-07-1473, Los Alamos National Laboratory (2008). 
\bibitem{Kai05} T. Kai {\it et al.}, Nucl. Inst. Meth. A {\bf 550}, 329 (2005). 
\bibitem{Prosper} H. B. Prosper, Nucl. Inst. Meth. A {\bf 238}, 500 (1985); Phys. Rev. D {\bf 37}, 1153 (1988); Phys. Rev. D {\bf 38}, 3584 (1988).
\bibitem{Lone80} M. A. Lone, D. C. Santry, and W. M. Inglis, Nucl. Inst. Meth. {\bf 174}, 521 (1980).
\bibitem{MARS} N. V. Mokhov, The MARS Code System User's Guide, Version 13(95), Fermilab-FN-628, (1995); N. V. Mokhov, Recent MARS15 Developments: Nuclide Inventory, DPA and Gas Production, Fermilab-Conf-10-518-APC (2010); http://www-ap.fnal.gov/MARS/
\bibitem{Ridikas07} D. Ridikas {\it et al.}, Eur. Phys. J. A {\bf 32}, 1 (2007); D. Ridikas {\it et al.}, AIP Conf. Proc. {\bf 798}, 277 (2005).
\bibitem{Kronfeld13} A. S. Kronfeld {\it et al.}, FERMILAB-TM-2557, arXiv:1306.5009 [hep-ex].
\bibitem{Boldyrev12} A. S. Boldyrev {\it et al.}, Inst. and Exp. Techniques {\bf 55}, 323 (2012).
\bibitem{Vankooten13} R. J. Van Kooten (Indiana University), {\it Private Communication} (2013).  
\bibitem{Cabrera02} S. Cabrera {\it et al.}, Nucl. Inst. Meth. A (Proc. Suppl.) {\bf 494}, 416 (2002).
\bibitem{Mkrtchyan13} H. Mkrtchyan {\it et al}., Nucl. Inst. Meth. A {\bf 719}, 85 (2013).
\bibitem{Eskut97} E. Eskut {\it et al}., Nucl. Inst. Meth. A {\bf 401}, 7 (1997).
\bibitem{Mcfarland06} K. S. McFarland (MINER$\nu$A Collaboration), Nucl. Phys. B (Proc. Suppl.) {\bf 159}, 107 (2006).
\bibitem{Babura1993} D. Babura {\it et al.},  Nucl. Inst. Meth. A {\bf 332}, 444 (1993). 
\bibitem{Michael08} D. G. Michael {\it et al}., Nucl. Inst. Meth. A {\bf 596}, 190 (2008).
\bibitem{Fratina09} S. Fratina {\it et al}., ATL-INDET-PUB-2009-002, ATL-COM-INDET-2009-042, http://cds.cern.ch/record/1229213
\bibitem{Iljinov82} A. S. Iljinov, V. I. Nazaruk, and S. E. Chigrinov, Nucl. Phys. A {\bf 382}, 378 (1982).
\bibitem{Minor90} E. D. Minor {\it et al}.,  Z. Phys. A {\bf 336}, 461 (1990).
\bibitem{Hofmann90} P. Hofmann {\it et al}., Nucl. Phys. A {\bf 512}, 669 (1990).
\bibitem{Golubeva96} E. S. Golubeva, A. S. Ilinov and L. A. Kondratyuk, Proceedings of the International Workshop on Future Prospects of Baryon Instability Search in $p$ decay and $n - \bar n$ Oscillation Experiments {\bf C96-03-28}, ORNL-6910, 295 (1996).  
\bibitem{Golubeva92} E. S. Golubeva {\it et al}., Nucl. Phys. A {\bf 537}, 393 (1992).
\bibitem{Iljinov94} A. S. Iljinov, M. V. Kazarnovsky, E. Ya. Paryev, Intermediate Energy Nuclear Physics, CRC Press, 1994.
\bibitem{Botvina90} A. S. Botvina, A. S. Iljinov and I. N. Mishustin, Nucl. Phys. A {\bf 507}, 649 (1990).
\bibitem{Ghesquiere74} C. Ghesqui\`{e}re, Symp. on Antinucleon Nucleon Interactions, Liblice, CERN Yellow Report {\bf 74-18}, 436 (1974).
\bibitem{Minor89} E. D. Minor {\it et al}., Preprint PSU LEPS 89/15, Pennsylvania State University (1989).
\bibitem{Andreopoulos10} C. Andreopoulos {\it et al}., Nucl. Inst. Meth. A {\bf 614}, 87 (2010).
\bibitem{Geant13} S. Agostinelli {\it et al}., Nucl. Inst. Meth. A {\bf 506}, 250 (2003); J. Allison {\it et al}., IEEE Trans. Nucl. Sci. {\bf 53}, 270 (2006).
\bibitem{BaldoCeolin91} M. Baldo-Ceolin {\it et al}., IEEE Trans. Nuc. Sci. {\bf 38-2}, 471 (1991).
\bibitem{Iverson2014} E. B. Iverson, D. V. Baxter, G. Muhrer, S. Ansell, R. Dalgliesh, F. X. Gallmeier, H. Kaiser, and W. Lu, Nucl. Intr. Meth. A {\bf 762}, 31 (2014). 
%\bibitem{Baxter12} D. V. Baxter {\it et al}., Physics Procedia {\bf 26}, 153 (2012).
\bibitem{Cubitt10} R. Cubitt {\it et al}., Nucl. Inst. Meth. A {\bf 622}, 182 (2010).
\bibitem{Nesvishevsky082} V. V. Nesvishevsky {\it et al}., Nucl. Inst. Meth. A {\bf 595}, 631 (2008).
\bibitem{Nesvishevsky10} V. V. Nesvishevsky {\it et al}., Materials {\bf 3}, 1768 (2010).
\bibitem{Kerbikov03} B. O. Kerbikov, Phys. Atom. Nucl. {\bf 66}, 2178 (2003).
\bibitem{Kerbikov04} B. O. Kerbikov, A. E. Kudryavtsev, and V. A. Lensky, J. Exp. Theo. Phys. {\bf 98}, 417 (2004).
\bibitem{Yoshiki89} H. Yoshiki and R. Golub, Nucl. Phys. A {\bf 501}, 869 (1989).
\bibitem{Golub92} R. Golub and H. Yoshiki, Nucl. Phys. A {\bf 536}, 648 (1992).
\bibitem{Wang09} Z. Wang {\it et al.}, Nucl. Inst. Meth. A {\bf 605}, 430 (2009).
\bibitem{Engl2010} A. Engl {\it et al.}, Nucl. Inst. Meth. A {\bf 623}, 91 (2010).
\bibitem{Leuschner07} M. B. Leuschner {\it et al.}, Nucl. Inst. Meth. B {\bf 261}, 956 (2007).
\bibitem{Windsor81} C. G. Windsor, {\it Pulsed Neutron Scattering}, Halsted Press, New York (1981).
\bibitem{Difilippo91} F. C. Difilippo, Nucl. Sci. Eng. {\bf 107}, 82 (1991).
\bibitem {Lovesey86} S. W. Lovesey, {\it Theory of Neutron Scattering from Condensed Matter}, vol. 1, Oxford Science Publications, Oxford, UK (1986).
\bibitem{Duderstadt76} J. J. Duderstadt and L. J. Hamilton, {\it Nuclear Reactor Analysis}, Wiley (1976). 
\bibitem{Davison57} B. Davison and J. B. Sykes, {\it Neutron Transport Theory}, Clarendon Press, Oxford (1957). 
\bibitem{Foderaro71} A. Foderaro, {\it The Elements of Neutron Interaction Theory}, MIT Press (2003).
\bibitem{1991GolubBook} R. Golub, D. Richardson and S. K. Lamoreaux, {\it Ultra-Cold Neutrons}, IOP Publishing Ltd (1991).
\bibitem{Dingenen62} W. Van Dingenen, Nucl. Inst. Meth. {\bf 16}, 116 (1962). 
\bibitem{Butterworth57} I. Butterworth, P. A. Egenstaff, H. London {\it et al.}, Phil. Mag. {\bf 2}, 917 (1957).
\bibitem{Utsuro73} M. Utsuro, J. Nucl. Sci. and Tech. {\bf 10}, 428 (1973). 
\bibitem{Shin10} Y. Shin, W. M. Snow, C-Y Liu, C. M. Lavelle, and D. V. Baxter,  Nucl. Instr. Meth. A {\bf 620} 382 (2010).
\bibitem{Baxter05} D. V. Baxter {\it et al}., Nucl. Inst. Meth. B {\bf 241}, 209 (2005).
\bibitem{Lavelle08} C. M. Lavelle {\it et al}., Nucl. Inst. Meth. A {\bf 587}, 324 (2008).
\bibitem{Sears89} V. F. Sears, {\it Neutron Optics}, Oxford University Press, Oxford (1989). 
\bibitem{Peshkin71} M. Peshkin and G. R. Ringo, Am. J. Phys. {\bf 39}, 324 (1971). 
\bibitem{Aleksejev98} A. Aleksejev {\it et al.}, Z. Naturforsch. {\bf 53 }, 855 (1998).
\bibitem{Koester91} W. Koester, H. Rauch and E. Seymann, Atomic Data and Nucl. Data Tables {\bf 49}, 65 (1991). 
\bibitem{Rauch00b} H. Rauch and W. Waschkowski, {\it Landolt-B\"{o}rnstein: Low Energy Neutrons and their Interaction with Nuclei and Matter}, Springer, Berlin (2000).
\bibitem{Maier63} H. Maier-Leibnitz and T. Springer, Reactor Sci. and Technol. {\bf 17}, 217 (1963).

\bibliographystyle{apsrev4-1} \bibliography{nnbar/refs}

\bibliographystyle{apsrev4-1} \bibliography{nnbar/refs}
%% not want to use model1b-num-names.bst.





% \end{thebibliography}

\end{thebibliography}

\end{document}